\documentclass[11pt,fleqn]{article}
\usepackage{amsmath,amsthm,amsfonts,amssymb,mathrsfs,environ}
\usepackage{geometry}
\usepackage{bm}
\usepackage{bbm}
\usepackage{dsfont}
\usepackage{calrsfs}
\usepackage{rotating}
\usepackage{hyperref}
\usepackage{longtable}
\usepackage{url}
\usepackage{enumerate}
\usepackage[round,longnamesfirst]{natbib}
\usepackage{setspace}
\usepackage[tableposition=top,singlelinecheck=off,labelfont=bf]{caption}
\usepackage{tabularx}
\usepackage{graphicx}
\usepackage{subfigure}
\usepackage{verbatim}
\usepackage{color}
\usepackage{pdflscape}
\usepackage{afterpage}
\usepackage{caption}
\usepackage{breqn}
\usepackage{fixltx2e}
\usepackage{caption}
\usepackage[title]{appendix}
\usepackage{algorithm,algpseudocode}
\usepackage[normalem]{ulem}
\usepackage{xr}
\DeclareMathOperator*{\argmin}{arg~min}

\DeclareMathOperator{\E}{\mathbb{E}}

\DeclareMathOperator{\Prob}{\mathbb{P}}
\DeclareMathOperator{\sign}{\text{sign}}

\DeclareMathOperator{\rank}{\text{rank}}
\DeclareMathOperator{\diag}{\text{diag}}

\newcommand{\abs}[1]{\left\lvert#1\right\rvert}
\newcommand{\norm}[1]{\left\lVert#1\right\rVert}

\newcommand{\indep}{\perp \!\!\! \perp}
\newcommand{\bA}{\bm A}
\newcommand{\ba}{\bm a}
\newcommand{\bB}{\bm B}
\newcommand{\bb}{\bm b}
\newcommand{\bC}{\bm C}
\newcommand{\bc}{\bm c}
\newcommand{\bD}{\bm D}
\newcommand{\bd}{\bm d}
\newcommand{\bE}{\bm E}
\newcommand{\be}{\bm e}

\newcommand{\bof}{\bm f}

\newcommand{\bH}{\bm H}
\newcommand{\bh}{\bm h}
\newcommand{\bI}{\bm I}

\newcommand{\bM}{\bm M}

\newcommand{\bP}{\bm P}

\newcommand{\bQ}{\bm Q}

\newcommand{\bR}{\bm R}

\newcommand{\bS}{\bm S}
\newcommand{\bs}{\bm s}
\newcommand{\bT}{\bm T}
\newcommand{\bt}{\bm t}
\newcommand{\bU}{\bm U}
\newcommand{\bu}{\bm u}
\newcommand{\bV}{\bm V}
\newcommand{\bv}{\bm v}
\newcommand{\bW}{\bm W}
\newcommand{\bw}{\bm w}

\newcommand{\bx}{\bm x}

\newcommand{\by}{\bm y}
\newcommand{\bZ}{\bm Z}
\newcommand{\bz}{\bm z}

\newcommand{\balpha}{\bm \alpha}
\newcommand{\bbeta}{\bm \beta}
\newcommand{\bgamma}{\bm \gamma}
\newcommand{\bdelta}{\bm \delta}
\newcommand{\bepsilon}{\bm \epsilon}

\newcommand{\bzeta}{\bm \zeta}
\newcommand{\boeta}{\bm \eta}
\newcommand{\btheta}{\bm \theta}
\newcommand{\biota}{\bm \iota}

\newcommand{\blambda}{\bm \lambda}
\newcommand{\bmu}{\bm \mu}

\newcommand{\bpi}{\bm \pi}

\newcommand{\bsigma}{\bm \sigma}
\newcommand{\btau}{\bm \tau}

\newcommand{\bphi}{\bm \phi}

\newcommand{\bomega}{\bm \omega}
\newcommand{\bLambda}{\bm \varLambda}

\newcommand{\bPi}{\bm \varPi}
\newcommand{\bSigma}{\bm \varSigma}

\newcommand{\bPhi}{\bm \varPhi}

\newcommand{\bOmega}{\bm \varOmega}

\theoremstyle{definition}

\newtheorem{assumption}{Assumption}

\newtheorem{remark}{Remark}
\theoremstyle{plain}
\newtheorem{theorem}{Theorem}
\newtheorem{lemma}{Lemma}
\newtheorem{corollary}{Corollary}

\title{An Automated Approach Towards Sparse Single-Equation Cointegration Modelling%
\thanks{The first author was financially supported by the Netherlands Organization for Scientific Research (NWO) under grant number 452-17-010. Previous versions of this paper were presented at CFE-CM Statistics 2017, NESG 2018 and (EC)$^2$ 2018. We gratefully acknowledge the comments by participants at these conferences. In addition, we thank the editor and two anonymous referees as well as Robert Ad\'amek, Alain Hecq, Luca Margaritella, Alexei Onatski, Hanno Reuvers, Sean Telg, Ines Wilms and Qiwei Yao for valuable comments and feedback, and Caterina Schiavoni for help with the data collection. All remaining errors are our own. Corresponding author:
Etienne Wijler. Department of Quantitative Economics, Maastricht University, P.O. Box 616, 6200 MD Maastricht, The Netherlands. E-mail: \href{mailto:e.wijler@maastrichtuniversity.nl}{\textcolor{blue}{e.wijler@maastrichtuniversity.nl}}}}
\author{Stephan Smeekes \and Etienne Wijler}
\date{Maastricht University\\
Department of Quantitative Economics\\
\today}

\begin{document}

\maketitle

\begin{abstract}
In this paper we propose the Single-equation Penalized Error Correction Selector (SPECS) as an automated estimation procedure for dynamic single-equation models with a large number of potentially (co)integrated variables. By extending the classical single-equation error correction model, SPECS enables the researcher to model large cointegrated datasets without necessitating any form of pre-testing for the order of integration or cointegrating rank. Under an asymptotic regime in which  both the number of parameters and time series observations jointly diverge to infinity, we show that SPECS is able to consistently estimate an appropriate linear combination of the cointegrating vectors that may occur in the underlying DGP. In addition, SPECS is shown to enable the correct recovery of sparsity patterns in the parameter space and to posses the same limiting distribution as the OLS oracle procedure. A simulation study shows strong selective capabilities, as well as superior predictive performance in the context of nowcasting compared to high-dimensional models that ignore cointegration. An empirical application to nowcasting Dutch unemployment rates using Google Trends confirms the strong practical performance of our procedure.
\\


\textit{JEL-Codes}: C32, C52, C55
\end{abstract}

\onehalfspacing
\section{Introduction}

In this paper we propose the Single-equation Penalized Error Correction Selector (SPECS) as a tool to perform automated modelling of a potentially large number of time series of unknown order of integration. In many economic applications, datasets will contain possibly (co)integrated time series, which has to be taken into account in the statistical analysis. Traditional approaches include modelling the full system of time series as a vector error correction model (VECM), estimated by methods such as maximum likelihood estimation \citep{Johansen1995}, or transforming all variables to stationarity before performing further analysis. However, both methods have considerable drawbacks when the dimension of the dataset increases. 

While the VECM approach allows for flexible modelling of potentially cointegrated series, these estimators suffer from the curse of dimensionality due to the large number of parameters to estimate. In practice they therefore quickly become difficult to interpret and computationally intractable on even moderately sized datasets. As such, to reliably apply such full-system estimators requires non-trivial a priori choices on the relevance of specific variables to keep the dimension manageable. Moreover, often one only has a single variable of interest, and estimating the parameter-heavy full system is not necessary. 

On the other hand, the alternative strategy of prior transformations to stationarity is more easily compatible with single variables of interest and larger dimensions, but requires either a priori knowledge of the order of integration of individual variables, or pre-testing for unit roots, which is prone to errors in particular if the number of variables is large \citep[cf.][]{Smeekes2020}. Additionally, this approach ignores the presence of cointegration among the variables, which may have detrimental effects on the subsequent analysis. 

SPECS is a form of penalized regression designed to sparsely estimate a conditional error correction model (CECM). We demonstrate that SPECS possesses the oracle property as defined in \citet{Fan2001}; in particular, SPECS simultaneously allows for consistent estimation of the non-zero coefficients and the correct recovery of sparsity patterns in the single-equation model. It therefore provides a fully data-driven way of selecting the relevant variables from a potentially large dataset of (co)integrated time series. Moreover, due to the flexible specification of the single-equation model, SPECS is able to take into account cointegration in the dataset without requiring any form of pre-testing for unit roots or testing for the cointegrating rank, and can thus be applied ``as is'' to any dataset containing an (unknown) mix of stationary and integrated time series. As a companion to this paper, an \textit{R} package is made available that implements a fast and easy-to-interpret algorithm for SPECS estimation, and provides immediate access to the dataset used in the empirical application.\footnote{\href{https://github.com/wijler/specs}{\textcolor{blue}{https://github.com/wijler/specs}}}.

Single-equation error correction models are frequently employed in tests for cointegration \citep[e.g.][]{Engle1987,Phillips1990b,Boswijk1994,Banerjee1998} as well as in forecasting applications \citep[e.g.][]{EngleY1987,Chou1996}, but require a weak exogeneity assumption for asymptotically efficient inference \citep{Johansen1992}. Weak exogeneity entails the existence of a single cointegrating vector that only appears in the marginal equation for the variable of interest. If this assumption holds, our procedure can be interpreted as an alternative to cointegration testing in the ECM framework \citep{Boswijk1994,Palm2010}. However, weak exogeneity may not be realistic in large datasets and we provide detailed illustrations of the implications of failure of this assumption and demonstrate that absent of weak exogeneity our procedure consistently estimates a linear combination of the true cointegrating vectors. While this impedes inference on the cointegrating relations, when the main aim of the model is nowcasting or forecasting, our procedure remains theoretically justifiable and provides empirical researchers with a simple and powerful tool for automated analysis of high-dimensional non-stationary datasets. In addition, for modeling a single variable of interest using a large set of potential regressors, SPECS provides a variable selection mechanism, allowing the researcher to discard variables that are irrelevant for this particular analysis. Our simulation results demonstrate strong selective capabilities in both low and high dimensions. Furthermore, a simulated nowcasting application highlights the importance of incorporating cointegration in the data as our proposed estimators obtain higher nowcast accuracies in comparison to a penalized autoregressive distributed lag (ADL) model. This finding is confirmed in an empirical application, where SPECS is employed to nowcast Dutch unemployment rates with the use of a dataset containing Google Trends series.

The use of penalized regression in time series analysis has gained in popularity, with a wide range of variants showing promising performance in applications \citep[see][for a recent overview]{Smeekes2018b}. Recent literature has also seen the development of methods for analyzing high-dimensional (co)integrated time series.

\citet{Kock2016} proposes the adaptive lasso to estimate an augmented Dickey-Fuller regression. While this univariate model is inherently different from ours, it provides an insightful demonstration of how the lasso may be used as an alternative to testing for non-stationarity, paralleling our suggestion to consider SPECS as an alternative for cointegration testing under the assumption of weak exogeneity.

For VECM systems, \citet{Wilms2016} propose a penalized maximum likelihood approach, with shrinkage performed on the cointegrating vectors, the coefficients regulating the short-run dynamics and the covariance matrix. While their method is shown to obtain forecast gains relative to the traditional Johansen method, no theoretical results are provided. \citet{Liao2015} provide an automated method of joint rank selection and parameter estimation with the use of an adaptive penalty and derive oracle properties in a fixed-dimensional framework. Next to this theoretical limitation on its applicability to large datasets, practical implementation is further complicated due to reliance on the eigenvalue decomposition of an asymmetric matrix, which introduces complex values into the corresponding objective function. As noted by \citet[][p. 424]{Liang2019}, this results in a non-standard harmonic function optimization problem. \citet{Liang2019} propose joint 
parameter estimation and rank determination by employing a penalty that makes use of the $QR$-decomposition of the long-run coefficient matrix. This method possesses oracle-like properties under a high-dimensional asymptotic regime, but it requires the availability of an initial OLS estimator, thereby preventing applications on datasets in which the number of variables exceeds, or is close to, the number of available time series observations. Additionally, estimation of the long-run and short-run dynamics is performed sequentially rather than simultaneously, necessitating a two-step procedure.

In a single-equation setting, \citet{Lee2018} derive fixed-dimensional oracle properties for the adaptive lasso applied to predictive regressions where the regressors are allowed to be of mixed orders of integration. However, as a consequence of their model formulation in which all variables enter in levels, their estimator appears to be susceptible to spurious regression when the regressors are not cointegrated.

Finally, outside the penalized regression framework, \citet{Zhang2018b} propose an eigenvalue decomposition to estimate the cointegrating space in the presence of any integer and fractional order of integration of the variables. However, the estimation procedure proposed by \citeauthor{Zhang2018b} does not perform variable selection, nor does it provide explicit estimates of the transient dynamics in a VECM. \citet{Onatski2019} develop a novel inference procedure for the cointegrating rank in high dimensions. Similar to the Johansen procedure, their test is based on the squared canonical correlations, for which they derive the limit spectral distribution under joint asymptotics with the use of arguments from random matrix theory.

Our proposed method provides several contributions to this existing literature. First, our theoretical results are derived in a high-dimensional framework where the number of parameters is allowed to grow with the sample size. This requires non-standard theoretical results on bounds of the smallest eigenvalue of a matrix of (co)integrated regressors, similar to those in \citet{Zhang2018b}, which are further developed in this paper. Second, unlike many of the penalized regression methods surveyed above, the practical implementation of SPECS is straightforward for large datasets, including cases where the number of parameters is larger than the time dimension. Third, our method completely removes the need for pre-testing for the order of integration or cointegrating rank, and is not sensitive to spurious regression. Fourth, to the best of our knowledge, our paper is the first to explicitly allow for the presence of deterministic components in the theory, a crucial feature for many applications.

The paper is structured as follows. In Section \ref{sec:model} we discuss the data generating process. Section \ref{Sec:SPECS} describes the SPECS estimator. The main theoretical results of the paper are presented in Section \ref{Sec:Theory}. Section \ref{sec: simulations} contains several simulation studies, followed by an empirical application in Section \ref{Sec:application}. We conclude in Section \ref{Sec:Conclusion}. The main proofs and preliminary lemmas needed for them are contained in Appendix \ref{App:main}, while Appendix \ref{App:eigenvalues} contains results on minimum eigenvalue bounds. Finally, Appendix \ref{App:supp} contains supplementary material on proofs of preliminary lemmas and additional theorems, as well as further details on the empirical application.

A word on notation. For any an $N$-dimensional vector $\bx$, $\norm{\bx}_p = \left(\sum_{i=1}^N x_i^p\right)^{1/p}$ denotes the $\ell_p$-norm, while for any matrix $\bD$ with $N$ columns, $\norm{\bD}_p = \underset{\bx \in \mathbb{R}^N}{\text{max}} \frac{\norm{\bD x}_p}{\norm{\bx}_p}$ is the corresponding induced norm and $\norm{\bD}_F$ denotes the Frobenius norm. For an index set $S \subset \{1, \ldots, N\}$, let $\bx_{S}$ be the vector containing the elements of $\bx$ corresponding to $S$. Similarly, for a matrix $\bD$ with $N$ rows, $\bD_{S}$ is the sub-matrix containing the rows of $\bD$ indexed by $S$. The orthogonal complement of $\bD$ is denoted by $\bD_\perp$, such that $\bD_\perp^\prime \bD = \bm{0}$. When $\bD$ is a square matrix, we denote its $N$ ordered eigenvalues by $\lambda_1(\bD) \geq \ldots \geq \lambda_N(\bD)$ and we use $\bD \succ 0$ to denote that the matrix is positive definite. We use $\biota_N$ to denote a vector of ones of length $N$ and $\bI_N$ to denote the $N$-dimensional identity matrix. We use $\overset{p}{\to}$ ($\overset{d}{\to}$) to denote convergence in probability (distribution) and $\overset{d}{=}$ denotes equivalence in distribution. Finally, we frequently make use of an arbitrary positive and finite constant $K$ whose value may change throughout the paper, but is always independent of the time and cross-sectional dimensions.

\section{The High-Dimensional Error Correction Model}\label{sec:model}

In this section we first discuss the data generating process for the vector time series along with the assumptions made. Next we transform the multivariate model to a single equation describing our variable of interest.

\subsection{Data Generating Process}
Assume one is interested in modelling a single variable of interest, say $y_t$, based on an $N$-dimensional time series $\bz_t = (y_t,\bx_t^\prime)$ observed at $t=1,\ldots,T$. Let $\bz_t$ be described by 
\begin{equation}\label{eq:z1}
\bz_t = \bmu + \btau t + \bzeta_t,
\end{equation}
with the stochastic component given by
\begin{equation}\label{eq:z2}
\begin{split}
\Delta\bzeta_t &= \bA\bB^\prime\bzeta_{t-1} + \sum_{j=1}^p \bPhi_j \Delta \bzeta_{t-j} + \bepsilon_t,
\end{split}
\end{equation}
where $\bA$ and $\bB$ are $(N \times r)$-dimensional matrices containing the adjustment rates and cointegrating vectors, respectively. The innovations $\bepsilon_t = (\epsilon_{1,t},\bepsilon_{2,t}^\prime)^\prime$ satisfy the following assumptions:
\begin{assumption}\label{Ass:moments}
The sequence of innovations $\lbrace\bepsilon_t\rbrace_{t\geq 1}$ is an $N$-dimensional martingale difference sequence (m.d.s.)  with $\E{(\bepsilon_t \bepsilon_t^\prime)}=\bSigma_\epsilon$. Furthermore, we assume that
\begin{enumerate}[(1)]
\item There exists an $m > 2$, such that $\max_{1 \leq i \leq N, 1 \leq t \leq T} \E\abs{\epsilon_{i,t}}^{2m} \leq K_m$, and
\item There exist constants $\phi_\min,\phi_\max > 0$, such that $\phi_\min \leq \lambda_\min\left(\bSigma_\epsilon\right) < \lambda_\max\left(\bSigma_\epsilon\right) \leq \phi_\max$.
\end{enumerate}
\end{assumption}
This assumption implies that $\bepsilon_t$ is an martingale difference sequence with at least (a bit more than) four moments existing. The eigenvalue bounds in the second part place some restrictions on the dependence among the elements of $\bepsilon_t$, ruling out for instance a strong common factor affecting all errors. However, a wide range of contemporaneous dependence structures, such as spatial dependence, is still allowed.

The model can be rewritten into a VECM form by substituting \eqref{eq:z1} into \eqref{eq:z2} to obtain
\begin{equation}\label{eq:VECM}
\begin{split}
\Delta \bz_t &= \bA\bB^\prime \left(\bz_{t-1}-\bmu-\btau (t-1)\right) + \tau^* + \sum_{j=1}^p \bPhi_j\Delta \bz_{t-j} + \bepsilon_t,\\
\end{split}
\end{equation}
where $\tau^* = (I-\sum_{j=1}^p\bPhi_j)\btau$. From this representation, it can directly be observed that the presence of a constant in \eqref{eq:z1} results in a constant within the cointegrating relationship if $\bB^\prime \bmu \neq \bm{0}$. Furthermore, the linear trend in \eqref{eq:z1} appears as a constant in the differenced series and may additionally appear as a trend within the cointegrating vector if $\bB^\prime \btau \neq \bm{0}$, the latter implying that the equilibrium error $\bB^\prime \bz_t$ is a trend stationary process. 

The following assumption asserts that the process is (at most) I(1), and the Granger Representation Theorem \citep[e.g.][p. 49]{Johansen1995} can be applied.
\begin{assumption}\label{Ass: GRT}
Define $\bA(z):= (1-z)\bI_N-\bA\bB^\prime z - \sum_{j=1}^p \bPhi_j (1-z) z^j$.
\begin{enumerate}[(1)]
\item The determinantal equation $\abs{\bA(z)}$ has all roots on or outside the unit circle.
\item $\bA$ and $\bB$ are $N \times r$ matrices with  $1 \leq r \leq N$ and $\rank(\bA) = \rank(\bB) = r$.
\item The $\left((N-r) \times (N-r)\right)$ matrix $\bA_\perp^\prime \left(I_N - \sum_{j=1}^p \bPhi_j\right)\bB_\perp$ is invertible. 
\end{enumerate}
\end{assumption}
 
Assumption \ref{Ass: GRT} enables \eqref{eq:VECM} to be written as a vector moving average (VMA) process
\begin{equation}\label{eq:GRT}
\bz_t = \bC\bs_t + \bmu + \btau t + \bC(L)\bepsilon_t + \bC\bz_0,
\end{equation}
where $\bC=\bB_\perp\left(\bA_\perp^\prime \left(\bI_N - \sum_{j=1}^p\bPhi_j\right) \bB_\perp\right)^{-1}\bA_\perp^\prime$, $\bs_t = \sum_{s=1}^t \bepsilon_s$, $\bC(L)\bepsilon_t$ is a stationary linear process and $\bz_0$ are initial values. Without loss of generality, we assume henceforth that $\bz_0 = \bm{0}$. 

We need a further restriction on the dependence in the VMA representation in the form of the following assumption, which ensures norm-summability of the coefficients in the Beveridge-Nelson decomposition.
\begin{assumption}\label{Ass:Dependence}
There exists a $K<\infty$ such that $\bC$ in \eqref{eq:GRT} satisfies $\norm{\bC}_\infty \leq K$. In addition, the matrix lag polynomial $\bC(L)$ is given by $\bC(z) = \sum_{l=0}^\infty \bC_lz^l$ and satisfies $\sum_{l=0}^\infty l\norm{\bC_l}_\infty \leq K$.
\end{assumption}

\subsection{Single-Equation Representation}\label{sec:ser}

The number of parameters to estimate in \eqref{eq:VECM} is at least $2Nr + N^2p$, such that the system quickly grows too large to accurately estimate based on traditional methods. As we assume a single variable $y_t$ is of interest, we therefore instead consider the lighter parameterized single-equation model for $y_t$. To ensure that the variables modelling the variation in $y_t$ remain exogenous, we orthogonalize the errors driving the single-equation model, say $\epsilon_{y,t}$, from the errors driving the marginal equations of the endogenous variables $\bx_t$. This is achieved by decomposing $\epsilon_{1,t}$ into its best linear prediction based on $\bepsilon_{2,t}$ and the corresponding orthogonal prediction error. To this end, partition the covariance matrix of $\bepsilon_t$ as
\begin{equation}\label{eq:covariance partition}
\bSigma_\epsilon = \begin{bmatrix}
\E(\epsilon_{1,t})^2 & \E(\epsilon_{1,t}\bepsilon_{2,t}^\prime) \\
\E(\epsilon_{1,t}\bepsilon_{2,t}) & \E(\bepsilon_{2,t}\bepsilon_{2,t}^\prime) \\
\end{bmatrix} = \begin{bmatrix}
\sigma_{11} & \bsigma_{21}^\prime \\
\bsigma_{21} & \bSigma_{22} \\
\end{bmatrix},
\end{equation}
such that we obtain
\begin{equation}\label{eq:error decomposition}
\begin{split}
\epsilon_{1,t} &= (0,\bsigma_{21}^\prime\bSigma_{22}^{-1}) \bepsilon_t + \left(1, - \bsigma_{21}^\prime\bSigma_{22}^{-1}\right)\bepsilon_t = \hat{\epsilon}_{1,t} + \epsilon_{y,t}.
\end{split}
\end{equation}
Define $\pi_0 = \bSigma_{22}^{-1}\bsigma_{21}$. Then, writing out \eqref{eq:error decomposition} in terms of the observable time series results in the single-equation model
\begin{equation}\label{eq:CECM}
\begin{split}
\Delta \by_t &=  \left(1,-\bpi_0^\prime \right)\left(\bA\bB^\prime (\bz_{t-1} - \bmu - \btau(t-1)) + \btau^* +  \sum_{j=1}^p \bPhi_j^\prime \Delta \bz_{t-j} \right) + \bpi_0^\prime \Delta \bx_t  + \epsilon_{y,t}\\
&= \bdelta^\prime \bz_{t-1} + \bpi^\prime \bw_t + \mu_0 + \tau_0(t-1) + \epsilon_{y,t},
\end{split}
\end{equation}
where $\bdelta^\prime = \left(1,-\bpi_0^\prime \right)\bA\bB^\prime$, $\bpi = (\bpi_0^\prime,\ldots,\bpi_p^\prime)^\prime$ with $\bpi_j^\prime = (1,-\bpi_0^\prime)\bPhi_j$ for $j=1,\ldots,p$, $\mu_0 = (1,-\bpi_0^\prime)\left(\bA\bB^\prime\bmu + \btau^*\right)$ and $\tau_0 = (1,-\bpi_0^\prime)\btau^*$. Note that $\bdelta$ is a vector of length $N$, whereas $\bpi$ is a vector of length $M = N(p+1)-1$. Additionally, $\bw_t=(\Delta \bx_t^\prime,\Delta \bz_{t-1}^\prime,\ldots,\Delta \bz_{t-p}^\prime)^\prime$ and $\epsilon_{y,t} = (1 - \bpi_0^\prime)\bepsilon_t $. Finally, we write the single-equation model in matrix notation as
\begin{equation}\label{eq:CECM_matrix}
\begin{split}
\Delta \by &= \bZ_{-1}\delta + \bW\pi + \biota_T \mu_0 + \bt\tau_0 + \bepsilon_y = \bV\bgamma + \bD\btheta + \bepsilon_y,
\end{split}
\end{equation}
where $\bZ_{-1} = (\bz_0,\ldots,\bz_{T-1})^\prime$, $\bW = (\bw_t,\ldots,\bw_T)^\prime$, $\bt = (0,\ldots,T-1)^\prime$, $\bV=(\bZ_{-1},\bW)$, $\bD = (\biota_T,\bt)$, $\bgamma = (\bdelta^\prime,\bpi^\prime)^\prime$ and $\btheta = (\mu_0,\tau_0)^\prime$.

\begin{remark}
The single-equation model may similarly be derived under the assumption of normal errors. In this framework, $\epsilon_{y,t}$ has the conditional normal distribution from which \eqref{eq:CECM} can be obtained \citep[cf.][]{Boswijk1994}. A benefit of assuming normality is that, under the additional assumption of weak exogeneity, the OLS estimates based on \eqref{eq:CECM} are optimal in the mean-squared sense. However, the assumption of normality is unnecessarily restrictive when the, perhaps overly, ambitious goal of complete and correct specification is abandoned.
\end{remark}

\begin{remark}
An additional benefit of the conditional error-correction model, as opposed to the predictive regressions specified in levels considered in \citet{Lee2018}, is that the former avoids spurious regression. In the case where all variables in $\bz_t$ are integrated of order one and independent of one another, the left-hand side of \eqref{eq:CECM_matrix} would remain stationary. Intuitively, any ``best fitting'' linear combination between the stationary component $\Delta y_t$ and $(\bz_{t-1}^\prime,\bw_t^\prime)^\prime$ would seek to minimize the contribution of the variables in $\bz_t$, as their stochastically trending nature substantially inflates the fitting error. This behaviour is well-documented for the fixed-dimensional OLS estimator -- cf. \citet[][A.9]{Boswijk1994} in which $\hat{\bdelta}_{OLS}$ turns out to be superconsistent -- and carries over to SPECS in high-dimensions.
\end{remark}

In general, the implied cointegrating vector $\bdelta$ in the single-equation model for $y_t$ contains a linear combination of the cointegrating vectors in $\bB$ with their weights being given by $\left(1,-\bpi_0^\prime \right)\bA$. Since the marginal equations of $\bx_t$ contain information about the cointegrating relationship, efficient estimation within the single-equation model is only attained under an assumption of weak exogeneity. \citet{Johansen1992} shows that sufficient conditions for weak exogeneity to hold are (i) normality of $\bepsilon_t$, (ii) $\rank(\bA \bB^\prime) = 1$, i.e. there is a single cointegrating $N$-dimensional cointegrating vector $\bbeta$, and (iii) the vector of adjustment rates takes on the form $\balpha = (\alpha_1,\bm{0}^\prime)^\prime$. However, these conditions are rather restrictive when considering high-dimensional economic datasets that are likely to possess multiple cointegrating relationships and complex covariance structures across the errors. Therefore, we opt to derive our results without assuming weak exogeneity, while acknowledging that direct interpretation of the estimated cointegrating vector will only be valid in the presence of weak exogeneity. Furthermore, we believe that whether the potential loss of asymptotic efficiency in our more parsimonious single-equation model translates to inferior performance in finite samples ultimately remains an empirical question.

As we consider sparse estimation of this single-equation model, let us briefly touch upon the required sparsity. For measuring the sparsity, we work directly in the single-equation representation.\footnote{In absence of weak exogeneity, it may not be directly obvious how we obtain a sparse single-equation model from the VECM. We therefore provide a more detailed discussion of the interpretation of sparsity absent of weak exogeneity in Section \ref{Sec:Sparsity}. In this section we just take the single-equation model directly as starting point.} Let $S_\delta = \lbrace i \vert \delta_i \neq 0\rbrace$ denote the index set of the non-zero elements in $\bdelta$, with its cardinality denoted by $ \abs{S_\delta}$, and let $S_{\pi}$ be defined accordingly for $\bpi$. In addition, let $r^*$ denote the dimension of the cointegration space of $\bz_{S_\delta,t}$, i.e. the number of independent linear stationary combinations of $\bz_{S_\delta,t}$ (cf. Remark \ref{Rem:r}), and define $s_\delta = \abs{S_\delta} - r^*$ and $s_\pi = \abs{S_\pi}+r^*$ as the number of ``effective'' relevant non-stationary and stationary variables, respectively. Our estimation goal will then be to obtain estimates of $S_\delta$ and $S_{\pi}$, as well as estimate $\bdelta_{S_\delta}$ and $\bpi_{S_\pi}$. To obtain consistency, we need the following assumptions on the amount of sparsity.
\begin{assumption}\label{ass:sparsity}
Assume that
(1) $s_\delta = o(T^{1/4})$; (2) $s_\pi = o(\sqrt{T})$ and (3) $\max\{s_\delta, \sqrt{s_\pi}\} = o(\gamma_\min \sqrt{T})$, where $\gamma_{\min} = \min\{\abs{\gamma_i}: \gamma_i \neq 0\}$.
\end{assumption}
Parts (1) and (2) put restrictions on how fast the number of relevant parameters is allowed to grow. The ``effective'' number of relevant stationary variables ($s_\pi$) is allowed to grow faster than the ``effective'' number of integrated variables ($s_\delta$), as a result of the collinearity induced by the stochastic trends (cf. Remark \ref{Rem:min_eig_zero}). Part (3) puts an additional restriction on the number of relevant coefficients as a function of the smallest non-zero coefficient. Clearly, if all coefficients are assumed to be fixed, (3) is not binding. In fact, one can allow $\gamma_\min$ to shrink at a rate up to $T^{-1/4}$ before it becomes binding. This assumption may therefore be interpreted as determining the fastest rate at which the population coefficients are allowed to decrease, as a function of $T$, $s_\delta$ and $s_\pi$, to still ensure it can be consistently picked up by our estimation method.

\subsection{Rotations and Bounds on Eigenvalues} \label{sec:rot}

Bounds on eigenvalues play a crucial role in establishing consistency properties of lasso-type penalized regression methods. However, due the mixed integrated nature of our data, where parts of the regressors are stationary, and other parts are only stationary after rotation, the object of our assumptions is not the sample covariance matrix directly, but instead a carefully transformed version. Under Assumptions \ref{Ass:moments}-\ref{Ass:Dependence}, it is then possible to ensure eigenvalue conditions on the sample covariance matrices. Before we can state the assumption, we must therefore establish some further notation and rotations to be used later.

Let $\bgamma = (\bdelta^\prime, \bpi^\prime)^\prime$ and $S_{\gamma}$ its active set. Without loss of generality, we partition the data matrix as $\bV=(\bV_{S_\gamma},\bV_{S_\gamma^c})$, with $\bV_{S_\gamma} = (\bZ_{-1,S_\delta},\bW_{S_\pi})$ representing the time series carrying non-zero coefficients in the population single-equation model, henceforth referred to as the set of relevant variables. In the presence of cointegration, it follows from \eqref{eq:GRT} that the relevant lagged levels can be written as
\begin{equation}\label{eq:GRT_relevant}
\begin{split}
\bz_{S_\delta,t} &= \bC_{S_\delta}\bs_t + \bmu_{S_\delta} + \btau_{S_\delta}t + \bu_{S_\delta,t},\quad
\bC_{S_\delta} = \bB_{\perp,S_\delta}\left(\bA_\perp^\prime \left(\bI_N - \sum_{j=1}^p\bPhi_j\right) \bB_\perp\right)^{-1}\bA_\perp^\prime 
\end{split}
\end{equation}
where $\bB_{\perp,S_\delta}$ is an $(\abs{S_\delta} \times (N-r))$-dimensional matrix containing the rows of $\bB_\perp$ indexed by $S_\delta$ and $\bu_{S_\delta,t} = \bC_{S_\delta}(L)\bepsilon_t$. The left null space of $\bB_{\perp,S_\delta}$, defined as $\bB^* = \left\lbrace \bx \in \mathbb{R}^{\abs{S_\delta}} \vert \ \bB_{\perp,S_\delta}^\prime \bx = \bm{0}\right\rbrace$, contains the linear combinations that convert $\bz_{S_\delta,t}$ to a stationary process. Accordingly, we also refer to this null space as the cointegrating space of $\bz_{S_\delta,t}$. By construction, $\bdelta_{S_\delta} \in \bB^*$, such that this cointegrating space is non-empty whenever $\bdelta \neq \bm{0}$. In this case, we define $\bB_{S_\delta}$ as a $(\abs{S_\delta} \times r^*)$-dimensional basis matrix of $\bB^*$, with $r^* \leq \abs{S_\delta}$ representing the dimension of the cointegrating space.\footnote{The matrix $\bB_{S_\delta}$ is not uniquely defined. However, in most instances, including those contained in the current work, identification of the span of $\bB_{S_\delta}$ is sufficient.}

Similarly, we define $\bB_{S_\delta,\perp}$ as a basis matrix of the left null-space of $\bB_{S_\delta}$, i.e. a $\left(\abs{S_\delta}\times(\abs{S_\delta}-r^*)\right)$-dimensional matrix of full column rank with the property that $\bB_{S_\delta,\perp}^\prime \bB_{S_\delta} = \bm{0}$. Then, we are able to define a $\bQ$-transformation that decomposes the reduced system into a stationary and non-stationary contribution as
\begin{equation}\label{eq:Q}
\begin{split}
\bQ &= \begin{bmatrix}
\bB_{S_\delta}^\prime & \bm{0}\\
\bm{0} & \bI_{\abs{S_\pi}}\\
\bB_{S_\delta,\perp}^\prime & \bm{0}
\end{bmatrix}, \quad \bQ^{-1} = \begin{bmatrix}
\bB_{S_\delta}\left(\bB_{S_\delta}^\prime\bB_{S_\delta}\right)^{-1} & \bm{0} & \bB_{S_\delta,\perp}\left(\bB_{S_\delta,\perp}^\prime \bB_{S_\delta,\perp}\right)^{-1}\\
\bm{0} & \bI_{\abs{S_\pi}} & 0
\end{bmatrix}.
\end{split}
\end{equation}
For the case $\bdelta = \bm{0}$, we define $\bQ = \bI_{\abs{S_\pi}}$. Post-multiplication of the data matrix by $\bQ^\prime$ gives
\begin{equation}\label{eq:Q-transformed}
\bV_{S_\gamma}\bQ^\prime = \begin{bmatrix}
\bZ_{-1,S_\delta}\bB_{S_\delta} & \bW_{S_\pi} & \bZ_{-1,S_\delta}\bB_{S_\delta,\perp}
\end{bmatrix}
\end{equation}
which we refer to as the $\bQ$-transformed version of $\bV_{S_\gamma}$. The first $s_\pi = \abs{S_\pi} + r^*$ columns of \eqref{eq:Q-transformed}, corresponding to $(\bZ_{-1,S_\delta}\bB_{S_\delta},\bW_{S_\pi})$, contain independent stationary linear combinations of the variables that are relevant to $\Delta y_t$ in the single-equation model. The remaining $s_\delta = \abs{S_\delta}-r^*$ columns, given by $\bZ_{-1,S_\delta}\bB_{S_\delta,\perp}$, contain all linearly independent combinations that are integrated of order one.

\begin{remark}\label{Rem:r}
We may interpret $r^*$ as the ``effective'' cointegration rank, where ``effective'' relates to variable of interest $y_t$. Essentially, we remove all variables not relevant to $y_t$ in the long-run ($S_\delta^c$) and then reconstruct a VECM from the remaining variables, which now has rank $r^*$.\end{remark}

Finally, we construct a transformed version of the sample covariance matrix based on $\bV_{S_\gamma}$, which plays a crucial role in the development of our theory. First, to regress out the deterministic components of the observed time series in \eqref{eq:VECM}, we define the matrix $\bM = \bI_T - \bD\left(\bD^\prime\bD\right)^{-1}\bD^\prime$.\footnote{Note that $\bD$ may vary depending on the deterministic specification of the model; setting $\bD = (\biota_T,\bt)$ allows for both a non-zero constant and linear trend, while simply setting $\bM = \bI_T$ may be desired (although not required) when it is believed that $\bmu=\btau=\bm{0}$.} Then, after rotating by $\bQ$ and regressing out the deterministic components by $\bM$, the stationary and non-stationary components are scaled via the matrix $\bS_T = \diag(\sqrt{T}\bI_{s_\pi},\frac{T}{\sqrt{s_\delta}}\bI_{s_\delta})$. Hence, our transformed sample covariance matrix is defined as
\begin{align}
\label{eq:Sigma_hat}
&\hat{\bSigma} = \bS_T^{-1}\bQ\bV_{S_\gamma}^\prime\bM\bV_{S_\gamma}\bQ^\prime\bS_T^{-1} = \begin{bmatrix}
\hat{\bSigma}_{11}  & \hat{\bSigma}_{12}\\
\hat{\bSigma}_{21} & \hat{\bSigma}_{22}
\end{bmatrix},\\
&\text{with} \quad
\label{eq:Sigma_11}
\hat{\bSigma}_{11} = \frac{1}{T}\begin{bmatrix}
\bB_{S_\delta}^\prime\bZ_{-1,S_\delta}^\prime \bM \bZ_{-1,S_\delta}\bB_{S_\delta} & \bB_{S_\delta}^\prime\bZ_{-1,S_\delta}^\prime \bM \bW_{S_\pi}\\
\bW_{S_\pi}^\prime \bM \bZ_{-1,S_\delta}\bB_{S_\delta} & \bW_{S_\pi}^\prime \bM \bW_{S_\pi}
\end{bmatrix}
\end{align}
and $\hat{\bSigma}_{22} = \frac{s_\delta}{T^2}\bB_{S_\delta,\perp}^\prime\bZ_{-1,S_\delta}^\prime\bM\bZ_{-1,S_\delta}\bB_{S_\delta,\perp}$.
We can now state the eigenvalue assumptions.
\begin{assumption}\label{Ass:eigenvalues}
Assume that, on a set with probability converging to 1 as $T,N,p \to \infty$, there exists a constant $\phi>0$, such that 
$\underset{\bx \in \mathbb{R}^{s_\pi}}{\text{inf}} \frac{\bx^\prime \hat{\bSigma}_{11}\bx}{\bx^\prime \bx} \geq \phi$ 
and
$\underset{\bx \in \mathbb{R}^{s_\delta}}{\text{inf}} \frac{\bx^\prime \hat{\bSigma}_{22} \bx}{\bx^\prime \bx} \geq \phi$.
\end{assumption}
The first part of Assumption \ref{Ass:eigenvalues} applies to stationary data and is known to hold when the minimum eigenvalue of the corresponding population covariance matrix is bounded away from zero \citep[e.g.][Section B.2]{Medeiros2016}. The second part, however, applies to integrated variables and requires arguments that are unique to the non-stationary setting. In particular, we note the necessity of applying a scaling by $\frac{s_\delta}{T^2}$, rather than the usual $\frac{1}{T^2}$ one may expect from the fixed-dimensional literature, cf. Remark \ref{Rem:min_eig_zero}. In Appendix \ref{App:eigenvalues}, we show several cases under which Assumption \ref{Ass:eigenvalues} is satisfied.

\begin{remark}\label{Rem:min_eig_zero}
As an illustration of the problems with adopting the usual scaling by $T^{-2}$, consider the simple example of an $s$-dimensional white noise sequence $\bu_t \overset{i.i.d.}{\sim} \mathcal{N}(\bm{0},\bI_s)$ and define $\bh_t = \sum_{j=1}^t \bu_j$. Then, in Lemma \ref{Lemma:BM_int} in Appendix \ref{App:eigenvalues} we show that $\Prob\left(\lambda_\min\left(\frac{1}{T^2}\sum_{t=1}^T\bh_t\bh_t^\prime\right) > \phi \right) \to 0$, as $s,T \to \infty$, regardless of their relative rates. Hence, even in this simple case we cannot assume that the minimum eigenvalue is bounded away from zero if we stick to the $T^{-2}$ scaling.
\end{remark}

\begin{remark}
There are several noteworthy instances in which $\lambda_\min\left(\hat{\bSigma}_{22}\right)$ is bounded away from zero with arbitrarily high probability without the need for Assumption \ref{Ass:eigenvalues}. Assume that the dimension of the orthogonal complement of the cointegrating space in the subset of relevant non-stationary variables converges to a finite constant, i.e. $s_\delta \to K$. Then, based on a standard functional central limit theorem,
\begin{equation*}
\hat{\bSigma}_{22} \overset{d}{\to} K\bB_{S_\delta,\perp}^\prime\bC_{S_\delta}\left(\int_0^1 \tilde{\bB}(r)\tilde{\bB}^\prime(r)dr\right)\bC_{S_\delta}^\prime\bB_{S_\delta,\perp} \overset{d}{=} \int_0^1 \bB^*(r)\bB^{*\prime}(r)dr,
\end{equation*}
where $\tilde{\bB}(r)$ is an $s_\delta$-dimensional Gaussian process, described in the proof of Lemma A.2 in \citet{Phillips1990a}, and $\bB^*(r)$ is simply a linearly transformed version. By the same lemma, it follows that $\int_0^1 \bB^*(r)\bB^{*\prime}(r)dr$ is positive-definite almost surely. Then, for any $\epsilon>0$, we may choose $\phi(\epsilon) > 0$ such that 
\begin{equation*}
\Prob\left(\lambda_{\min}\left(\hat{\bSigma}_{22}\right) \leq \phi(\epsilon)\right) \to \Prob\left(\lambda_\min\left(\int_0^1 \bB^*(r)\bB^{*\prime}(r)dr\right) \leq \phi(\epsilon)\right) \leq \epsilon.
\end{equation*}
A straightforward case in which $s_\delta$ remains finite is to simply assume that the number of relevant integrated variables stays finite, i.e. $\abs{S_\delta} \leq K$. However, a more general example occurs when the dimension of the cointegrating space of $\bz_{S_\delta,t}$ diverges at the rate $\abs{S_\delta}$. This occurs in the case of a non-stationary factor model with stationary idiosyncratic components, as proposed by \citet{Banerjee2014}. Further illustrations are provided in Remark \ref{Rem:factor}.
\end{remark}

\section{The Single-Equation Penalized Error Correction Selector}\label{Sec:SPECS}

Despite the dimension reduction obtained from moving towards a single-equation representation, regularization remains a necessity in high dimensions. The single-equation model \eqref{eq:CECM} contains a total of $N(p+2) + 1$ parameters, compared to the $2N(r+1) + N^2p$ parameters in the full-system VECM in \eqref{eq:VECM}, resulting in a substantial reduction in dimensionality. However, the dimension may still grow large when either: (1) the number of potentially relevant variables is large or (ii) when the number of lagged differences required to appropriately model the short-run dynamics is large. Therefore, we consider the use of $\ell_1$-regularization to enable estimation in high dimensions. 

The resulting estimator, henceforth referred to as the Single-equation Penalized Error Correction Selector (SPECS), is defined as the minimizer of the following objective function:
\begin{equation}\label{eq:SPECS}
G_T\left(\bgamma,\btheta\right) = \norm{\Delta \by - \bV\bgamma - \bD\btheta}_2^2 + \lambda_I\sum_{i=1}^{N+M} \omega_i\abs{\gamma_i} + \lambda_G\norm{\bdelta}_2,
\end{equation}
where $M = (N+1)p-1$ refers to the number of transformed variables in $\bw_t$, i.e. the length of $\bpi$. We denote the minimizers of \eqref{eq:SPECS} by $\hat{\bgamma}$. The group penalty, regulated by $\lambda_G$, serves to promote exclusion of the lagged levels as a group when there is no cointegration present in the data. In this case, the model is effectively estimated in differences and corresponds to a conditional model derived from a vector autoregressive model specified in differences. The individual $\ell_1$-penalties, regulated by $\lambda_I$, serve to enforce sparsity in the coefficient vectors $\bdelta$ and $\bpi$ respectively.

The penalty of each coefficient $\gamma_i$ is weighted by $\omega_i$ to enable simultaneous estimation and selection consistency of the coefficients. Therefore, SPECS resembles a sparse group lasso \citep[e.g.][]{Simon2013} with adaptive weighting, applied to the conditional error correction model. The weights $\omega_i$ in \eqref{eq:SPECS} are typically derived from an initial estimation procedure such as OLS (if the number of variables is small enough), ridge, or lasso. In particular, let $\hat{\bgamma}_I$ denote initial estimates obtained for $\bgamma$ using one of the aforementioned methods. The weights can then be constructed as $\omega_i = \abs{\hat\gamma_{I,i}}^{-k}$ for some $k>0$. As the coefficients of the irrelevant variables tend to zero, this will ``blow up'' the weights for these coefficients, making them unlikely to be selected in the final estimation. On the other hand, the weights of the relevant coefficients converge to a positive constant leaving them unaffected. This wedge between the weights of relevant and irrelevant coefficients is exactly needed to achieve selection consistency. As demonstrated by \citet{Zou2006a}, under such assumptions on the weights, the adaptive lasso attains simultaneous selection and estimation consistency, without the necessity for the rather stringent irrepresentable condition in \citet{Zhao2006}.\footnote{In fact, as the adaptive lasso can be written as a regular lasso on a transformed design matrix, the irrepresentable condition, while still needed, operates on this transformed design matrix and becomes a weighted irrepresentable condition. This condition is then in turn implied by appropriate assumptions on the weights. In this paper we directly take this route rather than going via an irrepresentable condition. Section 7.5 of \citet{Buhlmann2011} provides details on the links between these assumptions.} To maintain generality we work with general weights without specifying how they are obtained, and therefore define appropriate assumptions directly on the weights. In Section \ref{Sec:Init_est} we then return to weight construction and propose a feasible way to construct weights that are theoretically shown to satisfy our assumptions.

\begin{assumption}\label{Ass:Regularization} Assume that the weights and regularization penalties satisfy:
\begin{enumerate}
\item\label{ass:reg_om_max} $\omega_{S_\gamma,\max} = o_p(T^\xi$) for some $\xi > 0$, where $\omega_{S, \max} = \max\{\omega_i: i \in S\}$.
\item\label{ass:reg_lambda} $\lambda_I = o\left(\frac{\left(s_\delta + \sqrt{s_\pi}\right)T^{1/2-\xi}} {\sqrt{s_\delta + s_\pi}} \right)$ and $\lambda_G = o(\sqrt{T})$.
\item\label{ass:reg_om_min} Let $\omega_{S, \min} = \min\{\omega_i: i \in S\}$. Then
\begin{align*}
\omega_{S_\delta^c,\min}^{-1} &= o_p \left(\min\left\{(s_\delta + s_\pi)^{-1/2} T^{-1/2 - \xi} N^{-1/2}, \lambda_I (s_\delta + \sqrt{s_\pi})^{-1} T^{-1} N^{-1/2} \right\} \right) ,\\
\omega_{S_\pi^c,\min}^{-1} &= o_p \left(\min\left\{(s_\delta + s_\pi)^{-1/2} T^{-\xi} (Np)^{-1/2}, \lambda_I (s_\delta + \sqrt{s_\pi})^{-1} (TNp)^{-1/2} \right\} \right).
\end{align*}
\end{enumerate}
\end{assumption} 
Part (\ref{ass:reg_om_max}) puts an upper bound on the rate at which the weights corresponding to the relevant variables diverge. Part (\ref{ass:reg_lambda}) restricts the maximum admissible growth rate of the penalty. Exceeding this rate would in an excess of shrinkage bias that impedes estimation consistency. Finally, part (\ref{ass:reg_om_min}) states that the weights of the irrelevant variables -- interacting with the penalty parameter $\lambda_I$ -- grow sufficiently fast in order to guarantee that irrelevant variables are removed from the model with probability converging to one. The required minimum growth rate of the penalty parameter is inversely related to the growth rate of the weights of the irrelevant variables; faster diverging weights require less penalization to identify irrelevant variables.

\begin{remark} \label{rem:group}
The only restriction that Assumption \ref{Ass:Regularization} imposes on the growth rate of the group penalty is that $\frac{\lambda_{G} }{\sqrt{T}} \to 0$, which is necessary for preventing shrinkage bias induced by the group penalty from impeding estimation consistency. Since $\lambda_{G} = 0$ is an admissible value, it follows that the theoretical results presented in the following section apply to the minimizer of
   $G^*_T(\bgamma,\btheta) = \norm{\Delta \by - \bV\bgamma - \bD\btheta}_2^2 + \lambda_I\sum_{i=1}^{N+M}\omega_i\abs{\gamma_i}$
as well, as long as the remaining conditions are satisfied.
\end{remark}

\begin{remark} \label{rem:dets}
Note that the deterministic components $\btheta$ are left unpenalized in \eqref{eq:SPECS}, as their inclusion in the model is desirable to enable identification of the limiting distribution of the estimators. Similar to the classical Frisch-Wraugh-Lovell Theorem, \citet{Yamada2017} show that the inclusion of unpenalized components is equivalent to performing the estimation after regressing out those components. In other words, we may define $\bM = \bI_T - \bD\left(\bD^\prime\bD\right)^{-1}\bD^\prime$ and note that
\begin{equation*}
    \hat{\bgamma} = \argmin_{\bgamma} \norm{\bM\left(\Delta \by - \bV\bgamma\right)}_2^2 + \lambda_I\sum_{i=1}^{N+M} \omega_i\abs{\gamma_i} + \lambda_G\norm{\bdelta}_2.
\end{equation*}
If one believes that the trend or constant are zero, one may reflect this knowledge in the construction of $\bM$, with the convention that $\bM=\bI_T$ when $\bmu=\btau=\bm{0}$.
\end{remark}

Two common data-driven ways to select the tuning parameters $\lambda_I$ and $\lambda_G$ are using cross-validation and information criteria. As standard $K$-fold cross-validation does not respect the time order of the data, we instead consider a time series cross-validation (TSCV) scheme as proposed by e.g.~\citet{Hyndman2018} and \citet{Wilms2017}, where for different values of $\blambda = (\lambda_I, \lambda_G)^\prime$ the model is estimated on the first part of the sample, and its prediction for the next observation is recorded. The sample is then recursively moved forward towards the end, and the $\blambda$ with the lowest mean squared prediction error is selected. We refer to \citet{Smeekes2018a} for details on the implementation and a comparison with traditional $K$-fold cross-validation.

While cross-validation works well for prediction \citep{chetverikov2016}, it tends to generally select fairly low penalty levels and therefore includes many variables. An alternative way to select $\blambda$ is using information criteria, where we find the value of $\blambda$ as
\begin{equation*}
\hat{\blambda}_{IC} = \argmin_{\blambda} \ln \left(\frac{1}{T} \norm{\Delta \by - \bV \hat{\bgamma}(\blambda) - \bD \hat\theta}_2^2 \right) + \frac{C_T \widehat{df}(\blambda)}{T},
\end{equation*}
where $\hat{\bgamma}(\blambda)$ and $\hat{\btheta}$ denote the minimizers of $G_T\left(\bgamma,\btheta\right)$ in \eqref{eq:SPECS} for a particular value of $\blambda$.\footnote{As explained in Remark \ref{rem:dets}, $\hat{\btheta}$ does not depend on $\blambda$.} In addition, $\widehat{df}(\blambda)$ is an estimate of the degrees of freedom and $C_T$ is the criterion-specific penalty; for the latter we use the Bayesian Information Criterion \citep[BIC]{Schwarz1978} with $C_T= \ln(T)$.

\citet{Zou2007} show that for the (adaptive) lasso the number of non-zero coefficients is an appropriate estimate for the degrees of freedom for model selection using information criteria. For group lasso penalties, estimating the degrees of freedom is more complicated. \citet{Yuan2006} propose a heuristic rule, but this requires the least squares estimator which is not available for large $N$. Alternative rules are provided by \citet{Breheny2009} and \citet{vaiter2012} among others, but none are theoretically valid in our setting. For this reason we propose a simple, heuristic rule where we set $\widehat{df}(\blambda)$ equal to the number of non-zero coefficients. Essentially this means we ignore the strength of the group penalty on the complexity of the model as long as the group is selected, thereby overestimating $df(\blambda)$. As a consequence, we will only choose non-zero values of $\lambda_G$ if they either improve the fit directly or result in setting the whole group to zero without affecting the fit too much. This is an intentional choice, consistent with our theoretical treatment of the group penalty. As discussed in Remark \ref{rem:group}, the group penalty is not necessary and consistency can be achieved even with $\lambda_G = 0$, and can therefore be seen as an optional add-on penalty.

Finally, we note that in practice both methods require the respective objective function to be minimized for a two-dimensional grid of values for $\blambda$. By choosing the lower and upper bounds of the grid carefully, one can ensure that the selected tuning parameters satisfy the assumptions listed in the next subsection. Of course, even though this ensures the theoretical validity of the selection method, its practical performance can still vary considerably. Therefore we investigate the practical performance of BIC and TSCV in the simulations and empirical application respectively.

\section{Theoretical Results}\label{Sec:Theory}

In this section we derive the asymptotic properties of SPECS, describe the construction of the weights and discuss implications for particular model specifications.

\subsection{Asymptotic Properties}\label{Sec:Asymptotics}
The first result that we pursue is that of selection consistency, i.e. the ability of an estimation procedure to select the correct set of relevant variables with probability converging to one. In fact, \citet{Zhao2006} define a stronger property referred to as sign consistency, which additionally requires the procedure to identify the correct signs of the non-zero coefficients with probability converging to one. In the following theorem, we derive sign consistency of SPECS.

\begin{theorem}\label{Thm:Selection_Consistency}
Under Assumptions \ref{Ass:moments}-\ref{Ass:Regularization}, as $T,N,p, \to \infty$ it holds that
$\Prob\left(\emph{sign}\left(\hat{\bgamma}\right) = \emph{sign}\left(\bgamma\right)\right) \to 1$.
\end{theorem}

Theorem \ref{Thm:Selection_Consistency} provides an asymptotic justification for implementing SPECS as a high-dimensional variable selection device. Furthermore, selection consistency is a crucial property when one aims to obtain interpretable solutions or even utilize the estimator as an alternative to classical tests for cointegration. An example of a traditional test for cointegration is the ECM-test by \citet{Banerjee1998} which looks at the $t$-ratio of the ordinary least squares coefficient of the lagged dependent variable. Alternatively, \citet{Boswijk1994} proposes to test for the joint significance of the least squares coefficients of all lagged variables with a Wald-type test. In our case, one could interpret exclusion of the lagged levels of the dependent variable, or the lagged levels of all variables, as evidence against the presence of cointegration. However, as discussed, an assumption of weak exogeneity is necessary when the aim is a direct interpretation of the estimated cointegration vector. Notwithstanding this caveat, selection consistency offers valuable insights when viewed as a screening mechanism that excludes irrelevant variables even in the absence of weak exogeneity. Moreover, since the set of variables included is strictly smaller than the time series dimension, it is possible to apply a traditional consistent estimator to the selected set of variables \citep[e.g.][]{Belloni2013}. However, ideally SPECS would contain desirable properties that omit the need of a second estimation procedure. For this reason, we establish the simultaneous consistency of the estimated coefficients in the following theorem.

\begin{theorem}\label{Thm:Estimation_Consistency}
Let $\bS_T = \emph{diag}\left(\sqrt{T}\bI_{s_\pi},\frac{T}{\sqrt{s_\delta}}\bI_{s_\delta}\right)$ and $\bQ$ as defined in \eqref{eq:Q}. Under the same assumptions as in Theorem \ref{Thm:Selection_Consistency}, it holds that
$\norm{\bS_T\bQ^{\prime -1}\left(\hat{\bgamma}_{S_{\gamma}} - \bgamma_{S_\gamma}\right)}_2 = O_p\left(s_\delta + \sqrt{s_\pi}\right)$.
\end{theorem}

The estimation consistency derived in Theorem \ref{Thm:Estimation_Consistency} does not place any restrictions on the relative growth rates of $T,N,p$, because it relies solely on high-level assumptions stated in the preceding section. However, when we derive sufficient conditions for the eigenvalue assumptions in Assumption \ref{Ass:eigenvalues} in Appendix \ref{App:eigenvalues} and provide a feasible method to construct weights that satisfy Assumption \ref{Ass:Regularization} in Section \ref{Sec:Init_est}, these restrictions do appear. We refer to Section \ref{sec:rates} for an explicit discussion.

\begin{remark}\label{Rem:derotate}
As an immediate consequence of Theorem \ref{Thm:Estimation_Consistency}, we have
$\norm{\hat{\bgamma}_{S_{\gamma}} - \bgamma_{S_\gamma}}_2 = O_p\left(\frac{s_\delta + \sqrt{s_\pi}}{\sqrt{T}}\right)$,
such that SPECS attains $\sqrt{T}$-consistency when $s_\delta$ and $s_\pi$ remain finite. To see this, note that by the assumption on $s_\delta$, it holds that $\frac{T}{\sqrt{s_\delta}} \geq \sqrt{T}$ for sufficiently large $T$. Then,
\begin{equation*}
\norm{\bS_T\bQ^{\prime -1}\left(\hat{\bgamma}_{S_{\gamma}} - \bgamma_{S_\gamma}\right)}_2 \geq \sqrt{T}\norm{\bQ^{\prime -1}\left(\hat{\bgamma}_{S_{\gamma}} - \bgamma_{S_\gamma}\right)}_2.
\end{equation*}
Moreover, since the basis matrices $\bB_{S_\delta}$ and $\bB_{S_\delta,\perp}$ are not uniquely defined, we may impose a normalization such that $\norm{\bQ}_2 \leq 1$. Then,
\begin{equation*}
\begin{split}
&\norm{\hat{\bgamma}_{S_{\gamma}} - \bgamma_{S_\gamma}}_2 = \norm{\bQ^\prime\bQ^{\prime -1}\left(\hat{\bgamma}_{S_{\gamma}} - \bgamma_{S_\gamma}\right)}_2
\leq \norm{\bQ}_2\norm{\bQ^{\prime -1}\left(\hat{\bgamma}_{S_{\gamma}} - \bgamma_{S_\gamma}\right)}_2 \leq \norm{\bQ^{\prime -1}\left(\hat{\bgamma}_{S_{\gamma}} - \bgamma_{S_\gamma}\right)}_2,
\end{split}
\end{equation*}
such that $\norm{\bS_T\bQ^{\prime -1}\left(\hat{\bgamma}_{S_{\gamma}} - \bgamma_{S_\gamma}\right)}_2 \geq \sqrt{T}\norm{\hat{\bgamma}_{S_{\gamma}} - \bgamma_{S_\gamma}}_2$.
\end{remark}

As a corollary to Theorem \ref{Thm:Estimation_Consistency}, it is possible to establish a relationship between the limit distribution of SPECS and the OLS estimator based on the subset of relevant variables.
\begin{corollary}\label{Cor:OLS_oracle}
Define the OLS oracle estimator as $\hat{\bgamma}_{OLS,S_\gamma} = \argmin_{\bgamma}\norm{\bM(\Delta \by - \bV_{S_\gamma}\bgamma)}_2^2$. Then, with $\xi>0$ as in Assumption \ref{Ass:Regularization}, under the same assumptions as Theorem \ref{Thm:Selection_Consistency} it holds that
\begin{equation}
    \norm{\bS_T\bQ^{\prime -1}\left(\hat{\bgamma}_{S_\gamma} - \hat{\bgamma}_{OLS,S_\gamma}\right)}_2 = o_p\left(\frac{\lambda_I(\sqrt{s_\delta} + \sqrt{s_\pi})}{T^{1/2 - \xi}}\right).
\end{equation}
\end{corollary}
The oracle results in Corollary \ref{Cor:OLS_oracle}, combined with the sign consistency from Theorem \ref{Thm:Selection_Consistency},  are suggestive of a post-selection inferential procedure. In particular, one may implement a two-step estimation procedure in which SPECS is used to perform variable selection in the first step and a regular OLS regression is performed on the selected variables in the second step. Then, after strengthening part \ref{ass:reg_lambda} of Assumption \ref{Ass:Regularization} to
$\lambda_I = o\left(\frac{T^{1/2-\xi}}{\sqrt{s_\delta} + \sqrt{s_\pi}}\right)$,
Corollary \ref{Cor:OLS_oracle} seems to validate the use of the regular OLS distribution for this two-step estimator, essentially ignoring the variable selection from the first stage. For example, in the case where $\abs{S_\gamma}$ remains finite, one could use the standard fixed-dimensional results \citep[e.g.][]{Boswijk1994} to perform inference. However, such a post-selection inferential procedure should be treated with caution, as it is well known that the selection step impacts the sampling properties of the estimator \citep[see][]{Leeb2005}. The convergence results of many selection procedures, SPECS included, hold pointwise only, i.e. the finite-sample distributions do not converge uniformly over the parameter space to their asymptotic distribution. The practical implication is that for certain values in the parameter space, relying on the oracle properties for post-selection test statistics may provide strongly misleading results. While developing a valid post-selection inference procedure to, for example, test for cointegration is certainly of interest, the field of valid post-selection inference is, despite its rapid development, still in its infancy. None of the currently existing methods, such as those considered in \citet{Berk2013}, \citet{Vandegeer2014}, \citet{Lee2016} or \citet{Chernozhukov2018}, can easily be adapted to - let alone validated in - our setting. Developing such a method therefore requires a full new theory which is outside the scope of the current paper.

\subsection{Initial Estimates}\label{Sec:Init_est}

In this section, we provide the reader with a directly implementable method to construct weights that satisfy Assumption \ref{Ass:Regularization}. As discussed in Section 2.2, we construct the weights as $\omega_i = \abs{\hat{\gamma}_{I,i}}^{-k}$. For our initial estimator we focus here on the ridge estimator, from which we can derive results for OLS as a special case, and comment on the lasso later on in the section. 

Note that the power $k$ gives one the flexibility to adjust how big the wedge between relevant and irrelevant variables is. To illustrate, assume that $\hat{\gamma}_{I,i} = \gamma_i + O_p\left(T^{-a}\right)$ for all $i$. Then, it is clear that $\omega_i = O_p(1)$ when $\gamma_i \neq 0$ and $\omega_i = O_p\left(T^{ka}\right)$ when $\gamma_i = 0$. Therefore, larger values of $k$ will increase the rate at which the weights corresponding to the irrelevant variables diverge. Based on this principle, the availability of a consistent initial estimator allows us to construct weights that satisfy the conditions in Assumption \ref{Ass:Regularization}. However, while the idea of adjusting divergence rates through imposing varying values of $k$ seems theoretically attractive, large values of $k$ result in substantial amplification of finite-sample estimation error. As a result, the finite-sample performance of the lasso becomes unstable for large $k$, such that in practice one may want to set the value for $k$ as low as theoretically admissible.

Regardless of the choice of $k$, the basic ingredient for good adaptive weights is a consistent initial estimator. Therefore, we derive the consistency of the ridge estimator. Recall that the ridge estimator is defined as the minimizer of the following objective function:
\begin{equation}\label{eq:ridge}
G_R(\bgamma,\btheta) := \norm{\Delta \by - \bV\bgamma - \bD\btheta}_2^2 + \lambda_R\norm{\bgamma}_2^2.
\end{equation}
The properties of the ridge estimator are well-studied in the stationary setting \citep[e.g][Section 3.4.1]{Hastie2008}. However, to the best of our knowledge, no explicit results are available in the high-dimensional non-stationary case considered here.

In order to derive consistency of the ridge estimator, we redefine the transformed sample covariance matrix from Section \ref{sec:rot} and the corresponding bound on its minimum eigenvalue. Let $N_\delta = N-r$, $M_\pi = M+r$ and define the new scaling and rotation matrices as $\bS_R = \diag\left(\sqrt{T}\bI_{M_\pi},\frac{T}{\sqrt{N_\delta}}\bI_{N_\delta}\right)$ and
\begin{equation*}
\bQ_R = \begin{bmatrix}
\left(\bB^\prime\bB\right)^{-1/2}\bB^\prime & 0\\
0 & \bI_M\\
\left(\bB_\perp^\prime\bB_\perp\right)^{-1/2}\bB_\perp^\prime & 0
\end{bmatrix},
\end{equation*}
respectively. The new transformed covariance matrix, based on the full dataset, is given by
\begin{equation}\label{eq:Sigma_R}
    \hat{\bSigma}_R = \bS_R^{-1}\bQ_R\bV^\prime\bM\bV\bQ_R\bS_R^{-1} = \begin{bmatrix}
    \hat{\bSigma}_{R,11} & \hat{\bSigma}_{R,12}\\
    \hat{\bSigma}_{R,21} & \hat{\bSigma}_{R,22}
    \end{bmatrix},\\
\end{equation}
with 
    $\hat{\bSigma}_{R,11} = \frac{1}{T}\begin{bmatrix}
    \bB^\prime\bZ_{-1}^\prime \bM \bZ_{-1}\bB & \bB^\prime\bZ_{-1}^\prime \bM \bW\\
    \bW^\prime \bM \bZ_{-1}\bB & \bW^\prime \bM \bW
    \end{bmatrix},$
    and $\hat{\bSigma}_{R,22} = \frac{N_\delta}{T^2}\bB_\perp^\prime\bZ_{-1}^\prime\bM\bZ_{-1}\bB_\perp$.
Then, we extend the minimum eigenvalue bound in Assumption \ref{Ass:eigenvalues} to \eqref{eq:Sigma_R} as follows.
\begin{assumption}\label{Ass:eig_ridge}
Assume that, on a set with probability converging to 1 as $T,N,p \to \infty$, there exists a constant $\phi_R>0$, such that 
$\underset{\bx \in \mathbb{R}^{M_\pi}}{\text{inf}} \frac{\bx^\prime \hat{\bSigma}_{R,11}\bx}{\bx^\prime \bx} \geq \phi_R$ 
and
$\underset{\bx \in \mathbb{R}^{N_\delta}}{\text{inf}} \frac{\bx^\prime \hat{\bSigma}_{R,22} \bx}{\bx^\prime \bx} \geq \phi_R.$
\end{assumption}

We now derive the convergence rate of the ridge estimator under a further restriction on the growth rates of $N,M$. The consistency of the ridge estimator is given in the following theorem.

\begin{theorem}\label{Thm:ridge}
Assume that $\frac{N_\delta}{T^{1/4}} \to 0$, $\frac{M_\pi}{\sqrt{T}} \to 0$, and $\lambda_R = O\left(\frac{\left(N_\delta + \sqrt{M_\pi}\right)\sqrt{T}}{\sqrt{\abs{S_\delta} + \abs{S_\pi}}}\right)$. Then, under Assumptions \ref{Ass:moments}-\ref{Ass:Dependence} and \ref{Ass:eig_ridge}, it holds that
$\norm{\bS_R\bQ_R^{\prime -1}\left(\hat{\bgamma}_R - \bgamma\right)}_2 = O_p\left(N_\delta + \sqrt{M_\pi}\right)$.
\end{theorem}
Similar to Remark \ref{Rem:derotate}, it follows from Theorem \ref{Thm:ridge} that
$\norm{\hat{\bgamma}_R - \bgamma}_2 = O_p\left(\frac{N_\delta + \sqrt{M_\pi}}{\sqrt{T}}\right)$.
Based on the assumption that $\frac{N_\delta}{T^{1/4}} \to 0$ and $\frac{M_\pi}{\sqrt{T}} \to 0$ in Theorem \ref{Thm:ridge}, it follows directly that $\norm{\hat{\bgamma}_R - \bgamma}_2 = o_p(1)$, and therefore ridge can be used to construct weights that satisfy our Assumption \ref{Ass:Regularization}. The exact values of $k$ that are needed theoretically vary depending on the number of (total and relevant) variables in the dataset; we return to this issue in Section \ref{sec:rates}.

The attentive reader may note that the admissible growth rates of $N_\delta,M_\pi$ in Theorem \ref{Thm:ridge} are the same as those initially assumed on the subsets of relevant variables, i.e. $s_\delta,s_\pi$, in Theorem \ref{Thm:Selection_Consistency}. The restriction imposed on the number of stochastic trends, $\frac{N_\delta}{T^{1/4}} \to 0$, corresponds closely to that of Corollary 2.1 in \citet{Liang2019}, who consider (co)integrated processes as well and roughly require that $\frac{N}{T^{1/4 - \nu}} \to 0$ for some $\nu > 0$. The growth rate of the total number of (implied) stationary variables is restricted to $\frac{M_\pi}{\sqrt{T}} \to 0$. While this may seem limited in comparison to the admissible (near) exponential growth in the stationary setting with i.i.d. Gaussian errors 
\citep[e.g.][Thm 3]{KockCallot2015}, we stress that our time series framework is more general, allowing not only for integrated processes, but also substantial dependence in the stationary component. Regarding the latter, our assumptions closely match those in the second row of Table 6 of \citet{Medeiros2016} with $\zeta = 1$, where our allowed growth rates are only slightly slower.

Ideally, we would like to allow for faster rates of divergence for the set of the irrelevant variables. A prospective strategy to attain this, would be to implement the lasso as an initial estimator, the consistency of which may be derived with the use of a compatibility condition \citep[see for example][Ch. 6]{Buhlmann2011}. While desirable, deriving the validity of an appropriate compatibility condition is a considerable task. In addition to the difficulty of showing the theoretical validity of a compatibility condition in the non-stationary setting considered here, the use of a compatibility condition is further complicated by the fact that the stochastic trends asymptotically dominate the variation. More specifically, in order to attain a non-singular limit matrix, a rotation similar to $\bQ$ is required that separates the stationary and non-stationary components in the full dataset. The standard compatibility condition would have to be adjusted in a non-trivial manner to account for such a rotation. Consequently, we leave the development of a suitable compatibility condition to future research, and instead focus on the ridge estimator under the more stringent growth rates on the number of variables. In the simulations we explore settings beyond these restrictive assumptions, and our adaptive weights continue to function in this case as well. We therefore conjecture that the suitability of the ridge estimator can be extended to a more general setting.

\begin{remark}
Theorem \ref{Thm:ridge} imposes no minimum growth rate of the penalty term $\lambda_R$ in \eqref{eq:ridge}. Therefore, in the case where $M+N < T$, the choice $\lambda_R = 0$ is both theoretically admissible and computationally feasible, such that consistency of the OLS estimator follows as a by-product of our result. Similarly, under the conditions imposed in Theorem \ref{Thm:ridge}, the lasso can also be shown to be a consistent initial estimator. In particular, Assumption \ref{Ass:eig_ridge} allows for the derivation of 
a minimum eigenvalue bound for the sample covariance matrix of the full data set, which enables application of standard proofs of consistency that are familiar from the fixed-dimensional setting.  Due to space consideration, we refrain from providing a full proof on this conjecture, but refer the interested reader to Theorem 3.1 in \citet{Liao2015}, the proof of which may be adjusted to fit the current setting.
\end{remark}

\subsection{Implications for Particular Model Specifications}

To fully appreciate the theoretical results in the preceding section, a detailed understanding of the generality provided by the set of imposed assumptions is helpful. For example, as the results are derived without requiring weak exogeneity, our set of assumptions allows for the presence of stationary variables in the data. However, in the absence of weak exogeneity, model interpretation becomes non-standard and the notion of sparsity carries non-trivial annotations. Therefore, in this section we elaborate on several relevant model specifications to demonstrate the flexibility of the single-equation model and highlight the practical implications of variable selection in such a general framework.

\subsubsection{Sparsity and Weak Exogeneity}\label{Sec:Sparsity}

The benefit of $\ell_1$-regularized estimation stems from its ability to identify sparse parameter structures. However, the concept of sparsity in the conditional models here considered merits additional clarification, as the potential absence of weak exogeneity obscures standard interpretability. Accordingly, in this section we comment on the interplay between weak exogeneity and sparsity and provide several illustrative examples of sparse DGPs. For simplicity of illustration, we assume in this and the following section that $\bmu=\btau=\bm{0}$.

In Section \ref{sec:model} we argue that the coefficients regulating the long-run dynamics in the conditional model are generally derived from linear combinations of the cointegrating vectors in the VECM representation \eqref{eq:VECM}. By decomposing the matrix with adjustment rates as $\bA = (
\balpha_1, \bA_2^\prime)^\prime$, we obtain the explicit construction
$\bdelta = \bB(\balpha_1 - \bA_2^\prime\bSigma_{\epsilon,22}^{-1}\bsigma_{\epsilon,21})$.
Hence, it follows that $\delta_i=0$ if the sparsity condition 
$\bbeta_i^\prime \left(\balpha_1 - \bA_2^\prime\bSigma_{\epsilon,22}^{-1}\bsigma_{\epsilon,21}\right) = 0$
is satisfied, where $\bbeta_i$ is the $i$-th row of $\bB$. While this condition may hold in a variety of non-trivial ways, specific cases of interest that lead to sparsity in $\bdelta$ can be derived. For example, an integrated variable $x_{i,t}$ that does not cointegrate with any of the variables in the system ($\bbeta_i = \bm{0}$), will carry a zero coefficient in the derived single-equation long-run equilibrium.

As a more general example, assume that the researcher observes the $N$-dimensional time series $\bz_t = (\bz_{1,t}^\prime,\bz_{2,t}^\prime)^\prime = (y_t,\bx^\prime_t)^\prime$, from time $t=1,\ldots,T$, where $\bz_{1,t} = (y_t,\bx_{1,t}^\prime)^\prime$ is an $N_1$-dimensional time series and $\bz_{2,t}$ is an $N_2$-dimensional time series. Moreover,
\begin{equation}\label{eq:DGP_SVECM}
\begin{split}
\begin{bmatrix}
\Delta \bz_{1,t}\\
\Delta \bz_{2,t}
\end{bmatrix} &= \begin{bmatrix}
\bPi_{11} & \bPi_{12}\\
\bPi_{21} & \bPi_{22}
\end{bmatrix} \begin{bmatrix}
\bz_{1,t-1}\\
\bz_{2,t-1}
\end{bmatrix} + \sum_{j=1}^p \begin{bmatrix}
\bPhi_{j,11} & \bPhi_{j,12}\\
\bPhi_{j,21} & \bPhi_{j,22}
\end{bmatrix} \begin{bmatrix}
\Delta \bz_{1,t-j}\\
\Delta \bz_{2,t-j}
\end{bmatrix} + \begin{bmatrix}
\bepsilon_{1,t}\\
\bepsilon_{2,t}
\end{bmatrix}\\ 
&= \bPi\bz_{t-1} + \sum_{j=1}^p\bPhi_j \Delta \bz_{t-j} + \bepsilon_t.
\end{split}
\end{equation}
In addition, assume that $\bSigma_\epsilon = \E\left(\bepsilon_t\bepsilon_t^\prime\right)$ satisfies Assumption \ref{Ass:moments} and can be decomposed as
\begin{equation}\label{eq:Sigma_eps_res}
\begin{split}
&\bSigma_\epsilon = \begin{bmatrix}
\bSigma_{\epsilon,11} & \bm{0}\\
\bm{0} & \bSigma_{\epsilon,22}
\end{bmatrix},\text{ with } \bSigma_{\epsilon,11} = \begin{bmatrix}
\sigma_{1,11} & \bsigma_{1,21}^\prime\\
\bsigma_{1,21} & \bSigma_{1,22}
\end{bmatrix}.
\end{split}
\end{equation}
Then, the quantities appearing in the single-equation model in \eqref{eq:CECM} take on the form
\begin{equation}\label{eq:CECM_coefs}
\begin{split}
\bpi_0 &= \begin{bmatrix}
\bSigma_{1,22}^{-1} & \bm{0}\\
\bm{0} & \bSigma_{\epsilon,22}^{-1}
\end{bmatrix}\begin{bmatrix}
\bsigma_{1,21}\\
\bm{0}
\end{bmatrix} = \begin{bmatrix}
\bpi_{0,1}\\
\bm{0}
\end{bmatrix},\\
\bdelta &= \begin{bmatrix}
\bPi_{11}^\prime & \bPi_{21}^\prime\\
\bPi_{12}^\prime & \bPi_{22}^\prime
\end{bmatrix}\begin{bmatrix}
1\\
-\bpi_0\\
\end{bmatrix} = \begin{bmatrix}
\bPi_{11}^\prime\\
\bPi_{12}^\prime
\end{bmatrix}\begin{bmatrix}
1\\
-\bpi_{0,1}
\end{bmatrix} = \begin{bmatrix}
\bdelta_1\\
\bdelta_2
\end{bmatrix},\\
\bpi_j &= \begin{bmatrix}
\bPhi_{j,11}^\prime & \bPhi_{j,21}^\prime\\
\bPhi_{j,12}^\prime & \bPhi_{j,22}^\prime
\end{bmatrix}\begin{bmatrix}
1\\
-\bpi_0
\end{bmatrix} = \begin{bmatrix}
\bPhi_{j,11}^\prime\\
\bPhi_{j,12}^\prime
\end{bmatrix}\begin{bmatrix}
1\\
-\bpi_{0,1}
\end{bmatrix} = \begin{bmatrix}
\bpi_{j,1}\\
\bpi_{j,2}
\end{bmatrix}.
\end{split}
\end{equation}
The definitions in \eqref{eq:CECM_coefs} demonstrate that, under the restriction that the errors driving $\bz_{1,t}$ and $\bz_{2,t}$ are uncorrelated, sparsity in the single-equation model arises when (a subset of) $\bz_{2,t}$ does not Granger-Cause $\bz_{1,t}$. For example, in the extreme case where $\bPi_{12} = \bm{0}$ and $\bPhi_{12} = \bm{0}$, we have $\bdelta_2 = \bm{0}$ and $\bpi_{j,2} = 0$, respectively.
Consequently, then the single-equation model reads as
\begin{equation}\label{eq:CECM_sparse}
\begin{split}
\Delta y_t &= \bdelta^\prime \bz_{t-1} + \bpi_0^\prime \Delta \bx_t + \sum_{j=1}^p \bpi_j^\prime\Delta \bz_{t-j} + \epsilon_{y,t}\\ 
&= \bdelta_1^\prime \bz_{1,t-1} + \bpi_{0,1}^{\prime}\Delta \bx_{1,t} + \sum_{j=1}^p \bpi_{1,j}^\prime \Delta \bz_{1,t-j} + \epsilon_{y,t}.
\end{split}
\end{equation}

As an interesting special case, consider the decomposition in \eqref{eq:DGP_SVECM} in which $z_{2,t}=\epsilon_{2,t}$ is scalar-valued with $\E(\epsilon_{2,t}\bepsilon_{1,t}) = \bm{0}$. Then, it is straightforward to see that $\bpi_{12}=\bpi_{21}=\bm{0}$, $\pi_{22} = -1$ and, consequently, $\delta_N = 0$. This finding highlights that stationary variables result in sparsity in $\bdelta$ only when they are fully exogenous, as said variables may enter the implied cointegrating vector through their correlation structure with the other variables in the system. This further demonstrates the difficulty of direct interpretation of $\bdelta$ without imposing additional restrictions on the DGP. From a prediction perspective, however, the model's ability to include stationary variables through their correlation structure is clearly a desirable feature.

Finally, we consider a DGP in which $\bSigma_\epsilon$ follows a Toeplitz structure with $\sigma_{\epsilon,ij} = \rho^{\abs{i-j}}$. After partitioning $\bSigma_\epsilon$ as in \eqref{eq:covariance partition}, we can rewrite
\begin{equation}\label{eq:Toeplitz proof}
\begin{split}
\bsigma_{\epsilon,21} = \begin{bmatrix}
\rho^1\\
\vdots \\
\rho^{N-1}
\end{bmatrix} &= \begin{bmatrix}
\rho^0 & \ldots & \rho^{N-2}\\
\vdots & \ddots & \vdots\\
\rho^{N-2} & \ldots & \rho^0
\end{bmatrix}\begin{bmatrix}
\rho^1\\
0\\
\vdots\\
0
\end{bmatrix} =\bSigma_{\epsilon,22}\bpi_0,
\end{split}
\end{equation}
thus showing that $\bpi_0 = \bSigma_{\epsilon,22}^{-1}\bsigma_{\epsilon,21} = (\rho, 0, \ldots, 0)^\prime$.\footnote{It is straightforward to show that this property carries over to covariance matrices with a block-diagonal Toeplitz structure, with each block $\bSigma_\epsilon^{(k)}$ having the form $\sigma^{(k)}_{i,j}=\rho_{(k)}^{\abs{i-j}}$. The number of non-zero elements in the resulting vector $\bpi_0$ will equal the number of blocks in the covariance matrix.} As $\bdelta^\prime = (1,-\bpi_0^\prime)\bA\bB^\prime$, this implies that only the  long-run equilibria that occur in the equations for $\Delta y_t$ or its cross-sectionally neighbouring variable will be part of the linear combination in the derived the single-equation model. Consequently, any variables in the dataset that are not contained in the equilibria occurring in these equations will induce sparsity in $\bdelta$.

\subsubsection{Mixed Orders of Integration}

One of the most prominent benefits of SPECS is the ability to model potentially non-stationary and cointegrated data without the need to adopt a pre-testing procedure with the aim of checking, and potentially correcting, for the order of integration or to decide on the appropriate cointegrating rank of the system. The assumptions under which our theory is developed are compatible with a wide variety of DGPs, including settings where the dataset contains an arbitrary mix of $I(1)$ and $I(0)$ variables. The researcher simply transforms the dataset according to \eqref{eq:CECM} and SPECS provides consistent estimation of the parameters and identification of the correct implied sparsity pattern. The purpose of this section is to demonstrate this feature by means of some illustrative examples.

The central idea underlying the above feature is that a single-equation model can be derived from any system admitting a finite order VECM representation. In a VECM system containing variables with mixed orders of integration, however, each stationary variable adds an additional trivial cointegrating vector. Such a vector corresponds to a unit vector that equals 1 on the index of the stationary variable. For illustrative  purposes, we consider the following general example. Define $\bz_t = (\bz_{1,t}^\prime,\bz_{2,t}^\prime)^\prime$, where $\bz_{1,t} \sim I(0)$ and $\bz_{2,t} \sim I(1)$ and possibly cointegrated. Let the dimensions of $\bz_{1,t}$ and $\bz_{2,t}$ be $N_1$ and $N_2$ respectively. Then, $\bz_t$ admits the representation
\begin{equation}\label{eq:VECM_stat}
\begin{split}
\begin{bmatrix}
\Delta \bz_{1,t}\\
\Delta \bz_{2,t}
\end{bmatrix} &= \begin{bmatrix}
-\bI_{N_1} & \bm{0}\\
\bm{0} & \bPi_{22}
\end{bmatrix}\begin{bmatrix}
\bz_{1,t-1}\\
\bz_{2,t-2}
\end{bmatrix} + \bPhi(L)\Delta \bz_{t-1} + \bepsilon_t= \bA\bB^\prime\bz_{t-1} + \bPhi(L)\Delta \bz_{t-1} + \bepsilon_t,
\end{split}
\end{equation}
where $\bPhi(L)$ corresponds to a $p$-dimensional matrix lag polynomial by Assumption \ref{Ass: GRT} and $\bepsilon_t$ satisfies the conditions in Assumption \ref{Ass:moments}. As long as the design of \eqref{eq:VECM_stat} conforms to Assumption \ref{Ass: GRT} and \ref{Ass:Dependence}, our main results apply to this setting and both selection and estimation consistency is maintained. For the extreme case in which all variables are integrated of order one, but none are cointegrate, we define $\bA=\bB = \bm{0}$. Clearly, it follows that $\bdelta=\bm{0}$, such that the single-equation model can be seen as a conditional model obtained from a VAR specified in differences. In the other extreme case, when the levels of all variables in the VECM are weakly stationary, decomposition \eqref{eq:VECM_stat} would simply lead to a VECM in which $-\bA=\bB=\bI_N$, thereby enabling the results in Section \ref{Sec:Asymptotics} to carry through.\footnote{When all variables are stationary, SPECS can also be shown to consistently estimate the parameters based on the well-documented properties of the adaptive lasso in stationary time series settings, such as those considered in \citet{Medeiros2016} and \citet{Masini2019}.}

\subsubsection{Rates of Convergence} \label{sec:rates}

We conclude our theoretical analysis with a detailed illustration of the attainable rates of convergence in different asymptotic frameworks. The rates of convergence of $\hat{\bgamma}_R$ and $\hat{\bgamma}$, as well as the specific construction of the initial weights, are dependent on the growth rates of $N,p,r,\abs{S_\delta}$ and $\abs{S_\pi}$. Because of the trade-off between the admissible dimension and the rate of convergence, the choice of the desired asymptotic framework is likely dependent on the specific application. For example, typical macro-economic applications are characterized by short panel datasets which would require a framework in which the cross-sectional dimension grows as fast as theoretically admissible. On the other hand, in applications with a large number of time series observations, such as forecasting based on high-frequency data, the assumption that the number of (potentially) relevant variables grows slow relative to the available time periods seems reasonable. Therefore, to aid interpretation of our results, we provide an overview with different asymptotic frameworks and the corresponding penalty parameters, weight constructions and convergence rates of the initial estimator in Table \ref{Tab4:settings}. The weights for $\delta_i$ and $\pi_j$ are constructed as $\omega_i = \abs{\hat{\delta}_{R,i}}^{-k_\delta}$ and $\omega_{N+j} =  \abs{\hat{\pi}_{R,j}}^{-k_\pi}$.

\begin{table}
\caption{Dimensions, Penalties, Weights and Convergence Rates}
\begin{tabularx}{\textwidth}{XXXXXllXX}
\hline
$N$ & $p$ & $r$ & $\abs{S_\delta}$ & $\abs{S_\pi}$ & $k_\delta$ & $k_\pi$ & $\lambda_R,\lambda_I$ & $\norm{\hat{\bgamma} - \bgamma}_2$ \tabularnewline
\hline
\hline 
fixed & fixed & fixed & fixed & fixed & 2 & 1 & $KT^{2/5}$ & $O_p\left(T^{-1/2}\right)$ \tabularnewline
$T^{1/4}$ & fixed & fixed & fixed & fixed & 3 & 1 & $KT^{2/5}$ & $O_p\left(T^{-1/2}\right)$ \tabularnewline
$T^{1/4}$ & $T^{1/4}$ & fixed & fixed & fixed & 3 & 2 & $KT^{2/5}$ & $O_p\left(T^{-1/2}\right)$ \tabularnewline
$T^{1/4}$ & $T^{1/4}$ & $T^{1/4}$ & fixed & fixed & 3 & 2 & $KT^{2/5}$ & $O_p\left(T^{-1/2}\right)$ \tabularnewline
$T^{1/4}$ & $T^{1/4}$ & $T^{1/4}$ & fixed & $T^{1/4}$ & 4 & 2 & $KT^{2/5}$ & $O_p\left(T^{-3/8}\right)$ \tabularnewline
$T^{1/4}$ & $T^{1/4}$ & fixed & $T^{1/4}$ & $T^{1/4}$ & 4 & 2 & $KT^{2/5}$ & $O_p\left(T^{-1/4}\right)$ \tabularnewline
$T^{1/4}$ & $T^{1/4}$ & $T^{1/4}$ & $T^{1/4}$ & $T^{1/4}$ & 4 & 2 & $KT^{2/5}$ & $O_p\left(T^{-3/8}\right)$ \tabularnewline
\hline 
\end{tabularx}
\label{Tab4:settings}
\caption*{This table displays possible settings for the weights ($k_\delta,k_\pi$) and penalty parameters ($\lambda_I,\lambda_R$) that satisfy Assumption \ref{Ass:Regularization} under a variety of asymptotic frameworks ($N,r,p,\abs{S_\delta},\abs{S_\pi}$). The convergence rate of SPECS is displayed in the last column.}
\end{table}

The first row of Table \ref{Tab4:settings} corresponds to the classic fixed-dimensional case. It is reassuring that, similar to the OLS estimator, SPECS obtains $\sqrt{T}$-convergence, with the additional benefit of allowing for consistent recovery of the sparsity pattern. In fact the next three rows highlight that when $N$, $p$ or $r$ diverge, while the number of relevant variables remains fixed, SPECS maintains its $\sqrt{T}$-convergence as long as the penalty weights $k_\delta$ and $k_\pi$ are adjusted appropriately. In the fifth row, we allow the number of relevant stationary variables, i.e. $\abs{S_\pi}$ to diverge as well. This setting may be preferred when the integrated time series remain persistent after being transformed to stationarity by differencing. We observe that consistency is maintained, although even sharper weights are required and the rate of convergence has reduced to $T^{3/8}$. In the sixth row we additionally allow the number of relevant non-stationary, i.e. $\abs{S_\delta}$, to increase, whereas the number of cointegrating vectors remains fixed. The increased number of non-zero coefficients corresponding to non-stationary variables reduces the rate of convergence to $T^{1/4}$. Interestingly, in the last row we let the dimension of the cointegrating subspace $r$ grow at the same rate. As illustrated in Remark \ref{Rem:factor}, this setting naturally occurs when the data is modelled by a non-stationary factor model with idiosyncratic components. In this framework, the number of stochastic trends driving the subset of relevant variables, i.e. $s_\delta$, remains fixed, which positively affects the convergence rate of SPECS.

\bigskip
We consider the theoretical results presented in this section to be of a double nature. On the one hand, it is reassuring that consistent estimation remains feasible in growing dimensions and that suitable weights are available. On the other hand, we acknowledge that the required restrictions on the growth rate of the number of variables seem to caution against application of penalized regression in very high-dimensional settings. However, it is worth noting that the restrictions on $N$ and $p$ largely result from the use of ridge regression as an initial estimator. Indeed, the availability of a novel compatibility condition could justify the use of the lasso as an initial estimator and will allow for generalization of our theoretical results to even higher dimensional asymptotic frameworks. We consider this an interesting avenue for future research.

\begin{remark}\label{Rem:factor}
The VECM \eqref{eq:DGP_SVECM} can be rewritten into a non-stationary factor model with stationary idiosyncratic components, similarly to \citet{Banerjee2014}. Based on the VMA representation of $\bz_t$ in \eqref{eq:GRT}, with $\bC$ a matrix of reduced rank, we can rewrite the process as
\begin{equation}\label{eq:ns_factors}
\bz_t = \bC\bs_t + \bmu + \btau t + \bu_t = \bLambda\bof_t+ \bmu + \btau t + \bu_t,
\end{equation}
where $\bLambda = \bB_\perp\left(\bA_\perp^\prime\left(\bI-\sum_{j=1}^p\bPhi_j\right)\bB_\perp\right)^{-1}$, $\bof_t = \bA_\perp^\prime\bs_t$ and $\bu_t = \bC(L)\bepsilon_t + \bz_0$. This representation is particularly relevant in relation to the growth rate of $N_\delta = N-r$. Typically, the theory for consistent estimation of \eqref{eq:ns_factors} is derived under the assumption that the $N_\delta$ factors remain fixed, while letting both $N$ and $T$ go to infinity. Hence, in this framework, noting that $s_\delta \leq N_\delta$, the assumptions that $\frac{s_\delta}{T^{1/4}} \to 0$ and $\frac{N_\delta}{T^{1/4}} \to 0$ in Theorems \ref{Thm:Selection_Consistency}-\ref{Thm:ridge} are automatically satisfied. Consequently, the convergence rates of the initial and final estimators are given by $\norm{\hat{\bgamma}_R - \bgamma}_2 = O_p\left(\sqrt{\frac{M_\pi}{T}}\right)$ and $\norm{\hat{\bgamma}-\bgamma}_2 = O_p\left(\sqrt{\frac{s_\pi}{T}}\right)$.
\end{remark}

\section{Simulations}\label{sec: simulations}

In this section we analyze the selective capabilities and predictive performance of SPECS by means of simulations. We estimate the single-equation model according to the objective function \eqref{eq:SPECS} with the following settings for the penalty rates:
\begin{enumerate}
\item Ordinary Least Squares (OLS: $\lambda_G=0$, $\lambda_I=0$),
\item Autoregressive Distributed Lag (ADL: $\lambda_G = 0$, $\lambda_I > 0$, $\omega_i = \infty$ for $i=1,\ldots,N$),
\item SPECS - no group penalty (SPECS$_1$: $\lambda_G = 0$, $\lambda_I > 0$),
\item SPECS - group penalty (SPECS$_2$: $\lambda_G > 0$, $\lambda_I > 0$)\footnote{As a helpful reminder, the reader may relate the subscript to the number of penalty categories included in the estimation; SPECS$_1$ only contains an individual penalty whereas SPECS$_2$ contains both a group penalty and and individual penalty.}.
\end{enumerate}
The OLS estimator is only included when feasible according to the dimension of the model to estimate and we additionally include a penalized autoregressive distributed lag model (ADL) with all variables entering in first differences. The latter model can be interpreted as the conditional model one would obtain when ignoring cointegration in the data and specifying a VAR in differences as a model for the full system. The resulting conditional model is the same as the CECM that we consider, but with the built-in restriction $\bdelta=\bm{0}$.

We estimate the solutions for a grid of penalty values and construct the weights from an initial ridge estimator as proposed in Section \ref{Sec:Init_est}. For ADL and SPECS$_1$, we consider 100 possible values for $\lambda_I$ and choose the final model based on the BIC criterion. Alternatively, for SPECS$_2$, the model selection takes place over a two-dimensional grid consisting of 100 values for $\lambda_I$ and 10 possible values for $\lambda_G$, with model selection again being based on the BIC criterion. The weights are defined by $\omega_i = \abs{\hat{\gamma}_{R,i}}^{-k}$, where $k=2$ for $i \in \lbrace 1,\ldots,N\rbrace$ and $k=1$ for $i \in \lbrace N+1,\ldots,N+M \rbrace$. 

We consider three different settings under which we analyze the performance of our estimators; the first setting aims to analyze the effects of dimensionality and weak exogeneity, the second setting explores the effect of the variables' orders of integration and the third setting considers the performance in non-sparse settings. Each setting is described in detail below.

\subsection{Dimensionality and Weak Exogeneity}\label{Sec:dim_we}

In the first part of our simulation study we focus on the effects of dimensionality and weak exogeneity  on a (co)integrated dataset. Our simulation DGP takes the form
\begin{equation}\label{eq:general DGP sim}
\Delta \bz_t = \bA\bB^\prime \bz_{t-1} + \bPhi_1 \Delta \bz_{t-1} + \bepsilon_t,
\end{equation}
with $t=1,\ldots, T=100$, $\bepsilon_t \sim \mathcal{N}(0,\bSigma)$ and $\sigma_{ij} = 0.8^{|i-j|}$. Furthermore, $\bPhi_1$, the coefficient matrix regulating the short-run dynamics is generated as $0.4 \cdot \bI_N$, where $N$ varies depending on the specific DGP considered. Based on this DGP, the single-equation model takes on the form
\begin{equation*}
\Delta y_t = \bdelta^\prime \bz_{t-1} + \bpi_0^\prime \Delta \bx_t + \bpi_1^\prime \Delta \bz_{t-1} + \epsilon_{y,t},
\end{equation*}
with $\bpi_0$ and $\bpi_1$ as defined in \eqref{eq:CECM}. We consider a total of four different settings, corresponding to different combinations of (i) dimensionality (low/high) and (ii) weak exogeneity (present/absent). The corresponding parameter settings and implied cointegrating vector $\bdelta$ are given in Table \ref{table:DGP WE and Dim}.

\begin{table}[t]
\caption{Simulation Design for the First Study (Dimensionality and Weak Exogeneity)}\label{table:DGP WE and Dim}
\begin{tabularx}{\textwidth}{cccc}
\hline 
Low Dimension & $\bA$ & $\bB$ & $\bdelta$ \tabularnewline
\hline 
WE  & $\alpha_{1}\cdot\begin{bmatrix}1\\
\bm{0}_{9\times 1}
\end{bmatrix}$ & $\begin{bmatrix}\tilde{\biota}\\
\bm{0}_{5\times 1}
\end{bmatrix}$ & $\alpha_{1}\cdot\bB$ \tabularnewline
No WE & $\alpha_{1}\cdot
\bB$
& 
$\begin{bmatrix}\tilde{\biota} & \bm{0}_{5\times 1}\\
\bm{0}_{5\times 1} & \tilde{\biota}
\end{bmatrix}$
& $(1+\rho)\alpha_{1}\cdot
\begin{bmatrix}\tilde{\biota}\\
\bm{0}_{5\times 1}
\end{bmatrix}$ \tabularnewline
\hline 
High Dimension & $\bA$ & $\bB$ & $\bdelta$\tabularnewline
\hline 
WE & $\alpha_{1}\cdot\begin{bmatrix}1\\
\bm{0}_{49 \times 1}
\end{bmatrix}$ & $\begin{bmatrix}\tilde{\biota}\\
\bm{0}_{45\times 1}
\end{bmatrix}$ & $\alpha_{1}\cdot\bB$ \tabularnewline
No WE & $\alpha_{1}\cdot\bB$ & $\begin{bmatrix}\tilde{\biota} &\bm{0}_{5\times 1} & \bm{0}_{5\times 1}\\
\bm{0}_{5\times 1} & \tilde{\biota} & \bm{0}_{5\times 1}\\
\bm{0}_{5\times 1} & \bm{0}_{5\times 1} & \tilde{\biota}\\
\bm{0}_{35\times 1} & \bm{0}_{35\times 1} & \bm{0}_{35\times 1}
\end{bmatrix}$ & $(1+\rho)\alpha_{1}\cdot\begin{bmatrix}\tilde{\biota}\\
\bm{0}_{45\times 1}
\end{bmatrix}$ \tabularnewline
\hline 
\end{tabularx}
\caption*{Notes: The low-dimensional (high-dimensional) design corresponds to a system with $N=10$ ($N=50$) unique time series and $N^\prime = 31$ ($N^\prime = 151$) parameters to estimate. Furthermore, $\tilde{\biota} = (1, -\biota_4^\prime)^\prime$ and $\alpha_1 = -0.5,-0.45,\ldots,0$ regulates the adjustment rate towards the equilibrium.}
\end{table}

We measure the selective capabilities based on three metrics. The pseudo-power of the models measures the ability to appropriately pick up the presence of cointegration in the underlying DGP. For the OLS procedure we perform the Wald test proposed by \citet{Boswijk1994}. When the OLS fitting procedure is unfeasible due to the high-dimensionality, we perform the Wald test on the subset of variables included after fitting SPECS$_1$ and refer to this approach as Wald-PS (where PS stands for post-selection). Despite the caveats of oracle-based post-selection inference discussed after Corollary \ref{Cor:OLS_oracle},  the inclusion of Wald-PS still offers valuable insights regarding the performance one may expect of such a procedure in light of the aforementioned limitation. SPECS is used as an alternative to this cointegration test by simply checking whether at least one of the lagged levels is included in the model. The percentage of trials in which cointegration is found is then reported as the pseudo-power. 

Second, for each trial the Proportion of Correct Selection (PCS) measures the proportion of correctly selected variables, while the Proportion of Incorrect Selection (PICS) describes, as the name may suggest, the proportion of incorrectly selected variables. They are given by
\begin{equation*}
PCS = \frac{\abs{\left\lbrace \hat{\gamma}_j \neq 0\right\rbrace \cap \left\lbrace\gamma_j \neq 0\right\rbrace}}{\abs{\left\lbrace \gamma_j \neq 0 \right\rbrace}}; \qquad
PICS = \frac{\abs{\left\lbrace \hat{\gamma}_j \neq 0\right\rbrace \cap \left\lbrace\gamma_j = 0\right\rbrace}}{\abs{\left\lbrace \gamma_j = 0 \right\rbrace}}.
\end{equation*}
The PCS and PICS are calculated for SPECS$_1$ and SPECS$_2$ and averaged over all trials.

Finally, we consider the predictive performance in a simulated nowcasting application, where we implicitly assume that the information on the latest realization of $\bx_T$ arrives before the realization of $y_T$. These situations frequently occur in practice, see \citet{Giannone2008} and the references therein for an overview as well as the empirical application considered in Section \ref{Sec:application}. Due to the construction of the single-equation model, in which contemporaneous values of the conditioning variables contribute to the contemporaneous variation in the dependent variable, our proposed method is particularly well-suited to this application. For any of the considered fitting procedures, the nowcast is given by $\hat{y}_T = \hat{\bdelta}^\prime \bz_{T-1} + \hat{\bpi}^\prime \Delta \bx_T + \hat{\bphi}^\prime \Delta \bz_{T-1}$, where by construction $\hat{\bdelta}=\bm{0}$ in the ADL model. For each method we record the root mean squared nowcast error (RMSNE) relative to the OLS oracle procedure fitted on the relevant variables.

\begin{sidewaysfigure}
\includegraphics[width=\textwidth]{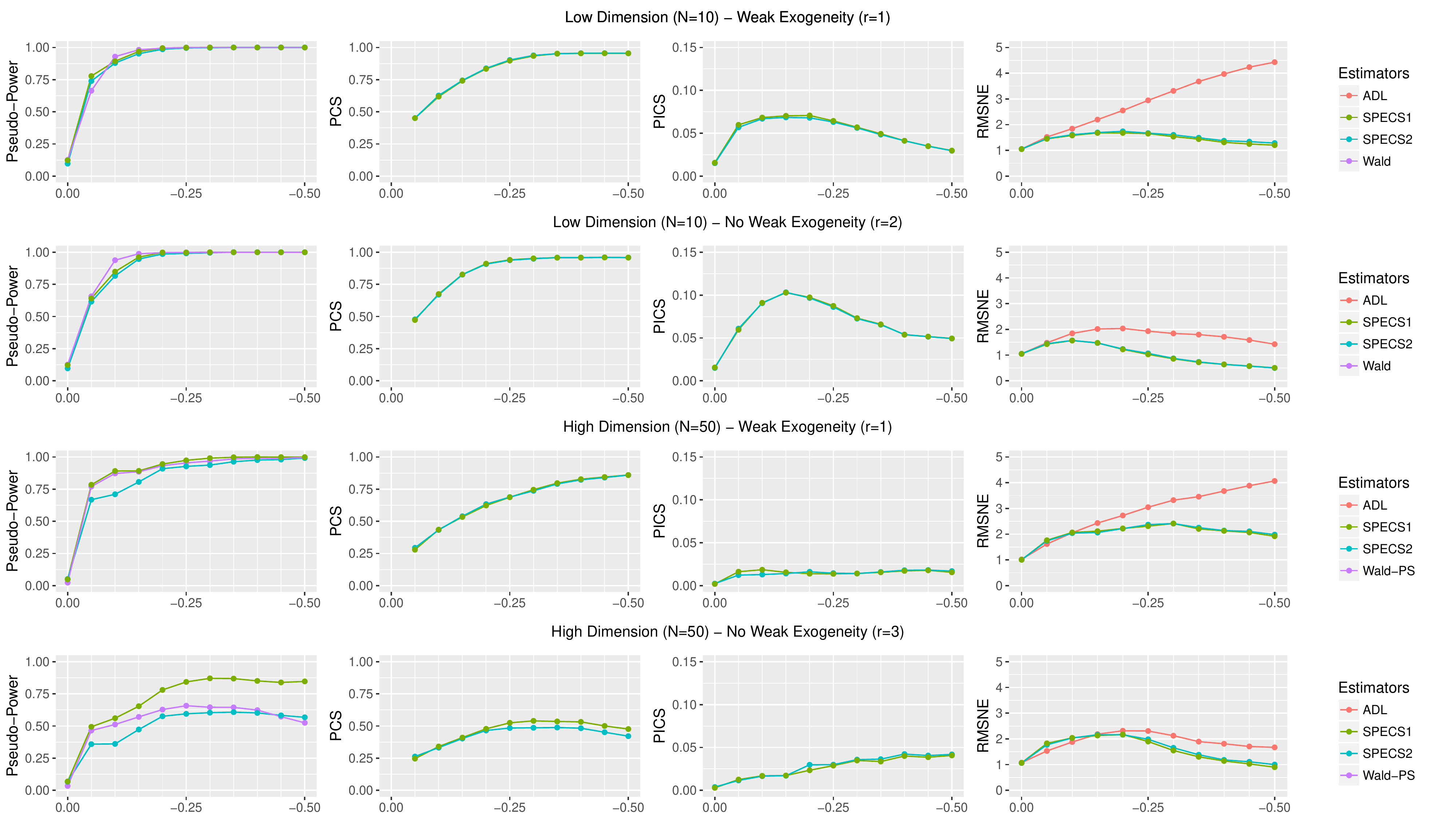}
    \caption{Pseudo-Power, Proportion of Correct Selection (PCS), Proportion of Incorrect Selection (PICS) and Root Mean Squared Nowcast Error (RMSNE) for Low- and High-Dimensional specifications. The adjustment rate multiplier $a$ is on the horizontal axis.}\label{Fig: metrics1}
\end{sidewaysfigure}

Figure \ref{Fig: metrics1} visually displays the evolution of our performance metrics over a range of values for $a$, representing increasingly faster rates of adjustment towards the long-run equilibrium. The first row of plots shows near-perfect performance of SPECS over all metrics. The pseudo-size is slightly lower than the size of the Wald test when the latter is controlled at 5\%, whereas the pseudo-power quickly approaches one. Following expectations, the pseudo-size for SPECS$_2$ is slightly lower as a result of the additional group penalty. Focussing on the selection of variables, we find that for faster adjustment rates, SPECS is able to exactly identify the sparsity pattern with very high frequency, as demonstrated by the PCS approaching 100\% and the PICS staying near 0\%. Furthermore, the MSNE obtained by our methods is close to the OLS oracle method and is substantially lower than the MSNE obtained by the ADL model for faster adjustment rates, while being almost identical absent of cointegration. The picture remains qualitatively similar when moving away from weak exogeneity while staying in a low-dimensional framework, although the gain in predictive performance over the ADL has decreased somewhat. We postulate that the ADL may benefit from a bias-variance tradeoff, given that the correctly specified single-equation model is sub-optimal in terms of efficiency absent of weak exogeneity compared to a full system estimator. Nonetheless, SPECS is clearly preferred.

The performance in the high-dimensional setting is displayed in rows 3 and 4 of Figure \ref{Fig: metrics1}. When the conditioning variables are weakly exogenous with respect to the parameters of interest, the selective capabilities remain strong. The pseudo-power demonstrates the attractive prospect of using our method as an alternative to cointegration testing, especially when taking into consideration that the traditional Wald test is infeasible in the current setting. In addition, the nowcasting performance remains far superior to that of the misspecified ADL. The last row depicts the performance absent of weak exogeneity. In this setting, exact identification of the implied cointegrating vector occurs less frequently, which seems to negatively impact the nowcasting performance. However, the misspecified ADL is still outperformed, despite the deterioration in the selective capabilities of our method.

\subsection{Mixed Orders of Integration}\label{sec:sim_mixed}

We next analyze the performance of SPECS on datasets containing variables with mixed orders of integration. The aim of this section is to gain an understanding of the relative performance of SPECS when not all time series are (co)integrated and to compare the performance of SPECS to traditional approaches that rely on pre-testing. The latter goal is attained by adding an additional penalized ADL model to the comparison, namely one in which the data is first corrected for non-stationarity based on a pre-testing procedure in which an Augmented Dickey-Fuller (ADF) test is performed on the individual series. We refer to this procedure as the ADL-ADF model. Based on the general DGP \eqref{eq:general DGP sim}, we distinguish four different cases, corresponding to (i) different orders of the dependent variable ($I(0)$/$I(1)$) and (ii) different degrees of persistence in the stationary variables (low/high). The choice to include varying degrees of persistence is motivated by the conjecture that the performance of the pre-testing procedure incorporated in the ADL-ADF model may deteriorate when the degree of persistence increases, which in turn translates to a decrease in the overall performance of the procedure. 

\begin{table}[t]
\caption{Simulation Design for the Second Study (Mixed Orders of Integration)}\label{table:DGP mixed order}
\begin{tabularx}{\textwidth}{cccc}
\hline
Mixed Order & $\bA$ & $\bB$ & $\bdelta$ \tabularnewline
\hline 
$y \sim I(0)$ & $\begin{bmatrix}
1 & 0 & \bm{0}_{1\times24}\\
\bm{0}_{15\times 1} & a\bB^* & \bm{0}_{15\times 24}\\
\bm{0}_{10\times 1} & \bm{0}_{10\times 3} & \bm{0}_{10 \times 24}\\
\bm{0}_{24 \times 1} & \bm{0}_{24 \times 3} & \bI_{24}
\end{bmatrix}$ & $\begin{bmatrix}
-b & 0 & \bm{0}_{1\times24}\\
\bm{0}_{15\times 1} & \bB^* & \bm{0}_{15\times 24}\\
\bm{0}_{10\times 1} & \bm{0}_{10\times 3} & \bm{0}_{10 \times 24}\\
\bm{0}_{24 \times 1} & \bm{0}_{24 \times 3} & -\bm{B}_{24\times 24}
\end{bmatrix}$ & $\begin{bmatrix}
-1\\
-\rho a\tilde{\biota}\\
\bm{0}_{44\times 1}
\end{bmatrix}$ \tabularnewline
$y \sim I(1)$ & $\begin{bmatrix}
a\bB^* & \bm{0}_{15\times 25}\\
\bm{0}_{10 \times 3} & \bm{0}_{10 \times 25}\\
\bm{0}_{25 \times 3} & \bI_{25}
\end{bmatrix}$ & $\begin{bmatrix}
\bB^* & \bm{0}_{15\times 25}\\
\bm{0}_{10 \times 3} & \bm{0}_{10 \times 25}\\
\bm{0}_{25 \times 3} & -\tilde{\bB}_{25 \times 25}
\end{bmatrix}$ & $(1+\rho)a \cdot \begin{bmatrix}
\tilde{\biota}\\
\bm{0}_{45\times 1}
\end{bmatrix}$, \tabularnewline
\hline 
\end{tabularx}
\caption*{Notes: see notes in Table \ref{table:DGP WE and Dim}. Additionally, we define $b=1$ ($b \sim U(0,0.2)$) and $\tilde{\bB}$ as a diagonal matrix with $b_{ii}=1$ ($b_{ii} \sim U(0,0.2)$) in the absence (presence) of persistence, and $\bB^* = (\bm{1}_{3\times 3} \otimes \tilde{\biota})$.}
\end{table}

\begin{sidewaysfigure}
\includegraphics[width=\textwidth]{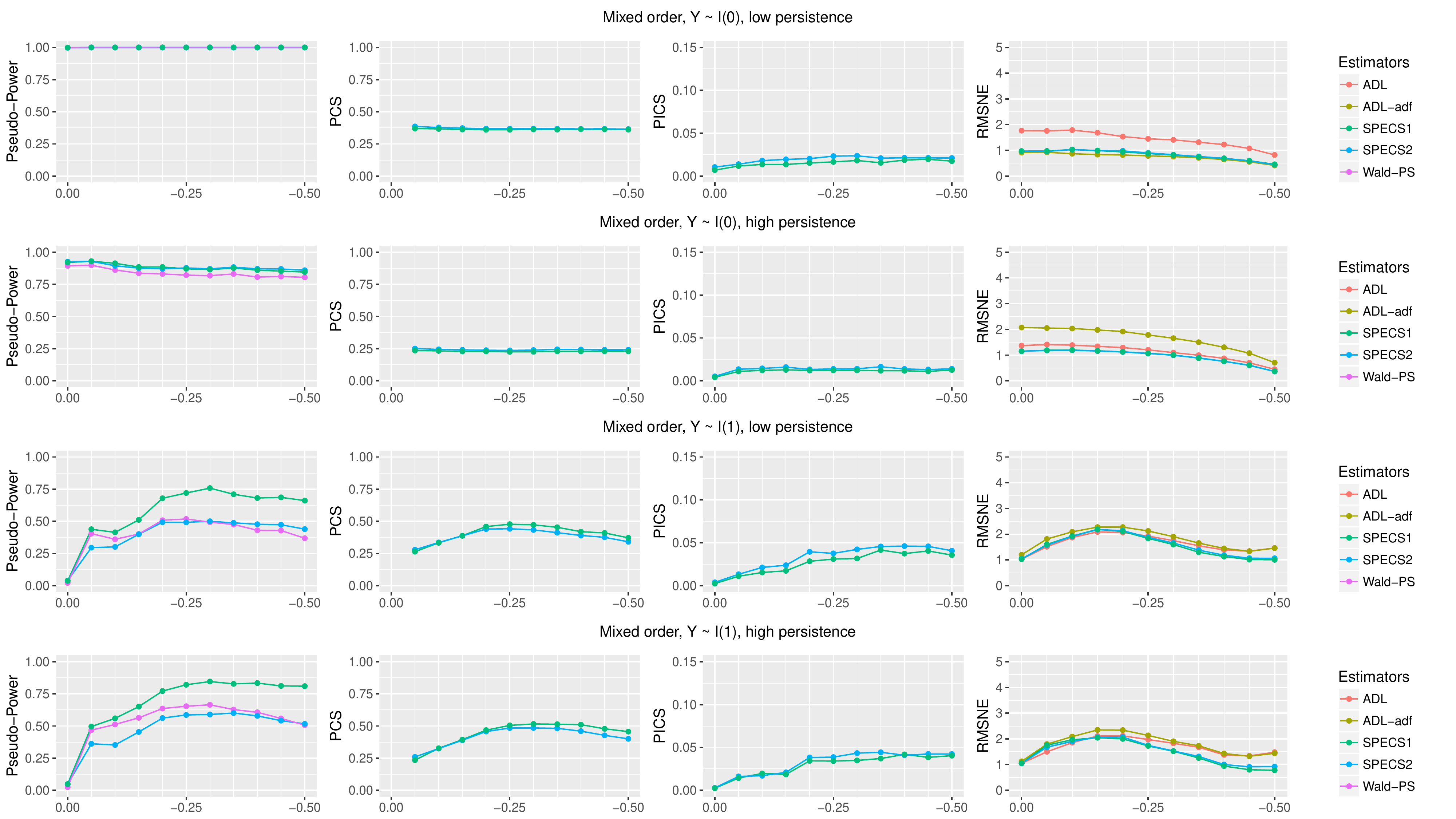}
        \caption{Pseudo-Power, Proportion of Correct Selection (PCS), Proportion of Incorrect Selection (PICS) and Root Mean Squared Nowcast Error (RMSNE) for four Mixed Order specifications. The adjustment rate multiplier $a$ is on the horizontal axis.}\label{Fig: metrics2}
\end{sidewaysfigure}

The parameter settings for the varying DGPs, displayed in Table \ref{table:DGP mixed order}, are chosen such that they allow for a subset of stationary variables in the system. In particular, we first consider a scenario in which the dependent variable itself admits a stationary autoregressive representation in levels. In addition, based on their cross-sectional ordering, the first 15 variables after $y$ are cointegrated based on three cointegrating vectors, the next 10 variables are non-cointegrated random walks, and the last 24 variables all admit a stationary autoregressive structure in levels. The degree of persistence in the stationary variables is regulated by the diagonal matrix $\tilde{\bB}$ in $\bB$, with elements $b_{ii} = 1$ in the low  persistence case and $b_{ii} \sim U(0,0.2)$ in the high persistence case. It can be seen from the last column in Table \ref{table:DGP mixed order}, that in line with the stationarity of the dependent variable, the first element in $\bdelta$ will always be equal to $-1$, whereas an additional five-dimensional cointegrating vector enters the single-equation model for positive values of $a$. For the scenario in which the dependent variable is integrated of order one, the first 15 variables (including $y$) are all cointegrated based on three cointegrating vectors, the next 10 variables are non-cointegrated random walks, whereas the last 15 variables all admit a stationary autoregressive representation. The persistence in the stationary variables is regulated similar to the previous case. Now, however, it is clear from the last column in Table \ref{table:DGP mixed order} that $\bdelta \neq \bm{0}$ only if $a > 0$, such that lagged levels only enter the single-equation when $y$ is cointegrated with its neighbouring variables. We display the performance of the models in Figure \ref{Fig: metrics2}.

In the first row of Figure \ref{Fig: metrics2}, corresponding to $y \sim I(0)$ and low persistence, SPECS correctly selects the lagged dependent variable in all simulation trials, such that the pseudo-power is always 1. Interestingly, PCS also seems constant around 35\%. Upon closer inspection, we find that SPECS chooses an alternative representation of the single-equation model in which the contribution of the non-trivial cointegrating vector seems to be absorbed in the lagged level of the dependent variable. While the resulting model differs from the implied oracle model, which is indeed accurately estimated by the OLS oracle procedure, the model choice seems motivated by a favourable bias-variance trade-off. In line with this conjecture, the nowcast performance of SPECS occasionally exceeds the OLS oracle procedure's where a larger number of parameters is estimated. The standard ADL nowcasts are again inferior, whereas the ADL-ADF model seems to benefit from correct identification of the stationarity of the dependent variable, which is particularly relevant given that the dependent variable itself is a main component in the optimal forecast. However, the nowcast accuracy of SPECS is almost identical to that of the ADL-ADF model, a finding that we interpret as reassuring and confirmatory of our claim that SPECS may be used without any pre-testing procedure. Moreover, the absence of strong persistence in the stationary variables idealizes the results of the ADL-ADF procedure. 

In typical macroeconomic applications many time series that are considered as I(0) display much slower mean reversion and, consequently, are more difficult to correctly identify as being stationary.\footnote{For example, the ten time series in the popular Fred-MD dataset which \citet{McCracken2016} propose to be I(0), i.e. the series corresponding to a tcode of one, all display strong persistence or near unit root behaviour, with the smallest estimated AR(1) coefficient exceeding 0.86.} Accordingly, in row 2 we display the result for a DGP where the stationary variables display more persistent behaviour. The performance of SPECS remains largely unaffected, whereas the nowcasting performance of the ADL-ADF model deteriorates drastically. We stress the relevance of this result, given that the estimation of ADL models after pre-testing for non-stationarity is fairly common practice. Somewhat surprisingly, the ADL model in differences nowcasts almost as well as SPECS here. Overall, however, the nowcast accuracy of SPECS remains the highest and, equally important, most stable across all specifications. 

Continuing the analysis of mixed order datasets, rows 3 and 4 of Figure \ref{Fig: metrics2} display the results for DGPs where the dependent variable is generated as being integrated of order one. The pseudo-power plot clearly reflects that $\bdelta \neq \bm{0}$ only when $a>0$. Furthermore, while SPECS performs well at removing the irrelevant variables, the relevant variables are not all selected correctly, resulting in somewhat lower values for the PCS metric. Nevertheless, the nowcast performance remains superior to that of the ADL model, especially in the presence of cointegration with fast adjustment rates.

\subsection{Non-sparse Data Generating Processes}\label{Sec:adverse}

To avoid idealizing the results through a choice of DGPs that suits our estimator, this section considers the performance of the penalized regression estimators in two different non-sparse settings. First, we consider an explicitly constructed VECM that contains many small, but non-zero coefficients. Second, we consider a DGP that contains a non-stationary factor structure on which the single-equation model is likely misspecified.

The non-sparse VECM is generated according to \eqref{eq:general DGP sim} with $\bB = \bI_3 \otimes \tilde{\biota}$, where $\tilde{\biota} = (1,-\biota_4^\prime)^\prime$, and $\bA = a\bB$ for $a=0,-0.05,\ldots,-0.5$. Hence, $N=15$ and the total number of parameters to estimate (including a constant and linear trend) is $N(p+2)+1=46$. A major difference with Section \ref{Sec:dim_we} is that we do not generate the covariance matrix of the errors as a Toeplitz-matrix, the latter being a crucial driver of sparsity in the preceding sections. Instead, we implement the procedure detailed in \citet[][p. 277-278]{Chang2004}, in which we 
    generate a $(N\times N)$ matrix $\bU$ with $u_{ij} \sim U(0,1)$ to construct the orthonormal matrix $\bH = \bU\left(\bU^\prime\bU\right)^{-1/2}$,
and generate a set of $N$ eigenvalues, $\lambda_1,\ldots,\lambda_N$, where $\lambda_1 = 0.01$, $\lambda_N=1$ and $\lambda_2,\ldots,\lambda_{N-1} \sim U(0.1,1)$ to construct $\bLambda = diag(\lambda_1,\ldots,\lambda_N)$. We then
construct the covariance matrix as $\bSigma = \bH\bLambda\bH^\prime$.
At each simulation trial, we generate a new $\bSigma$ such that the results cannot be attributed to a specific random draw of the covariance matrix. Based on this construction, $\bpi_0$, as defined below \eqref{eq:CECM}, and $\bdelta$ are non-sparse vectors with small elements; even in the setting with the strongest cointegration, i.e. $a=-0.5$, the median magnitude of the coefficients in $\bdelta$ across all trial is only 0.12. As before, we set $T=100$ and perform 1,000 simulation trials.

\begin{figure}
    \centering
    \includegraphics[width=\textwidth]{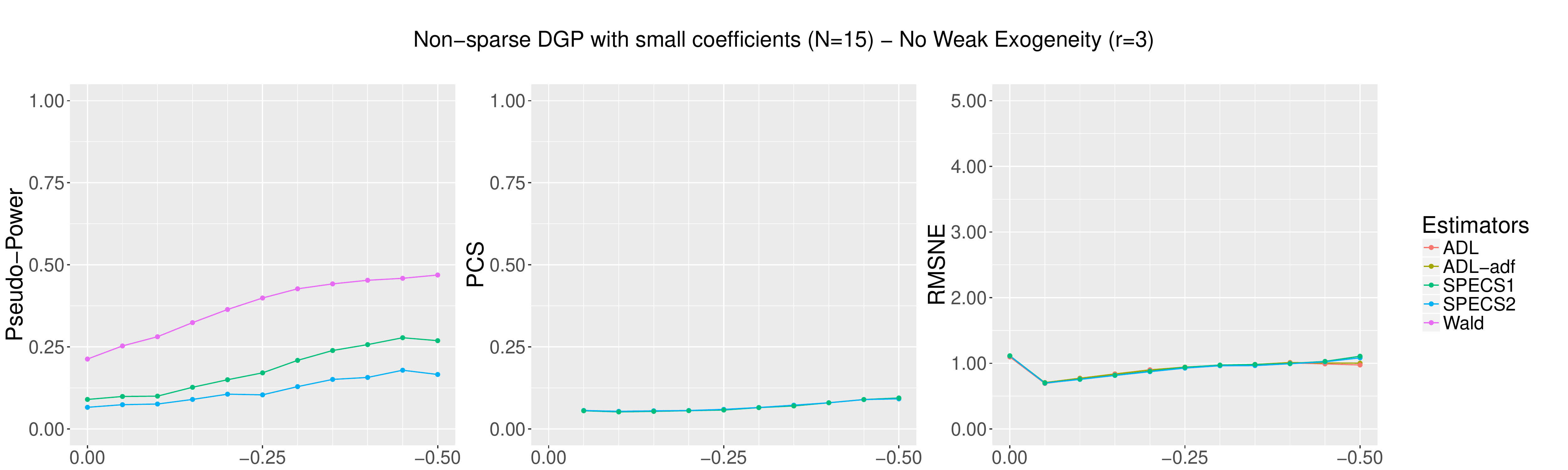}
    \caption{Pseudo-Power, Proportion of Correct Selection (PCS), Proportion of Incorrect Selection (PICS) and Root Mean Squared Nowcast Error (RMSNE). The adjustment rate multiplier $a$ is on the horizontal axis.}
    \label{Fig:non-sparse}
\end{figure}

The results are displayed in Figure \ref{Fig:non-sparse}, which contain a number of interesting results. Unsurprisingly, all estimators obtain a substantially lower (pseudo-)power in the current framework. The $\ell_1$-regularized estimators seem more sensitive to this than the traditional Wald estimator considered in \citet{Boswijk1994}. In line with the weak power, we observe that the PCS for both SPECS$_1$ and SPECS$_2$ is low, with on average only 0.75 out of 15 variables being included in levels.\footnote{The PICS is zero for all $a>0$, simply because the DGP is non-sparse, and is omitted accordingly.} Appropriate inference in the current setting is a difficult task and direct application of SPECS without alteration does not seem to be a feasible strategy. The development of a uniformly valid post-selection inference procedure, such as the desparsified lasso of \citet{Vandegeer2014}, may alleviate some of these issues. While we consider this an interesting avenue of research, it is outside the scope of the current paper. 

While these results may seem discouraging, the results on the nowcast accuracy display a different story. The mean-squared nowcast errors, relative to the OLS oracle procedure, are almost always below one and are similar for the SPECS and penalized ADL estimators. This highlights that the signal of the long-run component is so weak, that the estimation of a misspecified model which ignores cointegration benefits from a favourable bias-variance tradeoff. Therefore, the conclusion remains that SPECS obtains superior predictive performance relative to methods that ignore cointegration when the long-run component provides a strong signal, without sacrificing performance absent of cointegration or in the presence of very weak cointegration.

The second, and final, DGP that we consider contains a non-stationary factor structure and corresponds to setting III in \citet[][p. 92]{Palm2011}. We allow for contemporaneous correlation and dynamic structures in both the error processes driving the ``observable'' data and the idiosyncratic component in the factor structure. The DGP is given by $\bz_t = \blambda f_t + \bomega_t$, where $\bz_t$ is a $(50 \times 1)$ time series process, $f_t$ is a single scalar factor and
\begin{equation*}
f_t = \phi f_{t-1} + \zeta_t, \qquad \omega_{i,t} = \theta_i\omega_{i,t-1} + v_{i,t}.
\end{equation*}
Furthermore,
\begin{align*}
\bv_t &=  \bA_1\bv_{t-1} + \bepsilon_{1,t} + \bB_1 \bepsilon_{1,t-1},\qquad
\zeta_t = \alpha_2\zeta_{t-1} + \epsilon_{2,t} + \beta_2\epsilon_{2,t-1},
\end{align*}
where $\bepsilon_{1,t} \sim \mathcal{N}(\bm{0},\bSigma)$, with $\bSigma$ again generated as in \citet{Chang2004}, and $\epsilon_{2,t} \sim \mathcal{N}(0,1)$.

The comparison focuses exclusively on the nowcasting performance for a setting without dynamics ($\bA_1=\bB_1=\bm{0}$ and $\alpha_2=\beta_2=0$) and a setting with dynamics ($\alpha_2=\beta_2=0.4$). The construction of $\bA_1$ and $\bB_1$ is analogous to \citet[][p. 93]{Palm2011}. We report the RMSNEs of SPECS relative to the ADL in Table \ref{table:factor dgp}. Given that the single-equation model is misspecified in this setup, it is unreasonable to expect SPECS to outperform. Indeed, we observe that the RMSNEs are all very close to one and, while in most cases the ADL model performs slightly better, the difference seems negligible. Hence, the risk of using SPECS to estimate a misspecified model in the sense considered here, does not seem to be higher than the use of the alternative ADL model, whereas the relative merits of SPECS when applied to a wide range of correctly specified models are evident from the first part of the simulations.

\begin{table}
\caption{Nowcasting performance on a DGP with a non-stationary factor.}
\label{table:factor dgp}
\begin{tabularx}{\textwidth}{lXXX}
\hline 
\multicolumn{4}{c}{Root Mean Squared Nowcast Error}\tabularnewline
 & SPECS$_1$ & SPECS$_2$ & SPECS$_1$ - OLS\tabularnewline
\hline 
No Dynamics & 1.07 & 1.11 & 0.99\tabularnewline
Dynamics & 1.02 & 1.02 & 1.01\tabularnewline
\hline 
\end{tabularx}
\end{table}

\section{Empirical Application}\label{Sec:application}

Inspired by \citet{Choi2012}, we consider nowcasting Dutch unemployment with SPECS based on Google Trends data. Google Trends are time series consisting of normalized indices depicting the volume of search queries entered in Google, originating from a certain geographical area. The Dutch unemployment rates are made available by Statistics Netherlands, an autonomous administrative body focussing on the collection and publication of statistical information. These rates are published on a monthly basis with new releases being made available on the 15th of each new month. This misalignment of publication dates clearly illustrate a practically relevant scenario where improvements upon forward looking predictions of Dutch unemployment rates may be obtained by utilizing contemporaneous Google Trends series.

We collect a novel dataset containing seasonally unadjusted Dutch unemployment rates from the website of Statistics Netherlands\footnote{\href{http://statline.cbs.nl/StatWeb/publication/?VW=T&DM=SLEN&PA=80479eng&LA=EN}{\textcolor{blue}{http://statline.cbs.nl/StatWeb/publication/?VW=T\&DM=SLEN\&PA=80479eng\&LA=EN}}} and a set of manually selected Google Trends time series containing unemployment related search queries, such as ``Vacancy", ``Resume" and ``Unemployment Benefits". The dataset comprises of monthly observations ranging from January 2004 to December 2017. While the full dataset contains 100 unique search queries, a number of these contain zeroes for large sub-periods, indicating insufficient search volumes for those particular series. Consequently, we remove all series that are perfectly correlated over any sub-period consisting of 20\% of the total sample.\footnote{The dataset and corresponding \textit{R} package are available at \href{https://github.com/wijler/specs}{\textcolor{blue}{https://github.com/wijler/specs}}.}

The benchmark model we consider is an ADL model fitted to the differenced data. In detail, let $y_t$ and $\bx_t$ be the scalar unemployment rate and the vector of Google Trends series observed at time $t$, respectively, and define $\bz_t = (y_t,\bx_t^\prime)^\prime$. The benchmark ADL estimator fits
\begin{equation*}
\Delta y_t = \bpi_0^\prime \Delta \bx_t + \sum_{j=1}^p \bpi_j^\prime \Delta \bz_{t-j} + \mu_0 + \tau_0(t-1) + \epsilon_t.
\end{equation*}
However, this estimator ignores the order of integration of individual time series by differencing the whole dataset, while it is common practice to transform individual series to stationarity based on a preliminary test for unit roots. Hence, similar to Sections \ref{sec:sim_mixed} and \ref{Sec:adverse}, we include an additional ADL model where the decision to difference is based on a preliminary ADF test and refer to this method as ADL-ADF.\footnote{We note that none of the time series were found to be integrated of order 2. The outcome of the ADF test is reported for each time series in the online Appendix \ref{App: Variables}.} Finally, SPECS estimates
\begin{equation*}
\Delta y_t = \bdelta^\prime \bz_{t-1} + \bpi_0^\prime \Delta \bx_t + \sum_{j=1}^p \bpi_j^\prime \Delta \bz_{t-i} + \mu_0 + \tau_0(t-1) + \epsilon_t.
\end{equation*}
All tuning parameters are obtained by time series cross-validation and we use $k=1.1$ based on a preliminary analysis.\footnote{Comparing the nowcast accuracy for varying $k \in [0,4]$, we found the highest accuracy for $k = 1.1$.} The first nowcast is made by fitting the models on a window containing the first two-thirds of the complete sample, i.e. $t=1,\ldots,T_{c}$ with $T_c = \lceil \frac{2}{3}T \rceil$, based on which the nowcast for $\Delta y_{T_c+1}$ is produced. This procedure is repeated by rolling the window forward by one observation until the end of the sample is reached, producing a total of 54 pseudo out-of-sample nowcasts. Table \ref{Table:RMSNE Empirical} reports the MSNE relative to the ADL model for $p = 1,3,6$. 

\begin{table}
\begin{tabularx}{\textwidth}{XXXXX}
\hline 
$p$ & 
\# of parameters & ADL-ADF & SPECS$_1$ & SPECS$_2$\tabularnewline
\hline 
\hline 
1 & 262 & 1.27 & 0.99 & 1.07\tabularnewline
3 & 436 & 1.06 & 0.82{*} & 0.88\tabularnewline
6 & 697 & 0.90 & 0.90 & 0.84{*}\tabularnewline
\hline 
\end{tabularx}
\caption{Mean-Squared Nowcast Errora relative to the ADL model for varying number of lagged differences $p$. * denotes rejection by the Diebold-Mariano test at a 10\% significance level.}
\label{Table:RMSNE Empirical}
\end{table}

\begin{figure}[t]
\includegraphics[width=\textwidth]{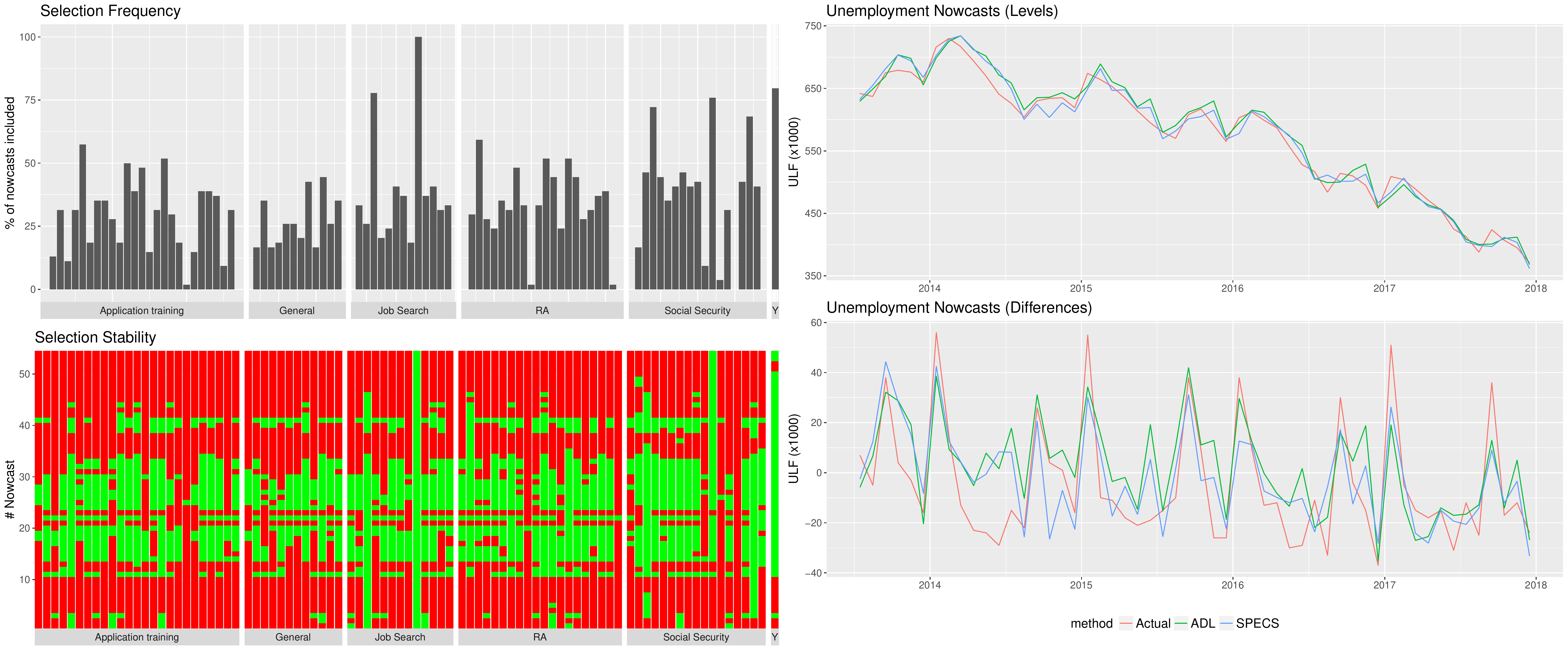}
\caption{\textit{Top-left}: Selection frequency, measured as the percentage of all nowcasts the variable was selected. \textit{Bottom-left}: Selection stability with green indicating a variable was included in the nowcast model and red indicating exclusion. \textit{Right}: Actual versus predicted unemployed labour force (ULF) in levels and differences.}
\label{Fig:empirical}
\end{figure}

The ADL-ADF estimator does not perform better than the regular ADL model for $p=1,3$, indicating that the potential for errors in pre-testing might lead to unfavourable results. SPECS performs well and is able to obtain smaller mean-squared nowcast errors than the ADL benchmark across almost all specifications, with the combination SPECS$_2$ and $p=1$ being the exception. Moreover, for SPECS$_1$ ($p=3$) and SPECS$_2$ ($p=6$), we find the differences in MSNE to be significant at the 10\% level according to the Diebold-Mariano test. The overall (unreported) MSNE is lowest for the SPECS$_1$ estimator based on $p=3$ lagged differences.  Given that the addition of lagged levels to the models improves the nowcast performance, the premise of cointegrating relationships between Dutch unemployment rates and Google Trends series seems likely. To further explore the presence of cointegration among our time series we group our variables in five categories; (1) Application Training, (2) General, (3) Job Search, (4) Recruitment Agencies (RA) and (5) Social Security. We narrow down our focus to the nowcasts of models with three lagged difference included, $p=3$, estimated by SPECS$_1$. In Figure \ref{Fig:empirical} we visually display the share of nowcasts in which the lagged levels of each variable are included in the estimated model. In addition, it depicts the selection stability of those variables, where a green colour indicates that a given variables is included in a given nowcast, and red vice versa. The figure also displays the actual unemployment rates compared to the nowcasted values.

Figure \ref{Fig:empirical} highlights that only few variables are consistently selected for all nowcasts, although in each category we can distinguish some variables that are included at higher frequencies. The variable whose lagged levels are always selected is ``Vakantiebaan", which is a search query for a temporary job during the summer holiday. We postulate that this variable is selected by SPECS to account for seasonality in the Dutch unemployment rates. In an unreported exercise we estimate the model with the addition of a set of eleven unpenalized dummies representing different months of the year. While the variable ``Vakantiebaan" is never selected, the mean squared nowcast error increases substantially. Hence, we opt to adhere to our standard model under the caveat that for at least one of the lagged levels included, seasonality effects rather than cointegration seem a more appropriate explanation for its inclusion. Other frequently included variables are queries for vacancies (``uwv.vacatures", 78\%), unemployment (``werkloos", 76\%) and social benefits (``ww uitkering", 72\%), where the stated percentages indicate the proportion of nowcast models in which the respective variables are selected. Furthermore, the last bar represents the frequency in which the lagged level of the Dutch unemployment rate is selected, which occurs for 43 out of 54 nowcasts (80\%). The frequent selection of the lagged level of unemployment rates in conjunction with the other lagged levels is indicative of the presence of cointegration among unemployment and Google Trends series. However, we do not attach any structural meaning to the found equilibria based on the difficulty of interpretation when one does not assume the presence of weak exogeneity.

In an attempt to gain insights into the temporal stability of our estimator, we visually display the selection stability in the bottom-left part of Figure \ref{Fig:empirical}. Generally, we see that for the early and later period of the sample very few time series enter the model in levels, whereas for the middle part of the sample the majority of variables are selected. The exact reason for these patterns to occur is unknown and raises questions on the stability of Google trends as informative predictors of Dutch unemployment rates. Standard feasible explanations concern structural instability in the DGP, seasonality effects or data idiosyncrasies. However, there are additional peculiarities specific to the use of Google trends such as normalization, data hubris and search algorithm dynamics, all of which might result in unstable performance \citep[cf.][]{Lazer2014}. Since the focus of this application is on the relative performance between our estimator and a common benchmark model, rather than on a structural analysis of the relation between Google Trends and unemployment rates, we leave this issue aside as it is outside the scope of the paper. Instead, we focus on the relative empirical performance of our methods, which, notwithstanding the aforementioned caveats, we deem convincingly favourable for SPECS. Finally, on the right of Figure \ref{Fig:empirical} we display the realized and predicted unemployment rates in levels and differences. Both the penalized ADL model and SPECS seem to follow the actual unemployment rates with reasonable accuracy, with the largest nowcast errors occurring in the first half of 2014. Prior to this period the unemployment rates had been steadily rising in the aftermath of the economic recession, whereas 2014 marks the start of a recovery period. Given that the models are fit on historical data, it is natural that the estimators overestimate the unemployment rate shortly after the start of the economic recovery. Perhaps not entirely coincidental, the start of the period over which the majority of lagged levels are included by SPECS coincides with this recovery period as well, thereby hinting towards structural instability in the DGP as a plausible cause for the observed selection instability.

\section{Conclusion}\label{Sec:Conclusion}

In this paper, we propose the use of SPECS as an automated approach for sparse single-equation error correction modelling in high-dimensional settings. SPECS is an intuitive estimator that applies penalized regression to a conditional error-correction model. We show that SPECS possesses the oracle property and is able to consistently select the long-run and short-run dynamics in the underlying DGP. These results are derived with the aid of a novel bound on the minimum eigenvalue of the sample covariance matrix containing integrated process, which may be of independent interest. Additionally, in pursuit of suitable weights that aid in the identification of the subset of relevant variables, we derive the consistency of the ridge estimator applied to the same model and demonstrate how ridge regression may be used to construct these weights.

We document favourable finite sample performance of SPECS by means of simulations and an empirical application. The simulation exercise confirms strong selective and predictive capabilities in both low and high dimensions with convincing gains over a benchmark penalized ADL model that ignores cointegration in the dataset. Furthermore, the simulation results demonstrate that the selective capabilities of SPECS remain adequate absent of weak exogeneity and the nowcasting performance remains superior to the benchmark. Finally, we consider an empirical application in which we nowcast the Dutch unemployment rate with the use of Google Trends series. Across all three different dynamic specifications considered, SPECS attains higher nowcast accuracy, thus confirming the findings from our simulation study. As a result, we believe that our proposed estimator, which is easily implemented with readily available tools at a low computational cost, offers a valuable tool for practitioners by enabling automated model estimation on relatively large and potentially non-stationary datasets and, most importantly, allowing to take into account potential (co)integration without requiring pre-testing procedures.

Finally, we highlight several important sources through which the assumptions and asymptotic framework may be generalized further. Sharper and more direct eigenvalue bounds can be utilized to cast SPECS into an even higher-dimensional setting. Similarly, a suitable compatibility condition can be used to validate the lasso as an initial estimator, resulting in improved weights and, again, a less restrictive asymptotic framework. These topics remain subject to our continuing investigation.


\label{Bibliography}
\bibliographystyle{apalike}
\bibliography{Bibliography}

\numberwithin{theorem}{section}
\numberwithin{lemma}{section}
\numberwithin{corollary}{section}
\numberwithin{proposition}{section}
\numberwithin{table}{section}

\begin{appendices}
\begin{small}
\section{Main Proofs} \label{App:main}

Before presenting the proofs, we start by defining several quantities of interest, some of which are simply repeated for the sake of convenience. First, recall that, under the assumption that $\bz_0 = \bm{0}$, the moving average representation of the observed time series is given by
\begin{equation}\label{eq:GRT_App}
\begin{split}
\bz_t &= \bC\bs_t + \bmu + \btau t +  \bC(L)\bepsilon_t,\qquad
\bZ_{-1} = \bS_{-1}\bC^\prime + \biota_T\bmu^\prime + \bt\btau^\prime + \bE_{-1}\bC^\prime(L),\\
\end{split}
\end{equation}
where $\bS_{-1} = (\bs_0,\ldots,\bs_{T-1})^\prime$, $\bC=\bB_\perp\left(\bA_\perp^\prime \left(\bI_N - \sum_{j=1}^p\bPhi_j\right) \bB_\perp\right)^{-1}\bA_\perp^\prime$, $\bt = (0,\ldots,T-1)^\prime$ and $\bE_{-1} = (\bepsilon_0,\ldots,\bepsilon_{T-1})^\prime$. Furthermore, by the Beveridge-Nelson decomposition $\bC(z) = \bC(1) + (1-z)\bC^*(z)$, where $\bC^*(z) = \sum_{l=0}^\infty \bC^*_l$ with $\bC^*_l = -\sum_{k=l+1}^\infty \bC_l$. Assumption \ref{Ass:Dependence} implies that,
\begin{equation*}
\sum_{l=0}^\infty \norm{\bC^*_l}_\infty = \sum_{l=0}^\infty \norm{\sum_{k=l+1}^\infty \bC_l}_\infty \leq \sum_{l=0}^\infty \sum_{k=l+1}^\infty \norm{\bC_l}_\infty = \sum_{l=1}^\infty l\norm{\bC_l}_\infty < \infty,
\end{equation*}
a property that is used to bound several quantities of interest in the proofs of our theoretical results. Letting $\bM = \bI_T - \bD\left(\bD^\prime\bD\right)^{-1}\bD^\prime$, we define
\begin{equation*}
    \tilde{\bZ}_{-1} = \bM\bZ_{-1} = \bM\bS_{-1}\bC^\prime + \bM\bE_{-1}\bC^\prime(L) = \tilde{\bS}_{-1}\bC^\prime + \tilde{\bE}_{-1}\bC^\prime(L),
\end{equation*}
where $\tilde{\bZ}_{-1} = \left(\tilde{z}_0,\ldots,\tilde{\bz}_{T-1}\right)^\prime$, and $\tilde{\bS}_{-1}$, $\tilde{\bE}_{-1}$ admitting a similar decomposition. From this representation, one can derive the stationary processes
\begin{equation*}\label{eq:GRT_coint}
\bB^\prime \tilde{\bz}_t = \bB^\prime\bC(L)\tilde{\bepsilon}_t = \bC^\beta(L)\tilde{\bepsilon}_t, \qquad \text{and} \qquad \Delta \tilde{\bz}_t = \bC\tilde{\bepsilon_t} + (1-L)\bC(L)\tilde{\bepsilon}_t = \bC^{\Delta}(L)\tilde{\bepsilon}_t.
\end{equation*}
Letting $\tilde{\bI} = (0\cdot \biota_{N-1},\bI_{N-1})$, we get the moving average representation
\begin{equation}\label{eq:GRT_w}
\tilde{\bw}_t = \begin{bmatrix}
\Delta \tilde{\bx}_t\\
\Delta \tilde{\bz}_{t-1}\\
\vdots\\
\Delta \tilde{\bz}_{t-p}
\end{bmatrix} = \begin{bmatrix}
\tilde{\bI}\bC^{\Delta}(L)\\
\bC^{\Delta}(L)L\\
\vdots\\
\bC^{\Delta}(L)L^p
\end{bmatrix}\tilde{\bepsilon}_t = \bC^w(L)\tilde{\bepsilon}_t, \qquad \tilde{\bW} = \tilde{\bE}C^{w\prime}(L),
\end{equation}
where $\tilde{\bE} = (\tilde{\bepsilon}_1,\ldots,\tilde{\bepsilon}_T)^\prime$. An additional useful representation follows from partitioning the data as $\bM\bV = \left(\bV_1,\bV_2\right)$, where $\bV_1 = \left(\tilde{\bZ}_{-1,S_\delta},\tilde{\bW}_{S_\pi}\right)$ contains the relevant variables. In congruence with Section \ref{sec:rot}, the $(\abs{S_\delta} \times r^*)$-dimensional matrix $\bB_{S_\delta}$ is defined as a basis matrix for the cointegration space of $\bz_{S_\delta,t}$ and $\bB_{S_\delta,\perp}$ is an $(\abs{S_\delta} \times \abs{S_\delta} - r^*)$-dimensional matrix for its left null space, i.e. $\bB_{S_\delta,\perp}^\prime\bB_{S_\delta} = \bm{0}$. Moreover, without loss of generality, we assume that the columns of $\bB_{S_\delta,\perp}$ are standardized to have unit $L_1$-norms. The $\bQ$-transformation is defined in \eqref{eq:Q} and the $\bQ$-transformed data are given by \eqref{eq:Q-transformed}. Denote the $t$-th row of $\bV_1\bQ^\prime$ by $\bv_t = \left(\bv_{1,t}^\prime,\bv_{2,t}^\prime\right)^\prime$, where 
\begin{equation*}
\begin{split}
\bv_{1,t} &= \begin{bmatrix}
\bB_{S_\delta}^\prime \tilde{\bz}_{S_\delta,t-1}\\
\tilde{\bw}_{S_\pi,t}
\end{bmatrix} = \begin{bmatrix}
\bB_{S_\delta}^\prime\bC_{S_\delta}(L)L\\
\bC^w_{S_\pi}(L)
\end{bmatrix}\tilde{\bepsilon}_t =: \bC^v(L)\tilde{\bepsilon}_t,\\
\bv_{2,t} &= \bB_{S_\delta,\perp}^\prime \tilde{\bz}_{S_\delta,t-1} = \bB_{S_\delta,\perp}^\prime\bC_{S_\delta}\tilde{\bs}_{t-1} + \bB_{S_\delta,\perp}^\prime\bC_{S_\delta}(L)\tilde{\bepsilon}_{t-1}.
\end{split}
\end{equation*}
Let $s_\pi = \abs{S_\pi} + r^*$ and $s_\delta = \abs{S_\delta}-r^*$ and define the scaling matrix $\bS_T = \diag\left(\sqrt{T}\bI_{s_\pi},\frac{T}{\sqrt{s_\delta}}\bI_{s_\delta}\right)$. Then, we define the appropriately scaled sample covariance matrix as
$\hat{\bSigma} = \bS_T^{-1}\left(\sum_{t=1}^T \bv_t\bv_t^\prime\right)\bS_T^{-1} = \begin{bmatrix}
\hat{\bSigma}_{11} & \hat{\bSigma}_{12}\\
\hat{\bSigma}_{21} & \hat{\bSigma}_{22}
\end{bmatrix}$.
Based on these quantities, we proceed to describe a set of lemmas and propositions that are required for the proofs of the main theorems in this paper.

\subsection{Preliminary Lemmas}\label{App:Lemmas}
In this section, we list a set of preliminary results that are used in the proofs of our main theorems in Section \ref{App:Theorems}. The proofs of all lemmas are delegated to the supplementary Appendix \ref{App:lemmas_proofs}. The first result will simplify the calculations on the stochastic components after regressing out $\bD$.
\begin{lemma}\label{Prop:idemp}
Let $\bA$ and $\bB$ denote arbitrary deterministic matrices of dimensions $(N_A \times d_A)$ and $(N_B \times d_B)$, respectively, with $\norm{\bA}_1 \leq K$ and $\norm{\bB}_1 \leq K$. Define two martingale difference sequences $\lbrace \epsilon^w_j\rbrace_{j=\infty}^\infty$ and $\lbrace \epsilon^u_j\rbrace_{j=\infty}^\infty$ of dimensions $N_A$ and $N_B$, respectively, where each sequence satisfies Assumption \ref{Ass:moments}. Any form of dependence between these two sequences is allowed and they may correspond to each other. Next, define a stationary $(T \times N_A)$ matrix $\bW = (\bw_1,\ldots,\bw_T)^\prime$ with $\bw_t = \bC^w(L)\bepsilon^w_{t-l}$, where $\bC^w(L)$ is an $(N_A \times N_A)$-dimensional matrix lag polynomial satisfying $\sum_{l=0}^\infty \norm{\bC^w_l}_\infty$. Similarly, let $\bU = (\bu_1,\ldots,\bu_T)^\prime$ with $\bu_t = \bC^u(L)\bepsilon^u_{t-l}$, where $\bC^u(L)$ is an $(N_B \times N_B)$-dimensional matrix lag polynomial satisfying $\sum_{l=0}^\infty \norm{\bC^u_l}_\infty$. Define the $(T \times N_A)$-dimensional partial sum matrix $\bS_{-1} = (\bm{0},\bs_1,\ldots,\bs_{T-1})^\prime$ with $\bs_t = \sum_{j=1}^{T-1} \bepsilon^w_j$. Then, letting $\bP = \bI_T - \bM$,
\begin{align*}
    (1)\; &\norm{\bA^\prime\bS_{-1}^\prime\bM\bS_{-1}\bA}_2 \leq \norm{\bA^\prime\bS_{-1}^\prime\bS_{-1}\bA}_2 \quad \text{and} \quad \norm{\bB^\prime\bU^\prime\bM\bU\bB}_2 \leq \norm{\bB^\prime\bU^\prime\bU\bB}_2,\\
    (2)\; &\norm{\bA^\prime\bS_{-1}^\prime\bP\bU\bB}_F = O_p\left(\sqrt{d_Ad_B}T\right)
    \quad
    \qquad (3)\; \norm{\bA^\prime\bW^\prime\bP\bU\bB}_F = O_p\left(\sqrt{d_Ad_B}\right).
\end{align*}
\end{lemma}

The second result describes a set on which SPECS obtains its selection consistency. 
\begin{lemma}\label{Prop:set_sufficiency}
Let $\bgamma_{S_\gamma}=(\bdelta_{S_\delta}^\prime,\bpi_{S_\pi}^\prime)^\prime$ denote the $\abs{S_\gamma}$-dimensional vector containing all non-zero coefficients and $\bomega = (\omega_1,\ldots,\omega_{N+M})^\prime$. Furthermore, define $\bs_1$  as the subgradient of $\norm{\hat{\bgamma}}_1$ and $\bs_2 = \left(\tilde{\bs}_2^\prime,\bm{0}^\prime\right)^\prime$, where $\tilde{\bs_2}$ is the subgradient of $\norm{\hat{\bdelta}}_2$. Then, $\Prob\left(\emph{sign}\left(\hat{\bgamma}\right) = \emph{sign}(\bgamma) \right) \geq \Prob(\mathcal{A}_T \cap \mathcal{B}_T)$, where
\begin{equation*}
\begin{split}
\mathcal{A}_T &= \bigcap_{i=1}^{\abs{S_\gamma}} \left\lbrace \abs{\left[\left(\bV_1^\prime\bV_1\right)^{-1}\bV_1^\prime \bepsilon_y\right]_i} < \abs{\gamma_{S_\gamma,i}} - \frac{\lambda_I}{2}\abs{\left[\left(\bV_1^\prime\bV_1\right)^{-1}\bOmega_1\bs_{1,S_\gamma}\right]_i} - \frac{\lambda_G}{2}\abs{\left[\left(\bV_1^\prime\bV_1\right)^{-1}\bs_{2,S_\gamma}\right]_i} \right\rbrace,\\
\mathcal{B}_T &= \bigcap_{i=1}^{\abs{S_\gamma^c}}\left\lbrace \abs{\left[\bV_2^\prime \bM_V\bepsilon_y\right]_i} < \frac{\lambda_I}{2} \left[\left(\bOmega_2\biota - \abs{\bV_2^\prime\bV_1\left(\bV_1^\prime\bV_1\right)^{-1}\bOmega_1\bs_{1,S_\gamma}}\right)\right]_i - \frac{\lambda_G}{2}\abs{\left[\bV_2^\prime\bV_1\left(\bV_1^\prime\bV_1\right)^{-1}\bs_{2,S_\gamma}\right]_i} \right\rbrace,
\end{split}
\end{equation*}
with $\bOmega_1 = \emph{diag}(\bomega_{S_\gamma})$, $\bOmega_2 =  \emph{diag}(\bomega_{S_\gamma^c})$, and $\bM_V = \bI_T - \bV_1\left(\bV_1^\prime\bV_1\right)^{-1}\bV_1^\prime$.
\end{lemma}

Next, we derive bounds on the empirical process that frequently appears throughout the proofs.
\begin{lemma}\label{Lemma:emp_proc}
Under Assumptions \ref{Ass:moments}-\ref{Ass:Dependence}, the stochastic order of the empirical process is
\begin{equation}
\norm{\bS_T^{-1}\bQ\bV_1^\prime\bepsilon_y}_2 = O_p\left(s_\delta + \sqrt{s_\pi}\right).
\end{equation}
\end{lemma}

Pursuing a minimum eigenvalue bound on $\hat{\bSigma}$, we show that its off-diagonal blocks converge to zero.
\begin{lemma}\label{Lemma:Sigma_12}
Under Assumptions \ref{Ass:moments}-\ref{ass:sparsity}, it holds that $\norm{\hat{\bSigma}_{12}}_2 \overset{p}{\to} 0$ as $T,s_\delta,s_\pi \to \infty$.
\end{lemma}

Combining Assumption \ref{Ass:eigenvalues} with Lemma \ref{Lemma:Sigma_12}, we obtain the following immediate result.
\begin{lemma}\label{Cor:eigenvalue}
Under Assumptions \ref{Ass:moments}-\ref{Ass:eigenvalues}, there exists a constant $\phi^* > 0$, such that, as $T,s_\delta,s_\pi \to \infty$, $\Prob\left(\lambda_1\left(\hat{\bSigma}\right) \geq \phi^*\right) \to 1$.
\end{lemma}

Finally, Lemma \ref{Lemma:emp_proc} and Lemma \ref{Cor:eigenvalue} have natural counterparts based on the full dataset.
\begin{lemma}\label{Cor:conversion}
Let $\hat{\bSigma}_{R}$ be as defined in \eqref{eq:Sigma_R}. Recall the definitions $N_\delta = N-r$, $M_\pi = M+r$ and assume that $\frac{N_\delta}{T^{1/4}} \to 0$ and $\frac{M_\pi}{\sqrt{T}} \to 0$. Then, under Assumptions \ref{Ass:moments}-\ref{Ass:Dependence} and \ref{Ass:eig_ridge},
\begin{enumerate}
\item $\Prob\left(\lambda_\min\left(\hat{\bSigma}_{R}\right) \geq \phi_R\right) \to 1$, as $T,N_\delta,M_\pi \to \infty$, and 
\item $\norm{\bS_R^{-1}\bQ_R\bV^\prime\bM\bepsilon_y}_2 = O_p\left(N_\delta + \sqrt{M_\pi}\right)$.
\end{enumerate}
\end{lemma}

\subsection{Proofs of Theorems \ref{Thm:Selection_Consistency} and  \ref{Thm:Estimation_Consistency}}\label{App:Theorems}

In this section we present the proofs of Theorems \ref{Thm:Selection_Consistency} and  \ref{Thm:Estimation_Consistency}. The proofs of Theorem \ref{Thm:ridge} and Corollary \ref{Cor:OLS_oracle} are delegated to the Supplementary Appendix \ref{Sec:App_cor1_thm3}.

\begin{proof}[\textbf{Proof of Theorem \ref{Thm:Selection_Consistency}}]
Based on Lemma \ref{Prop:set_sufficiency}, it suffices to show that $\Prob(\mathcal{A}_T \cap \mathcal{B}_T) \to 1$ as $T,N \to \infty$ or, equivalently, that $\Prob(\mathcal{A}_T^c) \rightarrow 0$ and $\Prob(\mathcal{B}_T^c) \rightarrow 0$. Thus, we start by deriving that $\Prob(\mathcal{A}_T^c) \rightarrow 0$.

Recall the definitions of $\bS_T = \diag\left(\sqrt{T}\bI_{s_\pi},\frac{T}{\sqrt{s_\delta}}\bI_{s_\delta}\right)$ and define $\bQ$ as in \eqref{eq:Q}, with $\norm{\bQ}_\infty \leq 1$ by the normalization on $\bB_{S_\delta}$ and $\bB_{S_\delta,\perp}$. Then, for $T$ large enough, we may write the set $\mathcal{A}_T^c$ as
\begin{equation}\label{eq:A_comp}
\begin{split}
&\mathcal{A}_T^c = \bigcup_{i=1}^{\abs{S_\gamma}} \left\lbrace \abs{\left[\bQ^\prime\bS_T^{-1}\left(\bS_T^{-1}\bQ\bV_1^\prime\bV_1\bQ^\prime\bS_T^{-1}\right)^{-1}\bS_T^{-1}\bQ\bV_1^\prime \bepsilon_y\right]_i}\right.\\
&\quad\quad \left. \geq \abs{\bgamma_{S_\gamma,i}} - \frac{\lambda_I}{2}\abs{\left[\bQ^\prime\bS_T^{-1}\left(\bS_T^{-1}\bQ\bV_1^\prime\bV_1\bQ^\prime\bS_T^{-1}\right)^{-1}\bS_T^{-1}\bQ\bOmega_1\bs_{1,S_\gamma}\right]_i} \right.\\
&\quad\quad -\left. \frac{\lambda_G}{2}\abs{\left[\bQ^\prime\bS_T^{-1}\left(\bS_T^{-1}\bQ\bV_1^\prime\bV_1\bQ^\prime\bS_T^{-1}\right)^{-1}\bS_T^{-1}\bQ\bs_{2,S_\gamma}\right]_i}\right\rbrace\\
&\quad= \bigcup_{i=1}^{\abs{S_\gamma}} \left\lbrace \abs{\left[\bQ^\prime\bS_T^{-1}\hat{\bSigma}^{-1}\bS_T^{-1}\bQ\bV_1^\prime \bepsilon_y\right]_i}\right.\\
&\quad\quad \left. \geq \abs{\bgamma_{S_\gamma,i}} - \frac{\lambda_I}{2}\abs{\left[\bQ^\prime\bS_T^{-1}\hat{\bSigma}^{-1}\bS_T^{-1}\bQ\bOmega_1\bs_{1,S_\gamma}\right]_i} - \frac{\lambda_G}{2}\abs{\left[\bQ^\prime\bS_T^{-1}\hat{\bSigma}^{-1}\bS_T^{-1}\bQ\bs_{2,S_\gamma}\right]_i}\right\rbrace\\
&\quad\subseteq \left\lbrace \norm{\bQ^\prime\bS_T^{-1}\hat{\bSigma}^{-1}\bS_T^{-1}\bQ\bV_1^\prime \bepsilon_y}_2 \right.\\
&\quad\quad \left.\geq \underset{1 \leq i \leq \abs{S_\gamma}}{\text{min}} \abs{\bgamma_{S_\gamma,i}} - \frac{\lambda_I}{2}\norm{\bQ^\prime\bS_T^{-1}\hat{\bSigma}^{-1}\bS_T^{-1}\bQ\bOmega_1\bs_{1,S_\gamma}}_2 - \frac{\lambda_G}{2}\norm{\bQ^\prime\bS_T^{-1}\hat{\bSigma}^{-1}\bS_T^{-1}\bQ\bs_{2,S_\gamma}}_2\right\rbrace
\end{split}
\end{equation}
We proceed by bounding the three quantities in \eqref{eq:A_comp} separately. First, by Assumption \ref{ass:sparsity}(1), $\frac{s_\delta}{T} \leq \frac{1}{\sqrt{T}} \Rightarrow \norm{\bS_T^{-1}}_2 = \frac{1}{\sqrt{T}}$ for large enough $T$. Moreover, letting $s = (s_\delta + s_\pi)$,
\begin{equation*}
\norm{\bS_T^{-1}\bQ\bOmega_1\bs_{1,S_\gamma}}_2 \leq \norm{\bS_T^{-1}}_2\norm{\bQ}_2\norm{\bOmega_1}_2\norm{\bs_{1,S_\gamma}}_2 \leq \frac{\sqrt{s}}{T^{1/2-\xi}}.
\end{equation*}
Then, by Assumption  \ref{Ass:eigenvalues}, it holds that
\begin{equation}\label{eq:bound_A1}
\begin{split}
&\norm{\bQ^\prime\bS_T^{-1}\hat{\bSigma}^{-1}\bS_T^{-1}\bQ\bV_1^\prime \bepsilon_y}_2 \leq \norm{\bS_T^{-1}}_2\norm{\bQ}_2\norm{\hat{\bSigma}^{-1}}_2\norm{\bS_T^{-1}\bQ\bV_1^\prime \bepsilon_y}_2 \leq \frac{\norm{\bS_T^{-1}\bQ\bV_1^\prime \bepsilon_y}_2}{\sqrt{T}\phi}
\end{split}
\end{equation}
on a set with probability converging to one. Furthermore, on the same set,
\begin{align}
&\norm{\bQ^\prime\bS_T^{-1}\hat{\bSigma}^{-1}\bS_T^{-1}\bQ\bOmega_1\bs_{1,S_\gamma}}_2 \leq \norm{\bS_T^{-1}\bQ}_2\norm{\hat{\bSigma}^{-1}}_2\norm{\bS_T^{-1}\bQ\bOmega_1\bs_{1,S_\gamma}}_2 \leq \frac{\sqrt{s}}{\phi T^{1-\xi}} \label{eq:bound_A2}\\
&\norm{\bQ^\prime\bS_T^{-1}\hat{\bSigma}^{-1}\bS_T^{-1}\bQ\bs_{2,S_\gamma}}_2 \leq \norm{\bS_T^{-1}\bQ}_2^2\norm{\hat{\bSigma}^{-1}}_2\norm{\bs_{2,S_\gamma}}_2 \leq \frac{1}{\phi T} \label{eq:bound_A3}
\end{align}

Based on \eqref{eq:bound_A1} and \eqref{eq:bound_A2}, we obtain probability bounds for $\mathcal{A}_T^c$ as follows:
\begin{equation}\label{eq:A complement}
\begin{split}
\Prob\left(\mathcal{A}_T^c\right)
& \leq \Prob\left( \frac{\norm{\bS_T^{-1}\bQ\bV_1^\prime \bepsilon_y}_2}{\sqrt{T}\phi} \geq \abs{\gamma_{\min}} - \frac{\lambda_I\sqrt{s}}{2\phi T^{1-\xi}} - \frac{\lambda_G}{2\phi T}\right) + o(1)\\
& = \Prob\left( \norm{\bS_T^{-1}\bQ\bV_1^\prime \bepsilon_y}_2 \geq \phi\abs{\gamma_{\min}}\sqrt{T} - \frac{\lambda_I\sqrt{s}}{2T^{1/2-\xi}} - \frac{\lambda_G}{2\phi\sqrt{T}}\right) + o(1).\\
\end{split}
\end{equation}
Then, to establish that $\Prob\left(\mathcal{A}_T^c\right) \to 0$, by Lemma \ref{Lemma:emp_proc} it suffices that $\frac{\abs{\gamma_\min}\sqrt{T}}{s_\delta + \sqrt{s_\pi}} \to \infty$, $\frac{\abs{\gamma_\min} T^{1-\xi}}{\lambda_I\sqrt{s}} \to \infty$ and $\frac{\abs{\gamma_\min}T}{\lambda_G} \to \infty$. The first condition corresponds to part (3) of Assumption \ref{ass:sparsity}. For the second and third condition, it follows from part (\ref{ass:reg_lambda}) of Assumption \ref{Ass:Regularization} that, for sufficiently large $T$,
\begin{equation*}
\frac{\abs{\gamma_\min} T^{1-\xi}}{\lambda_I\sqrt{s}} \geq \frac{\left(s_\delta + \sqrt{s_\pi}\right)T^{1/2-\xi}}{\lambda_I \sqrt{s}} \to \infty, \quad \text{and} \quad \frac{\abs{\gamma_\min}T}{\lambda_G} \geq \frac{\left(s_\delta + \sqrt{s_\pi}\right)\sqrt{T}}{\lambda_G} \to \infty,
\end{equation*}

Next, we show that $\Prob\left(\mathcal{B}_T^c\right) \to 0$. It follows from Lemma \ref{Prop:set_sufficiency} that $\mathcal{B}_T^c = \mathcal{B}_{z,T}^c \cup \mathcal{B}_{w,T}^c$, where
\begin{footnotesize}
\begin{equation}\label{eq:BZT_union}
\begin{split}
\mathcal{B}_{z,T}^c &= \bigcup_{i=1}^{\abs{S_\delta^c}}\left\lbrace \abs{\tilde{\bz}_{S_\delta^c,i}^\prime \bM_V\bepsilon_y} \geq \frac{\lambda_I}{2} \omega_{S_\delta^c,i} - \frac{\lambda_I}{2}\abs{\tilde{\bz}_{S_\delta^c,i}^\prime \bV_1\left(\bV_1^\prime\bV_1\right)^{-1}\bOmega_1\bs_{1,S_\gamma}} - \frac{\lambda_G}{2}\abs{\tilde{\bz}_{S_\delta^c,i}^\prime\bV_1\left(\bV_1^\prime\bV_1\right)^{-1}\bs_{2,S_\gamma}} \right\rbrace\\
\mathcal{B}_{w,T}^c&=\bigcup_{i=1}^{\abs{S_\pi^c}}\left\lbrace \abs{\tilde{\bw}_{S_\pi^c,i}^\prime \bM_V\bepsilon_y} \geq \frac{\lambda_I}{2} \omega_{S_\pi^c,i} - \frac{\lambda_I}{2}\abs{\tilde{\bw}_{S_\pi^c,i}^\prime \bV_1\left(\bV_1^\prime\bV_1\right)^{-1}\bOmega_1\bs_{1,S_\gamma}} - \frac{\lambda_G}{2}\abs{\tilde{\bw}_{S_\pi^c,i}^\prime\bV_1\left(\bV_1^\prime\bV_1\right)^{-1}\bs_{2,S_\gamma}}\right\rbrace
\end{split}
\end{equation}
\end{footnotesize}
and $\tilde{\bz}_{S_\delta^c,i}$ and $\tilde{\bw}_{S_\pi^c,i}$ represent the $i$-th columns of $\tilde{\bZ}_{-1,S_\delta^c}$ and $\tilde{\bW}_{S_\pi^c}$, respectively. For $\mathcal{B}_{z,T}^c$, note that
\begin{equation}\label{eq:BZT}
\begin{split}
\mathcal{B}_{z,T}^c &\subseteq \left\{ \norm{\tilde{\bZ}_{-1,S_\delta^c}^\prime \bM_V \bepsilon_y}_2 \geq \frac{\lambda_I}{2}\omega_{S_\delta^c,\min} - \frac{\lambda_I}{2}\norm{\tilde{\bZ}_{-1,S_\delta^c}^\prime\bV_1\left(\bV_1^\prime\bV_1\right)^{-1}\bOmega_1\bs_{1,S_\gamma}}_2\right.\\
&\left. \qquad\qquad\qquad\qquad\quad - \frac{\lambda_G}{2}\norm{\tilde{\bZ}_{-1,S_\delta^c}^\prime\bV_1\left(\bV_1^\prime\bV_1\right)^{-1}\bs_{2,S_\gamma}}_2\right\}.\\
\end{split}
\end{equation}
We proceed by bounding each individual term in \eqref{eq:BZT}. First, on a set with probability converging to 1,
\begin{equation}\label{eq:BZT_LHS}
\begin{split}
\norm{\tilde{\bZ}_{-1,S_\delta^c}^\prime \bM_V \bepsilon_y}_2 &\leq \norm{\tilde{\bZ}_{-1,S_\delta^c}^\prime\bepsilon_y}_2 + \norm{\tilde{\bZ}_{-1,S_\delta^c}^\prime \bV_1\left(\bV_1^\prime\bV_1\right)^{-1}\bV_1^\prime\bepsilon_y}_2\\
& \leq \norm{\tilde{\bZ}_{-1,S_\delta^c}^\prime\bepsilon_y}_2 + \frac{\norm{\tilde{\bZ}_{-1,S_\delta^c}}_2}{\sqrt{\phi}}\norm{\bS_T^{-1}\bQ\bV_1^\prime\bepsilon_y}_2,
\end{split}
\end{equation}
where the last inequality follows from the fact that
\begin{equation*}
\begin{split}
&\norm{\bV_1\left(\bV_1^\prime\bV_1\right)^{-1}\bV_1^\prime\bepsilon_y}_2 = \left(\bepsilon_y^\prime\bV_1\left(\bV_1^\prime\bV_1\right)^{-1}\bV_1^\prime\bepsilon_y\right)^{1/2}= \left(\bepsilon_y^\prime\bV_1\bQ^\prime\bS_T^{-1}\left(\bS_T^{-1}\bQ\bV_1^\prime\bV_1\bQ^\prime\bS_T^{-1}\right)^{-1}\bS_T^{-1}\bQ\bV_1^\prime\bepsilon_y\right)^{1/2}\\
&\quad = \norm{\left(\bS_T^{-1}\bQ\bV_1^\prime\bV_1\bQ^\prime\bS_T^{-1}\right)^{-1/2}\bS_T^{-1}\bQ\bV_1^\prime\bepsilon_y}_2 \leq \frac{\norm{\bS_T^{-1}\bQ\bV_1^\prime\bepsilon_y}_2}{\sqrt{\phi}}
\end{split}
\end{equation*}
by Lemma \ref{Cor:eigenvalue}. By the same argument, it follows that
\begin{align}
\norm{\tilde{\bZ}_{-1,S_\delta^c}^\prime\bV_1\left(\bV_1^\prime\bV_1\right)^{-1}\bOmega_1\bs_{1,S_\gamma}}_2 &\leq \frac{\norm{\tilde{\bZ}_{-1,S_\delta^c}}_2}{\sqrt{\phi}}\norm{\bS_T^{-1}\bQ\bOmega_1\bs_{1,S_\gamma}}_2 \leq \frac{\sqrt{s}\norm{\tilde{\bZ}_{-1,S_\delta^c}}_2}{\sqrt{\phi}T^{1/2-\xi}}, \label{eq:BZT_RHS1} \\
\norm{\tilde{\bZ}_{-1,S_\delta^c}^\prime\bV_1\left(\bV_1^\prime\bV_1\right)^{-1}\bs_{2,S_\gamma}}_2 &\leq \frac{\norm{\tilde{\bZ}_{-1,S_\delta^c}}_2}{\sqrt{\phi}}\norm{\bS_T^{-1}\bQ\bs_{2,S_\gamma}}_2 \leq \frac{\norm{\tilde{\bZ}_{-1,S_\delta^c}}_2}{\sqrt{\phi}T^{1/2}} \label{eq:BZT_RHS2}.
\end{align}
Then, plugging \eqref{eq:BZT_LHS}-\eqref{eq:BZT_RHS2} into \eqref{eq:BZT}, we obtain
\begin{equation}\label{eq:BZT2}
\begin{split}
\Prob\left(\mathcal{B}_{z,T}^c\right) &\leq \Prob\left(\norm{\tilde{\bZ}_{-1,S_\delta^c}^\prime\bepsilon_y}_2 \geq \frac{\lambda_I\omega_{S_\delta^c,\min}}{4} - \frac{\lambda_I\sqrt{s}\norm{\tilde{\bZ}_{-1,S_\delta^c}}_2}{4\sqrt{\phi}T^{1/2-\xi}} - \frac{\lambda_G\norm{\tilde{\bZ}_{-1,S_\delta^c}}_2}{4\sqrt{\phi}T^{1/2}}\right)\\
&\quad + \Prob\left(\norm{\bS_T^{-1}\bQ\bV_1^\prime\bepsilon_y}_2 \geq \frac{\sqrt{\phi}\lambda_I\omega_{S_\delta^c,\min}}{4\norm{\tilde{\bZ}_{-1,S_\delta^c}}_2} - \frac{\lambda_I\sqrt{s}}{4T^{1/2-\xi}} - \frac{\lambda_G}{4T^{1/2}}\right) + o(1).
\end{split}
\end{equation}

We proceed by deriving the stochastic order of the common term $\norm{\tilde{\bZ}_{-1,S_\delta^c}}_2$. Letting $\bU_{-1,S_\delta^c}$ denote the matrix containing the columns of $\bE_{-1}\bC^{w\prime}(L)$ indexed by $S_\delta^c$, and using that $\norm{\bM}_2 = 1$,
\begin{equation*}
\begin{split}
&\Prob\left(\norm{T^{-1}N^{-1/2}\tilde{\bZ}_{-1,S_\delta^c}}_2 \geq K_\epsilon\right) = \Prob\left(\norm{\bM\bS_{-1}\bC_{S_\delta^c}^\prime + \bM\bU_{-1,S_\delta}}_2 \geq K_\epsilon\right)\\
&\leq \Prob\left(\norm{\bC_{S_\delta^c}}_2\norm{T^{-1}N^{-1/2}\bS_{-1}}_2 \geq \frac{K_\epsilon}{2}\right)
+ \Prob\left(\norm{T^{-1}N^{-1/2}\bU_{-1,S_\delta^c}}_2 \geq \frac{K_\epsilon}{2}\right).
\end{split}
\end{equation*}
Furthermore, by Markov's inequality and Assumption \ref{Ass:moments}, for $K_\epsilon \geq \sqrt{\frac{4\norm{\bC_{S_\delta^c}}_2^2K}{\epsilon}}$,
\begin{equation*}
\begin{split}
\Prob\left(\norm{\bC_{S_\delta^c}}_2\norm{T^{-1}N^{-1/2}\bS_{-1}}_2 \geq \frac{K_\epsilon}{2}\right) &\leq \frac{4\norm{\bC_{S_\delta^c}}_2^2\sum_{i=1}^N\sum_{t=1}^{T-1} \E\left(s_{i,t}\right)^2}{K_\epsilon^2T^2N}
\leq \frac{4\norm{\bC_{S_\delta^c}}_2^2K}{K_\epsilon^2} \leq \epsilon,\\
\Prob\left(\norm{T^{-1}N^{-1/2}\bU_{-1,S_\delta^c}}_2 \geq \frac{K_\epsilon}{2}\right) &\leq \frac{4\sum_{i=1}^{\abs{S_\delta^c}}\sum_{t=1}^{T-1}\E\left(u_{S_\delta^c,i,t}\right)^2}{K_\epsilon^2T^2N} \leq \frac{4\phi_\max\sum_{i=1}^{\abs{S_\delta^c}}\sum_{l=0}^\infty \norm{\bc_{S_\delta^c,l,i}}_2^2}{K_\epsilon^2TN}\\
&\leq \frac{4\phi_\max\sum_{l=0}^\infty\norm{\bC_{S_\delta^c,l}}_2^2}{K_\epsilon^2T} \to 0.
\end{split}
\end{equation*}
Hence, 
for all $\epsilon>0$ there exist $K_\epsilon, T^*,N^* > 0$ such that $\Prob\left(\norm{\tilde{\bZ}_{-1,S_\delta^c}}_2 \geq T\sqrt{N}K_\epsilon \right) \leq \epsilon$ for all $T>T^*$ and $N>N^*$. Then, for sufficiently large $T,N$, the first RHS term of \eqref{eq:BZT2} is bounded by
\begin{equation}\label{eq:BZT2_1}
\begin{split}
&\Prob\left(\norm{\tilde{\bZ}_{-1,S_\delta^c}^\prime\bepsilon_y}_2 \geq \frac{\lambda_I\omega_{S_\delta^c,\min}}{4} - \frac{\lambda_I\sqrt{s}\norm{\tilde{\bZ}_{-1,S_\delta^c}}_2}{4\sqrt{\phi}T^{1/2-\xi}} - \frac{\lambda_G\norm{\tilde{\bZ}_{-1,S_\delta^c}}_2}{4\sqrt{\phi}T^{1/2}}\right)\\
&\leq \Prob\left(\norm{\bC_{S_\delta^c}\bS_{-1}^\prime \bM \bepsilon_y}_2 \geq \frac{\lambda_I\omega_{S_\delta^c,\min}}{8} - \frac{\lambda_IK_\epsilon\sqrt{s}T^{1/2+\xi}\sqrt{N}}{8\sqrt{\phi}} -  \frac{\lambda_GK_\epsilon \sqrt{TN}}{8\sqrt{\phi}}\right)\\ 
&\quad + \Prob\left(\norm{\bU_{-1,S_\delta^c}^\prime\bM\bepsilon_y}_2 \geq \frac{\lambda_I\omega_{S_\delta^c,\min}}{8} - \frac{\lambda_IK_\epsilon\sqrt{s}T^{1/2+\xi}\sqrt{N}}{8\sqrt{\phi}} -  \frac{\lambda_GK_\epsilon \sqrt{TN}}{8\sqrt{\phi}}\right) + \epsilon.
\end{split}
\end{equation}
As $\lbrace s_{i,t-1}\epsilon_{y,t}\rbrace$ is a m.d.s., it follows from 
Burkholder's inequality 
and the $C_r$-inequality that for $\epsilon > 0$,
\begin{equation}\label{eq:bound_Ze1}
\begin{split}
&\Prob\left(\frac{\norm{\bC_{S_\delta^c}\bS_{-1}^\prime\bepsilon_y}_2}{T\sqrt{N}} \geq K_\epsilon\right) \leq \frac{\norm{\bC_{S_\delta^c}}_2^2\sum_{i=1}^N\E\left(\sum_{t=2}^T s_{i,t-1}\epsilon_{y,t}\right)^2}{K_\epsilon^2T^2N} \\
&\leq \frac{K\norm{\bC_{S_\delta^c}}_2^2\sigma_y^2\sum_{i=1}^N\sum_{t=1}^{T-1}\E(s_{i,t})^2}{K_\epsilon^2T^2N} \leq \frac{K^*\norm{\bC_{S_\delta^c}}_2^2\sigma_y^2}{K_\epsilon^2} \leq \epsilon,\\
&\Prob\left(\frac{\norm{\bU_{-1,S_\delta^c}^\prime\bepsilon_y}_2}{T\sqrt{N}} \geq K_\epsilon\right) \leq \frac{\sum_{i=1}^{\abs{S_\delta^c}}\E\left(\sum_{t=2}^T\sum_{l=0}^\infty \bc_{S_\delta,l,i}^\prime\bepsilon_{t-1-l}\epsilon_{y,t}\right)^2}{K_\epsilon^2T^2N}\\
&\leq \frac{K\sigma_y^2\sum_{i=1}^{\abs{S_\delta^c}}\sum_{t=2}^T\sum_{l=0}^\infty \E\left(\bc_{S_\delta,l,i}^\prime\bepsilon_{t-1-l}\right)^2}{K_\epsilon^2T^2N} \leq \frac{K\sigma_y^2\phi_\max\sum_{l=0}^\infty \norm{\bC_{S_\delta,l}}_2^2}{K_\epsilon^2TN} \to 0,
\end{split}
\end{equation}
for $K_\epsilon \geq \sqrt{\frac{K^*\norm{\bC_{S_\delta^c}}_2^2\sigma_y^2}{\epsilon}}$. Then, 
part (2)-(3) of Lemma \ref{Prop:idemp}, it follows that $\norm{\tilde{\bZ}_{-1,S_\delta^c}^\prime\bepsilon_y}_2 = O_p(\sqrt{N}T)$. 
As $\omega_{S_\delta^c,\min}^{-1} = o_p\left(\frac{\lambda_I}{T\sqrt{N}}\right)$, $\omega_{S_\delta^c,\min}^{-1} = o_p\left(\frac{T^{\xi}}{\sqrt{s T N}} \right)$, and $\omega_{S_\delta^c,\min}^{-1} = o_p\left(\frac{\lambda_I}{\lambda_G\sqrt{TN}}\right)$ by Assumption \ref{Ass:Regularization}, we have that
\begin{equation*}
\Prob\left(\norm{\tilde{\bZ}_{-1,S_\delta^c}^\prime\bepsilon_y}_2 \geq \frac{\lambda_I\omega_{S_\delta^c,\min}}{4} - \frac{\lambda_I\sqrt{s}\norm{\tilde{\bZ}_{-1,S_\delta^c}}_2}{4\sqrt{\phi}T^{1/2-\xi}} - \frac{\lambda_G\norm{\tilde{\bZ}_{-1,S_\delta^c}}_2}{4\sqrt{\phi}T^{1/2}}\right) \to 0.
\end{equation*}

Next, we focus on the second RHS term of \eqref{eq:BZT2}. First, again using that $\norm{\tilde{\bZ}_{-1,S_\delta^c}}_2 = O_p(T\sqrt{N})$,
\begin{equation}\label{eq:BZT2_2}
\begin{split}
&\Prob\left(\norm{\bS_T^{-1}\bQ\bV_1^\prime\bepsilon_y}_2 \geq \frac{\sqrt{\phi}\lambda_I\omega_{S_\delta^c,\min}}{4\norm{\tilde{\bZ}_{-1,S_\delta^c}}_2} - \frac{\lambda_I\sqrt{s}}{4T^{1/2-\xi}} - \frac{\lambda_G}{4T^{1/2}}\right)\\
&\leq \Prob\left(\norm{\bS_T^{-1}\bQ\bV_1^\prime\bepsilon_y}_2 \geq \frac{\sqrt{\phi}\lambda_I\omega_{S_\delta^c,\min}}{4K_\epsilon T\sqrt{N}} - \frac{\lambda_I\sqrt{s}}{4T^{1/2-\xi}} - \frac{\lambda_G}{4T^{1/2}}\right) + \epsilon.\\
\end{split}
\end{equation}
Then, based on Lemma \ref{Lemma:emp_proc}, for the RHS of \eqref{eq:BZT2_2} to converge to zero, it is sufficient that
\begin{equation*}
\omega_{S_\delta^c,\min}^{-1} = o_p\left(\frac{\lambda_I}{(s_\delta + \sqrt{s_\pi})T\sqrt{N}}\right), \quad \omega_{S_\delta^c,\min}^{-1} = o_p\left(\frac{1}{\sqrt{s}T^{1/2 + \xi}\sqrt{N}}\right) \quad \text{and} \quad \omega_{S_\delta^c,\min}^{-1} = o_p\left(\frac{\lambda_I}{\lambda_G\sqrt{TN}}\right).
\end{equation*}
All three conditions are satisfied under Assumption \ref{Ass:Regularization}. Consequently, both RHS terms of \eqref{eq:BZT2} converge to zero, thereby concluding that $\Prob\left(\mathcal{B}_{z,T}^c\right) \to 0$.

\bigskip
It remains to prove that $\Prob\left(\mathcal{B}_{w,T}^c\right) \to 0$, where $\mathcal{B}_{w,T}^c$ is defined in \eqref{eq:BZT_union}. First, note that
\begin{equation*}
\begin{split}
\mathcal{B}_{w,T}^c \subseteq &\left\lbrace \norm{\tilde{\bW}_{S_\pi^c}^\prime \bM_V\bepsilon_y}_2 \geq  \frac{\lambda_I}{2} \omega_{S_\pi^c,\min} - \frac{\lambda_I}{2}\norm{\tilde{\bW}_{S_\pi^c}^\prime \bV_1\left(\bV_1^\prime\bV_1\right)^{-1}\bOmega_1\bs_{1,S_\gamma}}_2\right.\\
&\left. \qquad\qquad\qquad\qquad - \frac{\lambda_G}{2}\norm{\tilde{\bW}_{S_\pi^c}^\prime \bV_1\left(\bV_1^\prime\bV_1\right)^{-1}\bs_{2,S_\gamma}}_2 \right\rbrace.
\end{split}
\end{equation*}
Furthermore, on a set with probability converging to one,
\begin{align}
&\norm{\tilde{\bW}_{S_\pi^c}^\prime\bM_V \bepsilon_y}_2 \leq \norm{\tilde{\bW}_{S_\pi^c}^\prime\bepsilon_y}_2 + \frac{\norm{\tilde{\bW}_{S_\pi^c}}_2\norm{\bS_T^{-1}\bQ\bV_1^\prime\bepsilon_y}_2}{\sqrt{\phi}},\label{eq:bound_w1}\\
&\norm{\tilde{\bW}_{S_\pi^c}^\prime\bV_1\left(\bV_1^\prime\bV_1\right)^{-1}\bOmega_1\bs_{1,S_\gamma}}_2 \leq \frac{\norm{\tilde{\bW}_{S_\pi^c}}_2\norm{\bS_T^{-1}\bQ\bOmega_1\bs_{1,S_\gamma}}_2}{\sqrt{\phi}} \leq \frac{\sqrt{s}\norm{\tilde{\bW}_{S_\pi^c}}_2}{\sqrt{\phi}T^{1/2-\xi}} \label{eq:bound_w2}\\
&\norm{\tilde{\bW}_{S_\pi^c}^\prime\bV_1\left(\bV_1^\prime\bV_1\right)^{-1}\bs_{2,S_\gamma}}_2 \leq \frac{\norm{\tilde{\bW}_{S_\pi^c}}_2\norm{\bS_T^{-1}\bs_{2,S_\gamma}}_2}{\sqrt{\phi}} \leq \frac{\norm{\tilde{\bW}_{S_\pi^c}}_2}{\sqrt{\phi}T^{1/2}}\label{eq:bound_w3}.
\end{align}
Then, plugging \eqref{eq:bound_w1}-\eqref{eq:bound_w3} into $\mathcal{B}_{w,T}^c$ from \eqref{eq:BZT_union}, we obtain
\begin{equation}\label{eq:BWTs}
\begin{split}
\Prob\left(\mathcal{B}_{w,T}^c\right)
& \leq \Prob\left( \norm{\tilde{\bW}_{S_\pi^c}^\prime\bepsilon_y}_2 \geq \frac{\lambda_I\omega_{S_\pi^c,\min}}{4} - \frac{\lambda_I\sqrt{s}\norm{\tilde{\bW}_{S_\pi^c}}_2}{4\sqrt{\phi}T^{1/2-\xi}}  - \frac{\lambda_G\norm{\tilde{\bW}_{S_\pi^c}}_2}{4\sqrt{\phi}T^{1/2}}\right)\\
&\quad + \Prob\left( \norm{\bS_T^{-1}\bQ\bV_1^\prime\bepsilon_y}_2 \geq \frac{\lambda_I\sqrt{\phi}\omega_{S_\pi^c,\min}}{4\norm{\tilde{\bW}_{S_\pi^c}}_2} - \frac{\lambda_I\sqrt{s}}{4T^{1/2-\xi}} - \frac{\lambda_G}{4T^{1/2}}\right) + o(1)\\
&= \Prob\left(\mathcal{B}_{w_1,T}^c\right) + \Prob\left(\mathcal{B}_{w_2,T}^c\right) + o(1).
\end{split}
\end{equation}
Next, we derive the order of $\norm{\tilde{\bW}_{S_\pi^c}}_2$. From definition \eqref{eq:GRT_w}, and using that $\norm{\bM}_2=1$, it follows that
\begin{equation*}
   \norm{\tilde{\bW}_{S_\pi^c}}_2 = \norm{\bM\bE\bC_{S_\pi^c}^\prime(L)}_2 \leq \norm{\bE\bC_{S_\pi^c}^\prime(L)}_2 = \norm{\bW_{S_\pi^c}}_2.
\end{equation*}
Recalling that $w_{i,t} = \sum_{l=0}^\infty \bc^{w\prime}_{l,i}\bepsilon_{t-l}$, it holds that
\begin{equation*}
\E\left(w_{i,t}\right)^2 = \sum_{l=0}^\infty \bc^{w\prime}_{l,i}\bSigma_\epsilon \bc^w_{l,i} \leq \phi_\max \sum_{l=0}^\infty \norm{\bc^w_{l,i}}_2^2 \leq \phi_\max \sum_{l=0}^\infty \norm{\bC^w_l}_2^2,
\end{equation*}
by Assumption \ref{Ass:Dependence}. Then, for any $\epsilon > 0$, it follows that, for $K_\epsilon \geq \left(\phi_\max \sum_{l=0}^\infty \norm{\bC^w_l}_2^2\right)^{-1/2}$,
\begin{equation}\label{eq:W_bound}
\Prob\left(\frac{\norm{\tilde{\bW}_{S_\pi^c}}_2}{\sqrt{TM}} \geq K_\epsilon\right) \leq \frac{\sum_{i=1}^M\sum_{t=1}^T\E\left(w_{i,t}\right)^2}{K_\epsilon^2TM} \leq \frac{\phi_\max \sum_{l=0}^\infty \norm{\bC^w_l}_2^2}{K^2_\epsilon} \leq \epsilon.
\end{equation}
Furthermore, it is straightforward to verify that $\lbrace w_{i,t}\epsilon_{y,t} \rbrace$ is a martingale difference sequence. Thus, by the Markov bound and Burkholder's inequality,
\begin{equation}\label{eq:W_eps}
\begin{split}
&\Prob\left(\norm{\bW_{S_\pi^c}^\prime\bepsilon_y}_2 \geq K_\epsilon\sqrt{TM}\right) \leq \frac{\sum_{i=1}^M\E\left(\sum_{t=1}^T w_{i,t}\epsilon_{y,t}\right)^2}{K_\epsilon^2 TM} \leq \frac{K\sum_{i=1}^M\sum_{t=1}^T \E\left(w_{i,t}\epsilon_{y,t}\right)^2}{K_\epsilon^2TM}\\
&\quad \leq \frac{K\sum_{i=1}^M\sum_{t=1}^T \sum_{l_1,l_2=0}^\infty\sum_{j_1,j_2=1}^M \abs{c^w_{l_1,i,j_1}}\abs{c^w_{l_2,i,j_2}}\E\abs{\epsilon_{j_1,t-1_1}\epsilon_{j_2,t-1_2}\epsilon_{y,t}^2}}{K_\epsilon^2TM}\\
&\quad \leq \frac{K^*\sum_{i=1}^M \left(\sum_{l=0}^\infty \norm{\bc^w_{l,i}}_1\right)^2}{K_\epsilon^2 M} \leq \frac{K^* \left(\sum_{l=0}^\infty \norm{\bC^w_l}_\infty\right)^2}{K_\epsilon} \leq \epsilon,\\
\end{split}
\end{equation}
for $K_\epsilon \geq \left(\frac{K^* \left(\sum_{l=0}^\infty \norm{\bC^w_l}_\infty\right)^2}{\epsilon}\right)^{1/2}$. Then, 
part (3) of Lemma \ref{Prop:idemp} shows that $\norm{\bW_{S_\pi^c}^\prime\bM\bepsilon_y}_2 = O_p(\sqrt{TM})$. Using \eqref{eq:W_bound} to further simplify \eqref{eq:BWTs},
\begin{equation}\label{eq:BWT1}
\begin{split}
&\Prob\left( \norm{\tilde{\bW}_{S_\pi^c}^\prime\bepsilon_y}_2 \geq \frac{\lambda_I\omega_{S_\pi^c,\min}}{4} - \frac{\lambda_IK_\epsilon\sqrt{s}T^\xi\sqrt{M}}{4\sqrt{\phi}} - \frac{\lambda_GK_\epsilon\sqrt{M}}{4\sqrt{\phi}}\right) + \epsilon,
\end{split}
\end{equation}
such that \eqref{eq:W_eps} implies $\Prob\left(\mathcal{B}_{w_1,T}^c\right) \to 0$, if $\omega_{S_\pi^c,\min}^{-1} = o_p\left(\frac{\lambda_I }{\sqrt{TM}}\right)$, $\omega_{S_\pi^c,\min}^{-1} = o_p\left(\frac{1}{\sqrt{s}T^\xi\sqrt{M}}\right)$ and $\omega_{S_\pi^c,\min}^{-1} = o_p\left(\frac{\lambda_I \omega_{S_\pi^c,\min}}{\lambda_G\sqrt{M}}\right)$. 
as ensured by Assumption \ref{Ass:Regularization}. Similarly, 
for $\mathcal{B}_{w_2,T}^c$ as defined in \eqref{eq:BWTs} we get 
\begin{equation}\label{eq:BWT2}
\begin{split}
&\Prob\left(\mathcal{B}_{w_2,T}^c\right) \leq \Prob\left(\norm{\bS_T^{-1}\bQ\bV_1^\prime\bepsilon_y}_2 \geq \frac{\lambda_I\sqrt{\phi}\omega_{S_\pi^c,\min}}{4K_\epsilon\sqrt{TM}} - \frac{\lambda_I\sqrt{s}}{4T^{1/2-\xi}} - \frac{\lambda_G}{4T^{1/2}} \right) + \epsilon,
\end{split}
\end{equation}
such that, by Lemma \ref{Lemma:emp_proc}, sufficient conditions for $\Prob\left(\mathcal{B}_{w_2,T}^c\right) \to 0$ are given by
\begin{equation*}
\omega_{S_\pi^c,\min}^{-1} = o_p\left(\frac{\lambda_I}{(s_\delta + \sqrt{s_\pi})\sqrt{TM}}\right), \quad \omega_{S_\pi^c,\min}^{-1} = o_p\left(\frac{1}{\sqrt{s}T^\xi\sqrt{M}}\right) \quad \text{and} \quad \omega_{S_\pi^c,\min}^{-1} = \left(\frac{\lambda_I}{\lambda_G\sqrt{M}}\right).
\end{equation*}
All three conditions are satisfied under Assumption \ref{Ass:Regularization}. Hence, we may conclude that $\Prob\left(\mathcal{B}_{w,T}^c\right)\to 0$.
\end{proof}

\begin{proof}[\textbf{Proof of Theorem \ref{Thm:Estimation_Consistency}}]
First, we recall the definitions $\bV = \left(\bV_1,\bV_2\right)$, $\bV_1 = \left(\tilde{\bZ}_{-1,S_\delta},\tilde{\bW}_{S_\pi}\right)$, $\bOmega = \diag\left(\bomega\right)$ and $\bOmega_1 = \diag\left(\bomega_{S_\gamma}\right)$. Based on the first order conditions, it follows from \eqref{eq:FOC_rew1} that
\begin{equation}\label{eq:foc_est}
    \hat{\bgamma}_{S_\gamma} - \bgamma_{S_\gamma} = \left(\bV_1^\prime\bV_1\right)^{-1}\bV_1^\prime\bepsilon_y  - \frac{1}{2}\left(\bV_1^\prime\bV_1\right)^{-1}\left(\lambda_I\bOmega_1\bs_{1,S_\gamma} + \lambda_G\bs_{2,S_\gamma}\right),
\end{equation}
on a set with probability converging to one based on Theorem \ref{Thm:Selection_Consistency}. By pre-multiplying \eqref{eq:foc_est} by $\bS_T\bQ^{\prime -1}$ and taking the Euclidean norm on both sides, it follows that
\begin{equation}\label{eq:subgradient_bound}
\begin{split}
&\norm{\bS_T\bQ^{\prime -1}\left(\hat{\bgamma}_{S_\gamma}-\bgamma_{S_\gamma}\right)}_2\leq \norm{\left(\bS_T^{-1}\bQ\bV_1^\prime\bV_1\bQ^\prime\bS_T^{-1}\right)^{-1}}_2\norm{\bS_T^{-1}\bQ\left(\bV_1^\prime\bepsilon_y - \frac{\lambda_I}{2}\bOmega_1\bs_{1,S_\gamma} - \frac{\lambda_G}{2}\bs_{2,S_\gamma}\right)}_2\\
&\quad \leq \phi^{-1}\left(\norm{\bS_T^{-1}\bQ\bV_1^\prime\bepsilon_y}_2 + \frac{\lambda_I}{2}\norm{\bS_T^{-1}\bQ\bOmega_1\bs_{1,S_\gamma}}_2 +  \frac{\lambda_G}{2}\norm{\bS_T^{-1}\bQ\bs_{2,S_\gamma}}_2\right) + o_p(1),
\end{split}
\end{equation}
by Lemma \ref{Cor:eigenvalue}. We derive the stochastic order for the three RHS terms of \eqref{eq:subgradient_bound}. First,
$\norm{\bS_T^{-1}\bQ\bV_1^\prime\bepsilon_y}_2 = O_p\left(s_\delta + \sqrt{s_\pi}\right)$,
by Lemma \ref{Lemma:emp_proc}. By Assumption \ref{Ass:Regularization}, on a set 
with probability converging to one, the second term and third term on the RHS of \eqref{eq:subgradient_bound} 
are bounded by
\begin{align}
&\frac{\lambda_I}{2}\norm{\bS_T^{-1}\bQ\bOmega_1\bs_{1,S_\gamma}}_2 \leq \frac{\lambda_I}{2}\norm{\bS_T^{-1}}_2\norm{\bQ}_2\norm{\bOmega_1}_2\norm{\bs_{1,S_\gamma}}_2 \leq \frac{\lambda_I\sqrt{s}}{2T^{1/2-\xi}} = o\left(s_\delta + \sqrt{s_\pi}\right),\label{eq:Op_RHS2}\\
&\frac{\lambda_G}{2}\norm{\bS_T^{-1}\bQ\bs_{2,S_\gamma}}_2 \leq \frac{\lambda_G}{2}\norm{\bS_T^{-1}}_2\norm{\bQ}_2\norm{\bs_{2,S_\gamma}}_2 \leq \frac{\lambda_G}{\sqrt{T}} \to 0.\label{eq:Op_RHS3}
\end{align}
Hence, plugging these result into \eqref{eq:subgradient_bound}, we conclude that, as required,
\begin{equation*}
\norm{\bS_T\bQ^{\prime -1}\left(\hat{\bgamma}_{S_\gamma}-\bgamma_{S_\gamma}\right)}_2 = O_p\left(s_\delta + \sqrt{s_\pi}\right). \qedhere
\end{equation*}
\end{proof}

\section{Bounds on Minimum Eigenvalues}\label{App:eigenvalues}

In this Appendix, we provide sufficient conditions for Assumption \ref{Ass:eigenvalues}. We first present some preliminary results in Section \ref{App:eigenvalues_prelim} and main eigenvalue bounds in Section \ref{sec:minevbounds}. The proofs of these lemmas and theorems are delegated to the Supplementary Appendices \ref{App:lemmas_proofs} and \ref{sec:evproofs}, respectively.

\subsection{Preliminary Results}\label{App:eigenvalues_prelim}

We first present a general result linking the eigenvalues of two matrices together.

\begin{lemma}\label{Lemma:eig_transl}
Let $\bA$ and $\bB$ denote two $s$-dimensional square non-negative definite matrices. Then,
\begin{enumerate}[(1)]
\item for all $i=1,\ldots,s$, it holds that $\abs{\lambda_i\left(\bA\right) - \lambda_i(\bB)} \leq \norm{\bA-\bB}_2$,
\item if $\norm{\bA-\bB}_\max \leq \delta$, then $\lambda_\min(\bB) \geq \lambda_\min(\bA) - s\delta$.
\end{enumerate}
\end{lemma}

The following result demonstrates the issue of collinearity of integrated variables in high dimensions.

\begin{lemma}\label{Lemma:BM_int}
Define an $s$-dimensional white noise sequence $\bu_t \overset{i.i.d.}{\sim} \mathcal{N}(\bm{0},\bI_s)$ and let $\bh_t = \sum_{j=1}^t \bu_j$. Then, as $s,T \to \infty$, for any $\phi > 0$, 
\begin{equation}\label{eq:min_eig_conv}
\Prob\left(\lambda_\min\left(\frac{1}{T^2}\sum_{t=1}^T\bh_t\bh_t^\prime\right) > \phi \right) \to 0.
\end{equation}
\end{lemma}

\subsection{Main Results} \label{sec:minevbounds}

We first give a bound on the minimum eigenvalue of the covariance matrix of the stationary variables. This follows standard arguments in the literature, but is given for completeness.
\begin{theorem}\label{Lemma:min_eig1}
Define $\bSigma_{11} = \E\left(\bv_{1,t}\bv_{1,t}^\prime\right)$ and assume that $\lambda_\min\left(\bSigma_{11}\right) \geq 2\phi$ for some $\phi>0$. Then, under Assumptions \ref{Ass:moments}-\ref{Ass:Dependence} and \ref{ass:sparsity}(2), as $T,s_\delta,s_\pi \to \infty$ we have that
$\Prob\left(\lambda_\min\left(\hat{\bSigma}_{11}\right) \geq \phi\right) \to 1$.
\end{theorem}

Contrary to $\hat{\bSigma}_{11}$, the matrix $\hat{\bSigma}_{22} = \frac{s_\delta }{T^2}\bB_{S_\delta,\perp}^\prime\left(\sum_{t=1}^T\tilde{\bz}_{S_\delta,t}\tilde{\bz}_{S_\delta,t}^\prime\right)\bB_{S_\delta,\perp}$ does not converge in probability to a deterministic matrix. Accordingly we aim to bound $\hat{\bSigma}_{22}$ directly, under varying additional assumptions on the DGP and the growth rate of $s_\delta$. 
\begin{theorem}\label{Lemma:Zhang_LB}
Let $\hat{\bSigma}_{22}$ be as defined in Assumption \ref{Ass:eigenvalues} and assume that $\bepsilon_t \overset{i.i.d.}{\sim} \mathcal{N}\left(\bm{0},\bSigma_\epsilon\right)$. Then, under Assumptions \ref{Ass:moments}-\ref{Ass:Dependence}, there exists a constant $\zeta > 0$ such that, as $s_\delta,T \to \infty$ with $\frac{s_\delta}{T^{1/2}} \to 0$, we have that
$\Prob\left(\lambda_\min\left(\hat{\bSigma}_{22}\right) \geq \zeta\right) \to 1$.
\end{theorem}

It is possible to extend Theorem \ref{Lemma:Zhang_LB} to general distributions, based on an argument that relies on strong Gaussian approximations, at the additional cost of a further restriction on the growth rate of $s_\delta$.
\begin{theorem}\label{Lemma:Sigma_22_gen}
Let $\hat{\bSigma}_{22}$ be as defined in Assumption \ref{Ass:eigenvalues} and set $\bM=\bI_T$ assuming that $\bmu=\btau=\bm{0}$. 
Assume that $\bepsilon_t = \bD\bepsilon_{u,t}$, where $\bD$ is a $T \times T$-matrix with $\norm{\bD} \leq K < \infty$, and $\epsilon_{u,s,i} \indep \epsilon_{u,t,j}$ for all $i,j,s,t$ with $i \neq j$. Let $\bSigma_u = (\sigma_{u,ij})_{i,j=1}^N$ and assume that $\underset{1 \leq i \leq N}{\emph{max}} \E\abs{\sum_{t=1}^T \left(\epsilon_{u,t,i}^2 - \sigma_{u,ii}^2\right)}^2 = O\left(T^{1/2}\right)$. Then, under Assumptions \ref{Ass:moments}-\ref{Ass:Dependence}, a constant $\zeta>0$ exists, independent of $s_\delta$, $N$ and $T$, such that, as $s_\delta,N,T \to \infty$ with $\frac{s_\delta N}{T^{1/4}} \to 0$,
$\Prob\left(\lambda_\min\left(\hat{\bSigma}_{22}\right) > \zeta \right) \to 1.$
\end{theorem}

\section{Additional Proofs} \label{App:supp}

\subsection{Proofs of Lemmas}\label{App:lemmas_proofs}

\begin{proof}[\textbf{Proof of Lemma \ref{Prop:idemp}}]
\sloppy The first claim in Lemma \ref{Prop:idemp} follows directly from the fact that $(\bM\bS_{-1}\bA)^\prime(\bM\bS_{-1}\bA)$ and $(\bM\bS_{-1}\bA)(\bM\bS_{-1}\bA)^\prime$ are symmetric positive definite matrices that share the same non-zero eigenvalues. Hence,
\begin{equation*}
    \norm{\bA^\prime\bS_{-1}^\prime\bM\bS_{-1}\bA}_2 = \norm{\bM\bS_{-1}\bA\bA^\prime\bS_{-1}^\prime\bM}_2 \leq \norm{\bM}_2^2 \norm{\bS_{-1}\bA\bA^\prime\bS_{-1}^\prime}_2 = \norm{\bA^\prime\bS^\prime_{-1}\bS_{-1}\bA}_2,
\end{equation*}
where we have used that $\norm{\bM}_2=1$, as $\bM$ is an idempotent matrix. The same argument holds for $\norm{\bB^\prime\bU^\prime\bM\bU\bB}_2.$

To prove the remaining two claims in Lemma \ref{Prop:idemp}, we first derive the stochastic order of several quantities that appear frequently throughout the proof. Recall the definition of $\bt = (0,\ldots,T-1)$. Then, for $\bA$ and $\bB$ as defined in Lemma \ref{Prop:idemp} we claim that
\begin{equation}\label{eq:quantities}
\begin{split}
    &\norm{\bA^\prime\bS_{-1}^\prime\biota}_2 = O_p\left(d_A^{1/2}T^{3/2}\right), \quad \norm{\bA^\prime\bS_{-1}^\prime\bt}_2 = O_p\left(d_A^{1/2}T^{5/2}\right),\\
    &\norm{\bB^\prime\bU^\prime\biota}_2 = O_p\left(d_B^{1/2}T^{1/2}\right), \quad \norm{\bB^\prime\bU^\prime\bt}_2 = O_p\left(d_B^{1/2}T\right),\\
    &\norm{\bA^\prime\bW^\prime\biota}_2 = O_p\left(d_A^{1/2}T^{1/2}\right), \quad \norm{\bA^\prime\bW^\prime\bt}_2 = O_p\left(d_A^{1/2}T\right).
\end{split}
\end{equation}
To show \eqref{eq:quantities}, note that by Markov's and Minkowski's inequality,
\begin{equation*}
\begin{split}
    &\Prob\left(\norm{\bA^\prime\bS_{-1}^\prime\biota}_2 \geq K_\epsilon d_A^{1/2}T^{3/2}\right) \leq \frac{\sum_{j=1}^{d_A}\E\left(\sum_{i=1}^N\sum_{t=1}^T a_{i,j}s_{t,i}\right)^2}{K_\epsilon^2 d_AT^3}\\
    &\quad \leq \frac{\sum_{j=1}^{d_A}\left(\sum_{i=1}^N\sum_{t=1}^T \abs{a_{i,j}}\left(\E\left(s_{t-1,i}\right)^2\right)^{1/2}\right)^2}{K_\epsilon^2 d_AT^3} \leq \frac{K\left(\frac{1}{T}\sum_{t=1}^T\sqrt{\frac{t}{T}}\right)^2}{K_\epsilon^2} \leq \epsilon,\\
    &\Prob\left(\frac{\norm{\bA^\prime\bS_{-1}^\prime\bt}_2}{d_A^{1/2}T^{5/2}} \geq K_\epsilon\right) \leq \frac{\sum_{j=1}^{d_A}\left(\sum_{i=1}^N\sum_{t=1}^{T-1} \abs{a_{i,j}}t\left(\E\left(s_{t,i}\right)^2\right)^{1/2}\right)^2}{K_\epsilon^2 d_AT^5} \leq \frac{K\left(\frac{1}{T}\sum_{t=1}^T\left(\frac{t}{T}\right)^{3/2}\right)^2}{K_\epsilon^2} \leq \epsilon,
\end{split}
\end{equation*}
for $K_\epsilon \geq \frac{\sqrt{K}}{\epsilon}$. Let $\bb_j$ denote the $j$-th column of $\bB$ and define $\tilde{\bb}_{l,j} = \bC_l\bb_j$ with
\begin{equation*}
   \norm{\tilde{\bb}_j}_1 = \norm{\bC_l\bb_j} \leq \norm{\bC_l}_\infty\norm{\bb_j}_1 \leq K^\prime < \infty
\end{equation*}
by assumption. Then, for $K_\epsilon \geq \frac{\sqrt{K}\left(\sum_{l=0}^\infty \norm{\bC_l}_\infty\right)}{\epsilon}$,
\begin{equation*}
\begin{split}
     &\Prob\left(\frac{\norm{\bB^\prime\bU^\prime\biota}_2}{d_B^{1/2}T^{1/2}} \geq K_\epsilon\right) \leq \frac{\sum_{j=1}^{d_B}\E\left(\sum_{i=1}^N\sum_{l=0}^\infty\sum_{t=1}^T \tilde{b}_{l,j,i}\epsilon^u_{t-l,i}\right)^2}{K_\epsilon^2 d_B T}\\
     &\quad= \frac{\sum_{j=1}^{d_B}\sum_{i_1,i_2=1}^N\sum_{l_1,l_2=0}^\infty \tilde{b}_{l_1,j,i_1}\tilde{b}_{l_2,j,i_2}\sum_{t_1,t_2=1}^T\E\left(\epsilon^u_{t_1-l_1,i_1}\epsilon^u_{t_2-l_2,i_2}\right)}{K_\epsilon^2 d_B T} \leq \frac{K\left(\sum_{l=0}^\infty \norm{\bC_l}_\infty\right)^2}{K_\epsilon^2} \leq \epsilon,\\
     &\Prob\left(\frac{\norm{\bB^\prime\bU^\prime\bt}_2}{d_B^{1/2}T} \geq K_\epsilon\right) = \frac{\sum_{j=1}^{d_B}\sum_{i_1,i_2=1}^N\sum_{l_1,l_2=0}^\infty \tilde{b}_{l_1,j,i_1}\tilde{b}_{l_2,j,i_2}\sum_{t_1,t_2=1}^Tt_1t_2\E\left(\epsilon^u_{t_1-l_1,i_1}\epsilon^u_{t_2-l_2,i_2}\right)}{K_\epsilon^2 d_B T^2}\\
     &\quad\leq \frac{K\left(\sum_{l=0}^\infty \norm{\bC_l}_\infty\right)^2}{K_\epsilon^2} \leq \epsilon.
\end{split}
\end{equation*}
The proofs for the fifth and sixth term in \eqref{eq:quantities} are fully analogous and therefore omitted. Having established \eqref{eq:quantities}, we are equipped to prove the remaining two claims in Lemma \ref{Prop:idemp}. First, define $\text{det}\left(\bD^\prime\bD\right) = T\bt^\prime\bt - (\biota^\prime\bt)$ as the determinant of $\bD^\prime\bD$. Then, by explicitly writing out the definition of $\bP$ and using the stochastic orders derived in \eqref{eq:quantities}, it follows that
\begin{equation*}
    \begin{split}
        &\norm{\bA^\prime\bS_{-1}^\prime\bP\bU\bB}_F = \left(\text{det}(\bD^\prime\bD)\right)^{-1}\norm{\bA^\prime\left(\bt^\prime\bt\bS_{-1}^\prime\biota\biota^\prime\bU - \biota^\prime\bt\bS_{-1}^\prime\bt\biota^\prime\bU - \biota^\prime\bt\bS_{-1}^\prime\biota\bt^\prime\bU +T\bS_{-1}^\prime\bt\bt^\prime\bU \right)\bB}_F\\
        &\leq \left(\text{det}(\bD^\prime\bD)\right)^{-1}\left(\norm{\bA^\prime\bt^\prime\bt\bS_{-1}^\prime\biota\biota^\prime\bU\bB}_F + \norm{\bA^\prime\biota^\prime\bt\bS_{-1}^\prime\bt\biota^\prime\bU\bB}_F + \norm{\bA^\prime\biota^\prime\bt\bS_{-1}^\prime\biota\bt^\prime\bU\bB}_F + \norm{\bA^\prime \bT \bS_{-1}^\prime\bt\bt^\prime\bU\bB}_F\right)\\
        &= O_p\left(\sqrt{d_Ad_B}T\right) + O_p\left(\sqrt{d_Ad_B}T\right) + O_p\left(\sqrt{d_Ad_BT}\right)  + O_p\left(\sqrt{d_Ad_BT}\right) = O_p\left(\sqrt{d_Ad_B}T\right)
    \end{split}
\end{equation*}
and
\begin{equation*}
    \begin{split}
        &\norm{\bA^\prime\bW^\prime\bP\bU\bB}_F = \left(\text{det}(\bD^\prime\bD)\right)^{-1}\norm{\bA^\prime\left(\bt^\prime\bt\bW^\prime\biota\biota^\prime\bU - \biota^\prime\bt\bW^\prime\bt\biota^\prime\bU - \biota^\prime\bt\bW^\prime\biota\bt^\prime\bU +T\bW^\prime\bt\bt^\prime\bU \right)\bB}_F\\
        &\leq \left(\text{det}(\bD^\prime\bD)\right)^{-1}\left(\norm{\bA^\prime\bt^\prime\bt\bW^\prime\biota\biota^\prime\bU\bB}_F + \norm{\bA^\prime\biota^\prime\bt\bW^\prime\bt\biota^\prime\bU\bB}_F + \norm{\bA^\prime\biota^\prime\bt\bW^\prime\biota\bt^\prime\bU\bB}_F + \norm{\bA^\prime T\bW^\prime\bt\bt^\prime\bU\bB}_F\right)\\
        &= O_p\left(\sqrt{d_Ad_B}\right) + O_p\left(\sqrt{d_Ad_B}T^{-1/2}\right) + O_p\left(\sqrt{d_Ad_B}T^{-1/2}\right)  + O_p\left(\sqrt{d_Ad_B}T^{-1}\right) = O_p\left(\sqrt{d_Ad_B}\right).
    \end{split}
\end{equation*}
This completes the proof of Lemma \ref{Prop:idemp}.
\end{proof}

\begin{proof}[\textbf{Proof of Lemma \ref{Prop:set_sufficiency}}]
The proof follows from arguments similar to Proposition 1 in \citet{Zhao2006}. First, note that for $\hat{\bgamma}$ to be a minimizer of \eqref{eq:SPECS}, it must hold that
\begin{equation}\label{eq:foc}
    -2\bV^\prime\bM\left(\Delta \by - \bV\hat{\bgamma}\right) + \lambda_I\bOmega \bs_1 + \lambda_G \bs_2 = -2\bV^\prime\bM\left(\bepsilon_y - \bV\left(\hat{\bgamma} - \bgamma\right)\right) + \lambda_I\bOmega \bs_1 + \lambda_G \bs_2 = 0,
\end{equation}
where $\bOmega = \diag(\omega_1,\ldots,\omega_{N+M})$. Furthermore, $\bs_1 := s_1\left(\hat{\bgamma}\right)$ is the subgradient of the $\ell_1$-norm, with its $i$-th element defined by 
\begin{equation*}
    s_1\left(\hat{\bgamma}\right)_i = \begin{cases}
    \sign\left(\hat{\gamma}_i\right) & \text{if } \ \hat{\gamma}_i \neq 0\\
    x \in [0,1] & \text{if } \ \hat{\gamma}_i = 0
    \end{cases}
\end{equation*}
and $\bs_2 := s_2\left(\hat{\bgamma}\right) = \left(\tilde{s}_2\left(\hat{\bdelta}\right)^\prime,0\biota_M^\prime\right)^\prime$, where $\tilde{s}_2\left(\hat{\bdelta}\right)$ is the subgradient of the $\ell_2$-norm, defined by 
\begin{equation*}
    \tilde{s}_2\left(\hat{\bdelta}\right) = \begin{cases}
    \hat{\bdelta}\norm{\hat{\bdelta}}_2^{-1} & \text{if } \ \hat{\bdelta} \neq \bm{0}\\
    \hat{\bx} & \text{if } \ \hat{\bdelta} = \bm{0}
    \end{cases},
\end{equation*}
with $\norm{\hat{\bx}}_2=1$. Decompose the first-order conditions \eqref{eq:foc} into
\begin{align}
&\bV_1^\prime\bepsilon_y - \bV_1^\prime\bV_1\left(\hat{\bgamma}_{S_\gamma} - \bgamma_{S_\gamma}\right) - \bV_1^\prime\bV_2\hat{\bgamma}_{S_\gamma^c} = \frac{\lambda_I}{2}\bOmega_1\bs_{1,S_\gamma} + \frac{\lambda_G}{2}\bs_{2,S_\gamma}, \label{eq:FOC_real1}\\
& \bV_2^\prime\bepsilon_y - \bV_2^\prime\bV_1\left(\hat{\bgamma}_{S_\gamma} - \bgamma_{S_\gamma}\right) - \bV_2^\prime\bV_2\hat{\bgamma}_{S_\gamma^c} = \frac{\lambda_I}{2}\bOmega_2 \bs_{1,S_\gamma^c} + \frac{\lambda_G}{2}\bs_{2,S_\gamma^c}\label{eq:FOC_real2}.
\end{align}
We can rewrite \eqref{eq:FOC_real1} as
\begin{equation} \label{eq:FOC_rew1}
\begin{split}
(\bV_1^\prime\bV_1)^{-1}\bV_1^\prime\bepsilon_y &= \left(\hat{\bgamma}_{S_\gamma} - \bgamma_{S_\gamma}\right) + (\bV_1^\prime\bV_1)^{-1} \bV_1^\prime\bV_2\hat{\bgamma}_{S_\gamma^c}\\
&\quad + \frac{\lambda_I}{2} (\bV_1^\prime\bV_1)^{-1} \bOmega_1\bs_{1,S_\gamma} + \frac{\lambda_G}{2} (\bV_1^\prime\bV_1)^{-1} \bs_{2,S_\gamma}
\end{split}
\end{equation}
It then follows that
\begin{equation}
\begin{split}
\abs{(\bV_1^\prime\bV_1)^{-1}\bV_1^\prime\bepsilon_y} &\geq \abs{\hat{\bgamma}_{S_\gamma} - \bgamma_{S_\gamma}} - \abs{ (\bV_1^\prime\bV_1)^{-1} \bV_1^\prime\bV_2\hat{\bgamma}_{S_\gamma^c}}\\
&\quad - \frac{\lambda_I}{2}\abs{(\bV_1^\prime\bV_1)^{-1} \bOmega_1\bs_{1,S_\gamma}} - \frac{\lambda_G}{2} \abs{(\bV_1^\prime\bV_1)^{-1} \bs_{2,S_\gamma}}.
\end{split}
\end{equation}
Then on the set $\mathcal{A}_T$, we have that
\begin{align} \label{eq:ineq1}
\abs{\bgamma_{S_\gamma}} &> \abs{\hat{\bgamma}_{S_\gamma} - \bgamma_{S_\gamma}} - \abs{ (\bV_1^\prime\bV_1)^{-1} \bV_1^\prime\bV_2\hat{\bgamma}_{S_\gamma^c}}.
\end{align}
Next, let $\bP_V = \bV_1 (\bV_1^\prime \bV_1)^{-1} \bV_1^\prime$. Note that by multiplying \eqref{eq:FOC_rew1} by $\bV_1$ and rearranging we get
\begin{equation}
\begin{split}
\bV_1 \left(\hat{\bgamma}_{S_\gamma} - \bgamma_{S_\gamma}\right) &= \bP_V \bepsilon_y  - \bP_V \bV_2\hat{\bgamma}_{S_\gamma^c}\\
&\quad - \frac{\lambda_I}{2}\bV_1 (\bV_1^\prime\bV_1)^{-1} \bOmega_1\bs_{1,S_\gamma} - \frac{\lambda_G}{2}\bV_1 (\bV_1^\prime\bV_1)^{-1} \bs_{2,S_\gamma}.
\end{split}
\end{equation}
Plug this into \eqref{eq:FOC_real2} to get
\begin{equation}
\begin{split}
\bV_2^\prime \bM_V \bepsilon_y - \bV_2^\prime \bM_V \bV_2 \hat{\bgamma}_{S_\gamma^c} 
&= \frac{\lambda_I}{2}\left(\bOmega_2 \bs_{1,S_\gamma^c} - \bV_2^\prime \bV_1 (\bV_1^\prime\bV_1)^{-1} \bOmega_1\bs_{1,S_\gamma} \right)\\
&\quad + \frac{\lambda_G}{2}\left( \bs_{2,S_\gamma^c} - \bV_2^\prime \bV_1 (\bV_1^\prime\bV_1)^{-1} \bs_{2,S_\gamma} \right).
\end{split}
\end{equation}
It then follows that
\begin{equation}
\begin{split}
\abs{\bV_2^\prime \bM_V \bepsilon_y} &\geq \abs{\bV_2^\prime \bM_V \bV_2 \hat{\bgamma}_{S_\gamma^c} + \frac{\lambda_I}{2}\bOmega_2 \bs_{1,S_\gamma^c} + \frac{\lambda_G}{2}\bs_{2,S_\gamma^c}} \\
&\quad - \frac{\lambda_I}{2}\abs{\bV_2^\prime \bV_1 (\bV_1^\prime\bV_1)^{-1} \bOmega_1\bs_{1,S_\gamma}}_i - \frac{\lambda_G}{2}\abs{\bV_2^\prime \bV_1 (\bV_1^\prime\bV_1)^{-1} \bs_{2,S_\gamma}}.
\end{split}
\end{equation}
Then on the set $\mathcal{B}_T$, we have that
\begin{equation}\label{eq:set_B2}
\frac{\lambda_I}{2} \bOmega_2 \biota >
\abs{\bV_2^\prime \bM_V \bV_2 \hat{\bgamma}_{S_\gamma^c} + \frac{\lambda_I}{2}\bOmega_2 \bs_{1,S_\gamma^c} + \frac{\lambda_G}{2}\bs_{2,S_\gamma^c}}.
\end{equation}
Assume that $\hat{\bgamma}_{S_\gamma^c}^\prime{\bgamma}_{S_\gamma^c} \neq 0$. Then, pre-multiplying \eqref{eq:set_B2} by $\abs{\hat{\bgamma}_{S_\gamma^c}}^\prime$,
\begin{equation}\label{eq:B_cont}
\begin{split}
  \lambda_I\abs{\hat{\bgamma}_{S_\gamma^c}}^\prime\bOmega_2\biota &> \abs{\hat{\bgamma}_{S_\gamma^c}}^\prime\abs{2\bV_2^\prime \bM_V \bV_2 \hat{\bgamma}_{S_\gamma^c} + \lambda_I\bOmega_2 \bs_{1,S_\gamma^c} + \lambda_G\bs_{2,S_\gamma^c}}  \\
  &> \abs{2\hat{\bgamma}_{S_\gamma^c}^\prime\bV_2^\prime \bM_V \bV_2 \hat{\bgamma}_{S_\gamma^c} + \lambda_I\abs{\hat{\bgamma}_{S_\gamma^c}}^\prime\bOmega_2\biota + \lambda_G\hat{\bgamma}_{S_\gamma^c}^\prime\bs_{2,S_\gamma^c}},
\end{split}
\end{equation}
where we have used that $\hat{\bgamma}_{S_\gamma^c}^\prime\bOmega_2 \bs_{1,S_\gamma^c} = \abs{\hat{\bgamma}_{S_\gamma^c}}^\prime\bOmega_2\biota$. By positive semi-definiteness, it must hold that $\hat{\bgamma}_{S_\gamma^c}^\prime\bV_2^\prime \bM_V \bV_2 \hat{\bgamma}_{S_\gamma^c} \geq 0$. Furthermore, defining the index set $S_{\Delta \delta} := S_{\hat{\delta}} \setminus S_{\delta}$, it holds that
\begin{equation*}
    \hat{\bgamma}_{S_\gamma^c}^\prime\bs_{2,S_\gamma^c} = \begin{cases}
    \norm{\hat{\bdelta}}_2^{-1}\sum_{i \in S_{\Delta \delta}}\hat{\delta}_i^2 & \text{ if } S_{\Delta \delta} \neq \emptyset\\
    0 & \text{ otherwise}
    \end{cases}.
\end{equation*}
Hence, \eqref{eq:B_cont} results in a contradiction, thereby implying that $\hat{\bgamma}_{S_\gamma^c}=\bm{0}$. But if this is the case, then it follows from \eqref{eq:ineq1} that
\begin{equation} \label{eq:ineq3}
\abs{\gamma_{S_\gamma}} > \abs{\hat{\bgamma}_{S_\gamma} - \bgamma_{S_\gamma}}.
\end{equation}
For any $i \in S_\gamma$, take $\gamma_i > 0$ and $\hat{\gamma}_i \leq 0$. Then
\begin{equation}
\abs{\hat{\gamma}_i - \gamma_i} = \gamma_i - \hat{\gamma}_i \geq \gamma_i = \abs{\gamma}_i,
\end{equation}
which contradicts the previous inequality \eqref{eq:ineq3}. A similar argument holds for the case where $\gamma_i < 0$ and $\hat{\gamma}_i \geq 0$. Concluding, \eqref{eq:ineq3} only holds under sign consistency.
\end{proof}

\begin{proof}[\textbf{Proof of Lemma \ref{Lemma:emp_proc}}]
We show that $\norm{\bS_T^{-1}\bQ\bV_1^\prime\bepsilon_y}_2 = O_p\left(s_\delta + \sqrt{s_\pi}\right)$. First, by Lemma \ref{Prop:idemp}
\begin{equation*}
\begin{split}
\norm{\bS_T^{-1}\bQ\bV_1^\prime\bepsilon_y}_2 &\leq \norm{T^{-1/2}\sum_{t=1}^T\bC^v(L)\tilde{\bepsilon_t}\tilde{\epsilon}_{y,t}}_2 + \norm{\bB_{S_\delta,\perp}^\prime\bC_{S_\delta}\left(\frac{\sqrt{s_\delta}}{T}\sum_{t=1}^T\tilde{\bs}_{t-1}\tilde{\epsilon}_{y,t}\right)}_2\\
&\quad + \norm{\bB_{S_\delta,\perp}\left(\frac{\sqrt{s_\delta}}{T}\sum_{t=1}^T \bC_{S_\delta}(L)\tilde{\bepsilon}_{t-1}\tilde{\epsilon}_{y,t}\right)}_2,\\
 &= \norm{T^{-1/2}\sum_{t=1}^T\bC^v(L)\bepsilon_t\epsilon_{y,t}}_2 + \norm{\bB_{S_\delta,\perp}^\prime\bC_{S_\delta}\left(\frac{\sqrt{s_\delta}}{T}\sum_{t=1}^T\bs_{t-1}\epsilon_{y,t}\right)}_2 \\
 &\quad + \norm{\bB_{S_\delta,\perp}\left(\frac{\sqrt{s_\delta}}{T}\sum_{t=1}^T \bC_{S_\delta}(L)\bepsilon_{t-1}\epsilon_{y,t}\right)}_2 + O_p(s_\delta) =: \sum_{i=1}^3 \norm{\bd_i}_2 + O_p(s_\delta)
\end{split}
\end{equation*}
We proceed by deriving the stochastic orders of the three RHS terms separately. First, let $\eta_{i,t} = \sum_{l=0}^\infty \bc^{v\prime}_{l,i}\bepsilon_{t-l}$, where $\bc^{v}_{l,i}$ is the $i$-th row vector of $\bC^v_l$. Then, using that $\lbrace \eta_{i,t}\epsilon_{y,t}\rbrace$ is a martingale difference sequence, we sequentially apply Markov's and Burkholder's inequality to derive that
\begin{equation*}
\begin{split}
&\Prob\left(\norm{\bd_1}_2 > \frac{K_\epsilon(s_\delta + \sqrt{s_\pi})}{3}\right)  \leq \frac{9\sum_{i=1}^{s_\pi}\E\left(\sum_{t=1}^T\eta_{i,t}\epsilon_{y,t}\right)^2}{TK_\epsilon^2(s_\delta + \sqrt{s_\pi})^2} \leq \frac{K\sum_{i=1}^{s_\pi}\sum_{t=1}^T\E\left(\eta_{i,t}\epsilon_{y,t}\right)^2}{TK_\epsilon^2(s_\delta + \sqrt{s_\pi})^2}\\
&\leq \frac{K^*\left(\sum_{l=0}^\infty \norm{\bC^v_l}_1\right)}{K_\epsilon^2} \leq \epsilon,
\end{split}
\end{equation*}
for $K_\epsilon \geq \left(\frac{K^*\left(\sum_{l=0}^\infty \norm{\bC^v_l}_1\right)}{\epsilon}\right)^{1/2}$, where we have used that
\begin{equation*}
\begin{split}
\E\left(\eta_{i,t}\epsilon_{y,t}\right)^2 &\leq \sum_{l_1,l_2=0}^\infty\sum_{j_1,j_2=1}^N\abs{c^v_{l_1,i,j_1}}\abs{c^v_{l_2,i,j_2}}\E\left(\epsilon_{j_1,t-l_1}\epsilon_{j_2,t-l_2}\epsilon_{y,t}^2\right) \leq K\left(\sum_{l=0}^\infty \norm{\bC^v_{l}}_\infty\right)^2,
\end{split}
\end{equation*}
by Assumption \ref{Ass:moments}. Next, define $\ba_i = \bC_{S_\delta}\bbeta_{S_\delta,\perp,i}$. Using the fact that $\lbrace\ba_i^\prime\bs_{t-1}\epsilon_{y,t}\rbrace$ is a martingale difference sequence, it follows that
\begin{equation*}
\begin{split}
&\Prob\left(\norm{\bd_2}_2 > \frac{K_\epsilon(s_\delta + \sqrt{s_\pi})}{3}\right)  \leq \frac{s_\delta 9\sum_{i=1}^{s_\delta}\E\left(\sum_{t=1}^T\ba_i^\prime\bs_{t-1}\epsilon_{y,t}\right)^2}{T^2K_\epsilon^2(s_\delta + \sqrt{s_\pi})^2} \leq \frac{s_\delta K\sigma_y^2\sum_{i=1}^{s_\delta}\sum_{t=1}^T\E\left(\ba_i^\prime\bs_{t-1}\right)^2}{T^2K_\epsilon^2(s_\delta + \sqrt{s_\pi})^2} \\
&\quad \leq \frac{K\phi_\max\sigma_y^2\norm{\bC_{S_\delta}}^2_2}{K_\epsilon^2} \leq \epsilon,
\end{split}
\end{equation*}
for $K_\epsilon \geq \left(\frac{K\phi_\max\sigma_y^2\norm{\bC_{S_\delta}}^2_2}{\epsilon}\right)^{1/2}$, where we use the fact that
\begin{equation*}
\begin{split}
&\E\left(\ba_i^\prime\bs_{t-1}\right)^2 \leq \ba_i^\prime\E\left(\bs_{t-1}\bs_{t-1}^\prime\right)\ba_i = \ba_i^\prime\bSigma_\epsilon\ba_i(t-1) \leq \norm{\ba_i}_2^2\phi_\max(t-1) \leq \norm{\bC_{S_\delta}}^2_2\phi_\max(t-1),
\end{split}
\end{equation*}
by Assumption \ref{Ass:moments} and the normalization imposed on $\bB_{S_\delta,\perp}$. Finally, define $\xi_{i,t} = \bbeta_{S_\delta,\perp,i}^\prime\bC_{S_\delta}(L)\bepsilon_t$. Then, using that $\lbrace \xi_{i,t-1}\epsilon_{y,t}\rbrace$ is a martingale difference sequence, it follows by similar reasoning that
\begin{equation*}
\begin{split}
&\Prob\left(\norm{\bd_3}_2 > \frac{K_\epsilon(s_\delta + \sqrt{s_\pi})}{3}\right) \leq \frac{9s_\delta\sum_{i=1}^{s_\delta}\E\left(\sum_{t=1}^T\xi_{i,t-1}\epsilon_{y,t}\right)^2}{T^2K_\epsilon^2(s_\delta + \sqrt{s_\pi})^2} \leq \frac{ K^*\sum_{l=0}^\infty\norm{\bC_{S_\delta,l}}_2^2}{TK_\epsilon^2} \to  0,
\end{split}
\end{equation*}
where we use that
\begin{equation*}
\begin{split}
\E\left(\xi_{i,t-1}\right)^2 &= \sum_{l=0}^\infty \bbeta_{S_\delta,\perp,i}^\prime\bC_{S_\delta,l}\Sigma_\epsilon\bC_{S_\delta,l}^\prime\bbeta_{S_\delta,\perp,i} \leq \sum_{l=0}^\infty \norm{\bC_{S_\delta,l}^\prime\bbeta_{S_\delta,\perp,i}}_2^2\phi_\max \leq \sum_{l=0}^\infty \norm{\bC_{S_\delta,l}}_2^2\phi_\max,
\end{split}
\end{equation*}
with $\phi_\max$ being the upper bound on the maximum eigenvalue of $\bSigma_\epsilon$ from Assumption \ref{Ass:moments}. This completes the proof.
\end{proof}

\begin{proof}[\textbf{Proof of Lemma \ref{Lemma:Sigma_12}}]
First, define $\boeta_t = \bC^{v*}(L)\bepsilon_t$, where $\bC^{v*}(L)$ is based on the Beveridge-Nelson decomposition of $\bC^v(L) = \bC^v(1) + \bC^{v*}(L)(1-L)$. Then, with the use of summation by parts, we  decompose
\begin{equation}\label{eq:As in Sigma12}
\begin{split}
&\norm{\hat{\bSigma}_{21}}_2 = \norm{\frac{\sqrt{s_\delta}\sum_{t=2}^T\bv_{2,t}\bv_{1,t}^\prime}{T^{3/2}}}_2 = \norm{\frac{\sqrt{s_\delta}\bB_{S_\delta,\perp}^\prime\sum_{t=2}^T\tilde{\bz}_{S_\delta,t-1}\tilde{\bepsilon}_t^\prime\bC^{v\prime}(L)}{T^{3/2}}}_2\\
&\quad= \norm{\frac{\sqrt{s_\delta}\bB_{S_\delta,\perp}^\prime\bC_{S_\delta}\sum_{t=2}^T\tilde{\bs}_{t-1}\tilde{\bepsilon}_t^\prime\bC^{v\prime}(L)}{T^{3/2}}}_2 +  \norm{\frac{\sqrt{s_\delta}\bB_{S_\delta,\perp}^\prime\sum_{t=2}^T\tilde{\bu}_{S_\delta,t-1}\tilde{\bepsilon}_t^\prime\bC^{v\prime}(L)}{T^{3/2}}}_2\\
&\quad= \norm{\frac{\sqrt{s_\delta}\bB_{S_\delta,\perp}^\prime\bC_{S_\delta}\sum_{t=2}^T\bs_{t-1}\bepsilon_t^\prime\bC^{v\prime}(1)}{T^{3/2}}}_2 + \norm{\frac{\sqrt{s_\delta}\bB_{S_\delta,\perp}^\prime\bC_{S_\delta}\bs_{T-1}\boeta_T^\prime}{T^{3/2}}}_2\\
&\quad\quad + \norm{\frac{\sqrt{s_\delta}\bB_{S_\delta,\perp}^\prime\bC_{S_\delta}\sum_{t=2}^T\bepsilon_t\boeta_t^\prime}{T^{3/2}}}_2 +  \norm{\frac{\sqrt{s_\delta}\bB_{S_\delta,\perp}^\prime\sum_{t=2}^T\bu_{S_\delta,t-1}\bepsilon_t^\prime\bC^{v\prime}(L)}{T^{3/2}}}_2 + O_p\left(\frac{s_\delta\sqrt{s_\pi}}{T^{1/2}}\right)\\
&\quad=: \sum_{i=1}^4\bA_i + o_p(1),
\end{split}
\end{equation}
where the $O_p\left(\frac{s_\delta\sqrt{s_\pi}}{T^{1/2}}\right)$ term stems from application of Lemma \ref{Prop:idemp} and the last line uses that $\frac{s_\delta\sqrt{s_\pi}}{T^{1/2}} \to 0$ by assumption.

Hence, we proceed by showing that each $\norm{\bA_i}_2$ converges in probability to zero. First, let $\ba_i = \bC_{S_\delta}\bbeta_{S_\delta,\perp,i}$ and define $\bb_j = \bc^v_j(1)$, where $\bc^v_j(z) = \sum_{l=0}^\infty \bc^v_{l,j} z^l$, with $\bc^v_{l,j}$ being the $j$-th row of the $\bC^v_l$. Note that $\lbrace\ba_i^\prime\bs_{t-1}\bepsilon_t^\prime\bb_j\rbrace$ is a martingale difference sequence. Then, for any arbitrary constant $a>0$, we sequentially apply Markov's inequality, Burkholder's inequality  and the $C_r$-inequality to obtain
\begin{equation*}
\begin{split}
&\Prob\left(\norm{\bA_1}_2 \geq a\right) \leq \frac{s_\delta\sum_{i=1}^{s_\delta}\sum_{j=1}^{s_\pi}\E\left(\sum_{t=2}^T\ba_i^\prime\bs_{t-1}\bepsilon_t^\prime\bb_j\right)^2}{a^2T^3} \leq \frac{s_\delta K\sum_{i=1}^{s_\delta}\sum_{j=1}^{s_\pi}\sum_{t=2}^T\E\left(\ba_i^\prime\bs_{t-1}\bepsilon_t^\prime\bb_j\right)^2}{a^2T^3}\\
&\leq \frac{s_\delta K\phi_\max^2\sum_{i=1}^{s_\delta}\sum_{j=1}^{s_\pi}\norm{\ba_i}_2^2\norm{\bb_j}_2^2}{a^2T} \leq \frac{s_\delta^2s_\pi K\phi_\max^2\norm{\bC_{S_\delta}}_2^2\left(\sum_{l=0}^\infty\norm{\bC^v_l}_2\right)^2}{a^2T} \to 0,
\end{split}
\end{equation*}
as $\frac{s_\delta^2s_\pi}{T} \to 0$ by Assumption \ref{ass:sparsity}. 

\bigskip
Next, we focus on $\bA_2$. Define $\bb_{l,j} = \bc^{v*}_{l,j}$ as the $j$-th row of $\bC^{v*}_{l}$. Then, by sequentially applying Markov's inequality and Minkowski's inequality,
\begin{equation*}
\begin{split}
&\Prob\left(\norm{\bA_2}_2 \geq a\right) \leq \frac{s_\delta\sum_{i=1}^{s_\delta}\sum_{j=1}^{s_\pi}\E\left(\sum_{l=0}^\infty\ba_i^\prime\bs_{T-1}\bepsilon_{T-l}^\prime\bb_{l,j}\right)^2}{a^2T^3}\\
&= \frac{s_\delta\sum_{i=1}^{s_\delta}\sum_{j=1}^{s_\pi}\E\left(\sum_{l=0}^\infty\sum_{k_1,k_2=1}^N\sum_{s=1}^{T-1}a_{i,k_1}b_{l,j,k_2}\epsilon_{k_1,s}\epsilon_{k_2,T-l}\right)^2}{a^2T^3}\\
&\leq \frac{s_\delta\sum_{i=1}^{s_\delta}\sum_{j=1}^{s_\pi}\left(\sum_{l=0}^\infty\sum_{k_1,k_2=1}^N\sum_{s=1}^{T-1}\abs{a_{i,k_1}}\abs{b_{l,j,k_2}}\left(\E\left(\epsilon_{k_1,s}\epsilon_{k_2,T-l}\right)^2\right)^{1/2}\right)^2}{a^2T^3}\\
&\leq \frac{s_\delta K\sum_{i=1}^{s_\delta}\sum_{j=1}^{s_\pi}\norm{\ba_i}_1^2\left(\sum_{l=0}\norm{\bb_{l,j}}_1\right)^2}{a^2T} \leq \frac{s_\delta^2s_\pi K\norm{\bC_{S_\delta}}_\infty^2\left(\sum_{l=0}\norm{\bC^{v*}_l}_\infty\right)^2}{a^2T} \to 0.
\end{split}
\end{equation*}

Next, we focus on $\norm{\bA_3}_2$. Again, using a combination of Markov's inequality and Minkowski's inequality,
\begin{equation*}
\begin{split}
&\Prob\left(\norm{\bA_3}_2 \geq a\right) \leq \frac{s_\delta\sum_{i=1}^{s_\delta}\sum_{j=1}^{s_\pi}\E\left(\sum_{t=1}^{T-1}\sum_{l=0}^\infty\ba_i^\prime\bepsilon_t\bepsilon_{t-l}^\prime\bb_{l,j}\right)^2}{a^2T^3}\\
&= \frac{s_\delta\sum_{i=1}^{s_\delta}\sum_{j=1}^{s_\pi}\E\left(\sum_{t=1}^{T-1}\sum_{l=0}^\infty\sum_{k_1,k_2=1}^Na_{i,k_1}b_{l,j,k_2}\epsilon_{k_1,t}\epsilon_{k_2,t-l}\right)^2}{a^2T^3}\\
&\leq \frac{s_\delta\sum_{i=1}^{s_\delta}\sum_{j=1}^{s_\pi}\left(\sum_{t=1}^{T-1}\sum_{l=0}^\infty\sum_{k_1,k_2=1}^N\abs{a_{i,k_1}}\abs{b_{l,j,k_2}}\left(\E\left(\epsilon_{k_1,t}\epsilon_{k_2,t-l}\right)^2\right)^{1/2}\right)^2}{a^2T^3}\\
&\leq \frac{s_\delta K\sum_{i=1}^{s_\delta}\sum_{j=1}^{s_\pi}\norm{\ba_i}_1^2\left(\sum_{l=0}\norm{\bb_{l,j}}_1\right)^2}{a^2T} \leq \frac{s_\delta^2s_\pi K\norm{\bC_{S_\delta}}_\infty^2\left(\sum_{l=0}\norm{\bC^{v*}_l}_\infty\right)^2}{a^2T} \to 0,
\end{split}
\end{equation*}
as $\frac{s_\delta^2s_\pi}{T} \to 0$ by Assumption \ref{ass:sparsity}.

Finally, we consider $\norm{\bA_4}_2$. Define $\ba_{l,i} = \bC_{S_\delta,l}\bbeta_{S_\delta,\perp,i}$. Then,
\begin{equation*}
\begin{split}
&\Prob\left(\norm{\bA_4}_2 \geq a\right) \leq \frac{s_\delta\sum_{i=1}^{s_\delta}\sum_{j=1}^{s_\pi}\E\left(\sum_{t=1}^{T-1}\sum_{l_1,l_2=0}^\infty\ba_{l_1,i}^\prime\bepsilon_{t-1}\bepsilon_{t-l_2}^\prime\bb_{l_2,j}\right)^2}{a^2T^3}\\
&= \frac{s_\delta\sum_{i=1}^{s_\delta}\sum_{j=1}^{s_\pi}\E\left(\sum_{t=1}^{T-1}\sum_{l_1,l_2=0}^\infty\sum_{k_1,k_2}^Na_{l_1,i,k_1}b_{l_2,j,k_2}\epsilon_{k_1,t-1}\epsilon_{k_2,t-l_2}\right)^2}{a^2T^3}\\
&\leq \frac{s_\delta\sum_{i=1}^{s_\delta}\sum_{j=1}^{s_\pi}\left(\sum_{t=1}^{T-1}\sum_{l_1,l_2=0}^\infty\sum_{k_1,k_2}^N\abs{a_{l_1,i,k_1}}\abs{b_{l_2,j,k_2}}\left(\E\left(\epsilon_{k_1,t-1}\epsilon_{k_2,t-l_2}\right)^2\right)^{1/2}\right)^2}{a^2T^3}\\
& \leq \frac{s_\delta K\sum_{i=1}^{s_\delta}\sum_{j=1}^{s_\pi}\left(\sum_{l=0}^\infty\norm{\ba_{l,i}}_1\right)^2\left(\sum_{l=0}^\infty\norm{\bb_{l,j}}_1\right)^2}{a^2T}\\
& \leq \frac{s_\delta^2 s_\pi K\left(\sum_{l=0}^\infty\norm{\bC_{S_\delta,l}}_\infty\right)^2\left(\sum_{l=0}^\infty\norm{\bC^{v*}_l}_\infty\right)^2}{a^2T} \to 0,
\end{split}
\end{equation*}
as $\frac{s_\delta^2s_\pi}{T} \to 0$ by Assumption \ref{ass:sparsity}. This completes the argument.
\end{proof}

\begin{proof}[\textbf{Proof of Lemma \ref{Cor:eigenvalue}}]
Let $\tilde{\bSigma} = \diag\left(\hat{\bSigma}_{11},\hat{\bSigma}_{22}\right)$ and $\check{\bSigma} = \hat{\bSigma}-\tilde{\bSigma}$. Note that
\begin{equation*}
\lambda_\min\left(\hat{\bSigma}\right) \geq \lambda_\min\left(\tilde{\bSigma}\right) + \lambda_\min\left(\check{\bSigma}\right) \geq \lambda_\min\left(\tilde{\bSigma}\right) - \norm{\check{\bSigma}}_2.
\end{equation*}
Furthermore,
\begin{equation*}
\begin{split}
&\Prob\left(\lambda_\min\left(\tilde{\bSigma}\right) < \phi\right) = \Prob\left(\min\left(\lambda_\min\left(\hat{\bSigma}_{11}\right),\lambda_\min\left(\hat{\bSigma}_{22}\right)\right) < \phi\right)\\
&\quad \leq \Prob\left(\lambda_\min\left(\hat{\bSigma}_{11}\right) < \phi\right) + \Prob\left(\lambda_\min\left(\hat{\bSigma}_{22}\right) < \phi\right) \to 0,
\end{split}
\end{equation*}
by Assumption \ref{Ass:eigenvalues}. Thus, for any $\epsilon > 0$, we may choose a $T_1$ such that $\Prob\left(\lambda_\min\left(\tilde{\bSigma}\right) < \phi\right) \leq \frac{\epsilon}{2}$ for all $T>T_1$. Moreover, by Lemma \ref{Lemma:Sigma_12}, there exists a $T_2$ such that $\Prob\left(\norm{\check{\bSigma}}_2 \geq \frac{\phi}{2}\right) \leq \frac{\epsilon}{2}$, whenever $T > T_2$. Then, for all $T > \max(T_1,T_2)$, we have
\begin{equation*}
\begin{split}
&\Prob\left(\lambda_\min\left(\hat{\bSigma}\right) < \frac{\phi}{2}\right) \leq \Prob\left(\lambda_\min\left(\tilde{\bSigma}\right) - \norm{\check{\bSigma}}_2 < \frac{\phi}{2}\right)\\
&\leq \Prob\left(\lambda_\min\left(\tilde{\bSigma}\right) - \norm{\check{\bSigma}}_2 < \frac{\phi}{2}, \norm{\check{\bSigma}}_2 < \frac{\phi}{2}\right) + \Prob\left(\norm{\check{\bSigma}}_2 \geq \frac{\phi}{2}\right)\\
&\leq \Prob\left(\lambda_\min\left(\tilde{\bSigma}\right) < \phi\right) + \frac{\epsilon}{2} \leq \epsilon.
\end{split}
\end{equation*}
Since $\epsilon$ was chosen arbitrarily, the claim is shown for $\phi^* = \frac{\phi}{2}$. The same proof works for any $0<\phi^*<\phi$.
\end{proof}

\begin{proof}[\textbf{Proof of Lemma \ref{Cor:conversion}}]
First, note that $\bQ_R^\prime\bQ_R = \bI_{M+N}$ by construction. From the VMA representation \eqref{eq:GRT}, it follows that $\bM\bV\bQ_R^\prime = \begin{bmatrix}\bV_{R1},\bV_{R2}\end{bmatrix}$, where $\bV_{R1}$ is an $(T\times M_\pi)$-dimensional matrix containing stationary processes and $\bV_{R2}$ is an $(T\times N_\delta)$-dimensional matrix containing integrated processes. We denote the rows of $\bV_{R1}$ and $\bV_{R2}$ by $\bv_{R1,t}$ and $\bv_{R2,t}$, respectively. Then, after a set of suitable replacements, the proof of Lemma \ref{Cor:conversion} is entirely analogous to the proofs of Lemmas \ref{Lemma:emp_proc}-\ref{Cor:eigenvalue}. The required substitutions are summarized below.
\end{proof}

\noindent
\begin{tabularx}{\textwidth}{XXXXXXXXXX}
\hline 
Old & $s_\delta$ & $s_\pi$ & $\bB_{S_\delta}$ & $\bB_{S_\delta,\perp}$ & $\bQ$ & $\bv_t$ & $\bv_{1,t}$ & $\bv_{2,t}$ & $\bS_T$ \tabularnewline
New & $N_\delta$ & $M_\pi$ & $\bB$ & $\bB_\perp$ & $\bQ_R$ & $\bv_{R,t}$ & $\bv_{R1,t}$ & $\bv_{R2,t}$ & $\bS_R$ \tabularnewline
\hline 
\end{tabularx}

\bigskip
\begin{proof}[\textbf{Proof of Lemma \ref{Lemma:eig_transl}}]
Part (1) is a well-known consequence of the additive Weyl inequalities, see for example \citet[][Theorem 3.3.16]{Horn1994}.
Part (2) corresponds to Lemma 6.17 in \citet{Buhlmann2011} and is described in similar form in Lemma 3 in \citet{Medeiros2016}. For the sake of completion, we repeat the short proof here. Take $\bx \in \mathbb{R}^s\setminus \lbrace0\rbrace$. Then,
\begin{equation*}
\begin{split}
\bx^\prime\bA\bx - \bx^\prime\bB\bx \leq \abs{\bx^\prime\left(\bA-\bB\right)\bx} \leq \norm{\bx}_1\norm{\left(\bA-\bB\right)\bx}_\infty \leq \norm{\bx}_1^2\delta \leq \bx^\prime\bx s\delta,
\end{split}
\end{equation*}
from which clearly follows that $\frac{x^\prime \bB x}{x^\prime x} \geq \frac{x^\prime \bA x}{x^\prime x} - s\delta$. Taking the infimum on both sides completes the proof.
\end{proof}

\bigskip
\begin{proof}[\textbf{Proof of Lemma \ref{Lemma:BM_int}}.]
We show that $\lambda_\min\left(\frac{1}{T^2}\sum_{t=1}^T\bh_t\bh_t^\prime\right) \overset{p}{\to} 0$, as $T,s \to \infty$. Let $E = \lbrace \be_1,\ldots,\be_s\rbrace$ be the collection of basis vectors. Since $E \subset \mathbb{R}^s$, we have for any $\epsilon > 0$,
\begin{equation*}
\begin{split}
&\Prob\left(\lambda_\min\left(\frac{1}{T^2}\sum_{t=1}^T\bh_t\bh_t^\prime\right) > \phi \right) \leq \Prob\left(\underset{\bx \in E}{\text{min}} \bx^\prime\left(\lambda_\min\left(\frac{1}{T^2}\sum_{t=1}^T\bh_t\bh_t^\prime\right)\right)\bx > \phi\right)\\
&= \Prob\left(\underset{1 \leq i \leq s}{\text{min}} \frac{1}{T^2}\sum_{t=1}^T h_{i,t}^2 > \phi\right) = \Prob\left(\frac{1}{T^2}\sum_{t=1}^T h_{1,t}^2 > \phi\right)^s\\
&\leq \left\{\Prob\left(\int_0^1 W^2(r)dr > \phi\right) + \abs{\Prob\left(\frac{1}{T^2}\sum_{t=1}^Th_{1,t}^2 > \phi\right)-\Prob\left(\int_0^1 W^2(r)dr > \phi\right)} \right\}^s,
\end{split}
\end{equation*}
where $W(r)$ is a standard univariate Brownian Motion. First, assume that
\begin{equation}\label{eq:BM_int}
    \Prob\left(\int_0^1 W^2(r)dr > \phi\right) \leq 1 - 2\epsilon(\phi),
\end{equation}
for some $\epsilon(\phi)>0$. Then, by the functional central limit theorem, there exists a $T^*$ such that
\begin{equation*}
\abs{\Prob\left(\frac{1}{T^2}\sum_{t=1}^T h_{1,t}^2 > \phi\right)-\Prob\left(\int_0^1 W^2(r)dr > \phi\right)} \leq \epsilon(\phi)
\end{equation*}
for all $T>T^*$. Consequently, for large enough $T$,
\begin{equation*}
    \Prob\left(\lambda_\min\left(\frac{1}{T^2}\sum_{t=1}^T\bh_t\bh_t^\prime\right) > \phi \right) \leq (1 - \epsilon(\phi))^s \to 0
\end{equation*}
as $s,T \to \infty$, which is the claim of Lemma \ref{Lemma:BM_int}. Hence, all that is left is to verify the truth of \eqref{eq:BM_int}.

First, note that
\begin{equation*}
    \Prob\left(\int_0^1 W^2(r)dr \geq \phi\right) \leq \Prob\left(\underset{0\leq r \leq 1}{\text{sup}} \abs{W(r)} \geq \sqrt{\phi}\right) = 1 - \Prob\left(\underset{0\leq r \leq 1}{\text{sup}} \abs{W(r)} < \sqrt{\phi}\right).
\end{equation*}
We show that for every $\phi > 0$, it holds that $\Prob\left(\underset{0\leq r \leq 1}{\text{sup}} \abs{W(r)} < \sqrt{\phi} \right) \geq 2\epsilon(\phi) > 0$ for some $\epsilon(\phi)>0$. Let $W_1$ and $W_2$ denote two independent standard Brownian motions over the interval $[0,1]$ and note that we may construct an additional standard Brownian motion as $W_\Delta = (W_1 - W_2)/\sqrt{2}$. The sample paths of $W_1,W_2,W_\Delta$ lie in the function space $C([0,1])$, which contains all continuous functions from the unit interval to $\mathbb{R}$ and is equipped with the supremum norm $\norm{f}_\infty = \max \lbrace \abs{f(x)} \ | \ x \in [0,1]\rbrace$. It is well-known that $C([0,1])$ is a separable metric space \citep[e.g.][p. 438]{Davidson1994}. Then, define $B(y,\sqrt{2\phi}) = \lbrace x \in C([0,1]) \ | \ \norm{x-y}_\infty \leq \sqrt{2\phi}\rbrace$ and note that by Theorem 5.6 of \citet{Davidson1994} there exists a countable collection of elements $\lbrace x_1,x_2,\ldots\rbrace \subset C([0,1])$ such that $C([0,1]) \subseteq \bigcup_i B(x_i,\sqrt{2\phi})$. By countable additivity,
\begin{equation}\label{eq:C_cover}
    1 = \Prob\left(W_1 \in C([0,1]) \right) = \Prob\left(W_1 \in \bigcup_i B(x_i,\sqrt{2\phi}) \right) \leq \sum_i \Prob\left(W_1 \in B(x_i,\sqrt{2\phi}) \right).
\end{equation}
Furthermore, it must be true that there exists a $B(x_i,\sqrt{2\phi})$ with $\Prob\left(W_1 \in B(x_i,\sqrt{2\phi}) \right) = q > 0$, because otherwise the RHS of \eqref{eq:C_cover} would be zero, resulting in a contradiction. Since $W_1$ and $W_2$ are independent, we conclude that
\begin{equation*}
    \Prob\left(\underset{0 \leq r \leq 1}{\text{sup}}\abs{W_\Delta(r)} < \sqrt{\phi}\right) \geq \Prob\left(W_1 \in B(x_i,\sqrt{2\phi}), W_2 \in B(x_i,\sqrt{2\phi}) \right) = q^2 > 0.
\end{equation*}
Since $q$ depends only on $\phi$, we may write $2\epsilon(\phi)=q^2$, thereby completing the proof.
\end{proof}

\subsection{Proofs of Corollary \ref{Cor:OLS_oracle} and Theorem \ref{Thm:ridge}}\label{Sec:App_cor1_thm3}

\begin{proof}[\textbf{Proof of Corollary \ref{Cor:OLS_oracle}}]
The result in Corollary \ref{Cor:OLS_oracle} follows almost directly from the proof of Theorem \ref{Thm:Estimation_Consistency}. Noting that the first term on the RHS of \eqref{eq:foc_est} can be rewritten as $\hat{\bgamma}_{OLS,S_\gamma} - \bgamma_{S_\gamma}$, it follows that
\begin{equation*}
    \hat{\bgamma}_{S_\gamma} - \hat{\bgamma}_{OLS,S_\gamma} =  - \frac{1}{2}\left(\bV_1^\prime\bV_1\right)^{-1}\left(\lambda_I\bOmega_1\bs_{1,S_\gamma} + \lambda_G\bs_{2,S_\gamma}\right).
\end{equation*}
Then, after scaling, rotating and taking the Euclidean norm, we obtain
\begin{equation}\label{eq:subgradient_bound_ols}
\begin{split}
&\norm{\bS_T\bQ^{\prime -1}\left(\hat{\bgamma}_{S_\gamma}-\hat{\bgamma}_{OLS,S_\gamma}\right)}_2\leq \norm{\left(\bS_T^{-1}\bQ\bV_1^\prime\bV_1\bQ^\prime\bS_T^{-1}\right)^{-1}}_2\norm{\bS_T^{-1}\bQ\left(\frac{\lambda_I}{2}\bOmega_1\bs_{1,S_\gamma} + \frac{\lambda_G}{2}\bs_{2,S_\gamma}\right)}_2\\
&\quad \leq \phi^{-1}\left(\frac{\lambda_I}{2}\norm{\bS_T^{-1}\bQ\bOmega_1\bs_{1,S_\gamma}}_2 + \frac{\lambda_G}{2}\norm{\bS_T^{-1}\bQ\bs_{2,S_\gamma}}_2\right) + o_p(1),
\end{split}
\end{equation}
by Lemma \ref{Cor:eigenvalue}. The orders of the two RHS terms of \eqref{eq:subgradient_bound_ols} are given by \eqref{eq:Op_RHS2} and \eqref{eq:Op_RHS3}, thereby concluding the proof of Corollary \ref{Cor:OLS_oracle}.
\end{proof}

\begin{proof}[\textbf{Proof of Theorem \ref{Thm:ridge}}]
The analytic expression for the ridge estimator is given by
\begin{equation}\label{eq:ridge_1}
\begin{split}
\hat{\bgamma}_R &= \left(\bV^\prime \bM\bV + \lambda_R\bI_{N+M}\right)^{-1}\bV^\prime\bM \Delta \by =  \left(\bV^\prime\bM \bV + \lambda_R\bI_{N+M}\right)^{-1}\left(\bV^\prime\bM\bV\bgamma + \bV^\prime\bM\bepsilon_y\right)\\ 
&= \bgamma + \left(\bV^\prime\bM \bV + \lambda_R\bI_{N+M}\right)^{-1}\left(\bV^\prime\bM\bepsilon_y - \lambda_R\bgamma\right).\\
\end{split}
\end{equation}
After appropriate scaling, \eqref{eq:ridge_1} reads as
\begin{equation}\label{eq:ridge_scaled}
\begin{split}
\bS_{R}\bQ_R^{\prime -1}\left(\hat{\bgamma}_R - \bgamma\right) &= \left(\bS_R^{-1}\bQ_R\bV^\prime\bM \bV\bQ_R^\prime\bS_R^{-1} + \lambda_R\bS_R^{-2}\right)^{-1}\left(\bS_R^{-1}\bQ_R\bV^\prime\bM\bepsilon_y - \lambda_R\bS_R^{-1}\bQ_R\bgamma\right).
\end{split}
\end{equation}
We proceed by bounding the norms of the three RHS quantities in \eqref{eq:ridge_scaled} as
\begin{equation}\label{eq:ridge_normed}
\begin{split}
\norm{\bS_{R}\bQ_R^{\prime -1}\left(\hat{\bgamma}_R - \bgamma\right)}_2 &\leq \norm{\left(\bS_R^{-1}\bQ_R\bV^\prime \bM\bV\bQ_R^\prime\bS_R^{-1} + \lambda_R\bS_R^{-2}\right)^{-1}}_2\\
&\quad \times\left(\norm{\bS_R^{-1}\bQ_R\bV^\prime\bM\bepsilon_y}_2 + \lambda_R \norm{\bS_R^{-1}\bQ_R\bgamma}_2\right)\\
\end{split}
\end{equation}
Focussing on the first RHS term of \eqref{eq:ridge_normed}, we note that
\begin{equation*}
\begin{split}
&\norm{\left(\bS_R^{-1}\bQ_R\bV^\prime\bM \bV\bQ_R^\prime\bS_R^{-1} + \lambda_R\bS_R^{-2}\right)^{-1}}_2 \geq \frac{1}{\lambda_\min\left(\hat{\bSigma_R}\right) + \frac{\lambda_R}{T^2}} + o_p(1) \geq \frac{1}{\phi_R} + o_p(1),
\end{split}
\end{equation*}
by part 1 of Lemma \ref{Cor:conversion}. The stochastic order of the second RHS term is given by part 2 of Lemma \ref{Cor:conversion} as $\norm{\bS_R^{-1}\bQ_R\bV^\prime\bM\bepsilon_y}_2 = O_p\left(N_\delta + \sqrt{M_\pi}\right)$. The third and final RHS term is deterministically bounded by
\begin{equation*}
\begin{split}
&\lambda_R \norm{\bS_R^{-1}\bQ_R\bgamma}_2 \leq \frac{\lambda_R}{\sqrt{T}}\left(\norm{\left(\bB^\prime\bB\right)^{-1/2}\bB^\prime \bdelta}_2 + \norm{\bpi}_2\right)\\
&\quad \leq \frac{\lambda_R}{\sqrt{T}}\left(\sqrt{\abs{S_\delta}}\norm{\bdelta}_\infty + \sqrt{\abs{S_\pi}}\norm{\bpi}_\infty\right) = O\left(\frac{\lambda_R\left(\sqrt{\abs{S_\delta} + \abs{S_\pi}}\right)}{\sqrt{T}}\right).
\end{split}
\end{equation*}
As a result, we obtain the stochastic order of \eqref{eq:ridge_scaled} as
\begin{equation*}
\begin{split}
\norm{\bS_{R}\bQ_R^{\prime -1}\left(\hat{\bgamma}_R - \bgamma\right)}_2 &= O_p\left(N_\delta + \sqrt{M_\pi}\right) + O_p\left(\frac{\lambda_R\left(\sqrt{\abs{S_\delta} + \abs{S_\pi}}\right)}{\sqrt{T}}\right) = O_p\left(N_\delta + \sqrt{M_\pi}\right)
\end{split}
\end{equation*}
where the last equality follows from the assumption that $\lambda_R = O\left(\frac{\left(N_\delta + \sqrt{M_\pi}\right)\sqrt{T}}{\sqrt{\abs{S_\delta} + \abs{S_\pi}}}\right)$.
\end{proof}

\subsection{Proofs of Main Theorems in Section B.2} \label{sec:evproofs}
\begin{proof}[\textbf{Proof of Theorem \ref{Lemma:min_eig1}}]
We prove Theorem \ref{Lemma:min_eig1} by showing that $\norm{\hat{\bSigma}_{11}-\bSigma_{11}}_2 \overset{p}{\to} 0$ as $T,s_\pi \to \infty$, after which application of part (1) of Lemma \ref{Lemma:eig_transl} leads to the desired result. From (the proof of) part (3) of Lemma \ref{Prop:idemp}, it follows that we may assume that $\bmu=\btau=0$ and $\bM = \bI$ without loss of generality.

\citet{Chen2013} derive the convergence rates for thresholded estimates of high-dimensional covariance matrices, based on the functional dependence measure in \citet{Wu2005}. The key feature of this dependence measure is the construction of a coupled version of the stochastic process, which in our setting results in the process $\bv^*_{1,t} = \sum_{l=0}^\infty \bC^v_l \bepsilon_{t-l}^*$, where $\bepsilon_t^* = \tilde{\bepsilon}_t$ for $t\neq 0$ and $\bepsilon_0^*$ is an i.i.d. copy of $\tilde{\bepsilon}_0$. By Assumption \ref{Ass:moments}, for any $w \leq 2m$, the functional dependence measure for element $v_{1,j,t}$ is bounded by
\begin{equation*}
\begin{split}
\theta_{j,t,w} &= \abs{v_{1,j,t} - v_{1,j,t}^*}_w = \abs{\sum_{i=1}^N\sum_{l=0}^\infty c^v_{l,j,i}\left(\epsilon_{i,t-l} - \epsilon^*_{i,t-l}\right)}_w = \abs{\sum_{i=1}^N c^v_{t,j,i}\left(\epsilon_{i,0} - \epsilon^*_{i,0}\right)}_w\\
&\leq \sum_{i=1}^N \abs{c^v_{t,j,i}}\abs{\epsilon_{i,0} - \epsilon^*_{i,0}}_w \leq \abs{\epsilon_{i,0} - \epsilon^*_{i,0}}_w\norm{\bc^v_{t,j}}_1 \leq K\norm{\bc^v_{t,j}}_1,
\end{split}
\end{equation*}
where $\bc^v_{t,j}$ is the $j$-th row of $\bC^v_t$ and $\abs{\cdot}_w = \left(\E(\cdot)^w\right)^{1/w}$. Then, with the addition of Assumption \ref{Ass:Dependence},
\begin{equation*}
\Theta_{k,w} = \max_{1 \leq j \leq N} \sum_{l=k}^\infty \theta_{j,l,w}  \leq K\sum_{l=k}^\infty \norm{\bC^v_l}_\infty = O\left(k^{-1}\right),
\end{equation*}
for all $k>0$. Therefore, the conditions in Theorem 2.1 of \citet{Chen2013} are satisfied. From this theorem, it follows that $\E\norm{\hat{\bSigma}_{11}-\bSigma_{11}}_F^2 = O\left(\frac{s_\pi^2}{T}\right)$, by taking the limit of the threshold value as it approaches zero. Thus, application of the Markov inequality shows that $\norm{\hat{\bSigma}_{11}-\bSigma_{11}}_2 \overset{p}{\to} 0$ as $T,s_\pi \to \infty$ with $\frac{s_\pi}{\sqrt{T}}\to 0$. By part (1) of Lemma \ref{Lemma:eig_transl}, this implies that $\abs{\lambda_\min\left(\hat{\bSigma}_{11}\right) - \lambda_\min\left(\bSigma_{11}\right)} \overset{p}{\to} 0$. Then,
\begin{align*}
&\Prob\left(\lambda_\min\left(\hat{\bSigma}_{11}\right) \geq \phi\right) \geq \Prob\left(\lambda_\min\left(\bSigma_{11}\right) - \abs{\lambda_\min\left(\hat{\bSigma}_{11}\right) - \lambda_\min\left(\bSigma_{11}\right)} \geq \phi\right)
\\ 
&\quad
\geq \Prob\left(\abs{\lambda_\min\left(\hat{\bSigma}_{11}\right) - \lambda_\min\left(\bSigma_{11}\right)} \leq \phi\right) \to 1.\qedhere
\end{align*}
\end{proof}

\begin{proof}[\textbf{Proof of Theorem \ref{Lemma:Zhang_LB}}]
First, letting $\tilde{\bu}_{S_\delta,t} = \bC_{S_\delta}(L)\tilde{\bepsilon}_t$, we decompose $\hat{\bSigma}_{22}$ into
\begin{equation}\label{eq:As}
\begin{split}
\hat{\bSigma}_{22} &= \bB_{S_\delta,\perp}^\prime\left(\frac{s_\delta}{T^2}\sum_{t=1}^T \left(\bC_{S_\delta}\tilde{\bs}_{t-1} + \tilde{\bu}_{S_\delta,t-1}\right)\left(\bC_{S_\delta}\tilde{\bs}_{t-1} + \tilde{\bu}_{S_\delta,t-1}\right)^\prime\right)\bB_{S_\delta,\perp}\\
&= \bB_{S_\delta,\perp}^\prime\bC_{S_\delta}\left(\frac{s_\delta}{T^2}\sum_{t=1}^T \tilde{\bs}_{t-1}\tilde{\bs}_{t-1}^\prime\right)\bC_{S_\delta}^\prime\bB_{S_\delta,\perp}  + \bB_{S_\delta,\perp}^\prime\bC_{S_\delta}\left(\frac{s_\delta}{T^2}\sum_{t=1}^T \tilde{\bs}_{t-1}\tilde{\bu}_{S_\delta,t-1}^\prime\right)\bB_{S_\delta,\perp}\\
&\quad + \bB_{S_\delta,\perp}^\prime\left(\frac{s_\delta}{T^2}\sum_{t=1}^T \tilde{\bu}_{S_\delta,t-1}\tilde{\bs}_{t-1}^\prime\right)\bC_{S_\delta}^\prime\bB_{S_\delta,\perp} + \bB_{S_\delta,\perp}^\prime\left(\frac{s_\delta}{T^2}\sum_{t=1}^T\tilde{\bu}_{S_\delta,t-1}\tilde{\bu}_{S_\delta,t-1}^\prime\right)\bB_{S_\delta,\perp}\\
&=: \bA_1 + \bA_2 + \bA_2^\prime + \bA_3,
\end{split}
\end{equation}
such that $\lambda_\min\left(\hat{\bSigma}_{22}\right) \geq \lambda_\min(A_1) - 2\norm{\bA_2}_2 - \norm{\bA_3}_2$. We show that there exists a $\zeta>0$ such that $\Prob\left( \lambda_\min\left(\bA_1\right) > \zeta\right) \to 1$, whereas $\Prob\left(\norm{\bA_2}_2 > \zeta\right) \to 0 \text{ and } \Prob\left(\norm{\bA_3}_2 > \zeta\right) \to 0$ as $s_\delta,T \to \infty$. 

Define $\bU$ as a lower triangular matrix with ones on and below the diagonal, such that we may write $\bS = \bU\bE$. Then, by the assumption that $\bepsilon_t \overset{i.i.d.}{\sim} \mathcal{N}\left(\bm{0},\bSigma_\epsilon\right)$, it holds that
\begin{equation*}
\begin{split}
\bA_1 &= \bB_{S_\delta,\perp}^\prime\bC_{S_\delta}\left(\frac{s_\delta}{T^2}\sum_{t=1}^T \tilde{\bs}_{t-1}\tilde{\bs}_{t-1}^\prime\right)\bC_{S_\delta}^\prime\bB_{S_\delta,\perp} = \bB_{S_\delta,\perp}^\prime\bC_{S_\delta}\left(\frac{s_\delta}{T^2}\sum_{t=1}^T \tilde{\bs}_t\tilde{\bs}_t^\prime\right)\bC_{S_\delta}^\prime\bB_{S_\delta,\perp} + o_p(1)\\
&= \frac{s_\delta}{T^2} \bB_{S_\delta,\perp}^\prime\bC_{S_\delta}\bE^\prime\bU^\prime\bM\bU\bE\bC_{S_\delta}^\prime\bB_{S_\delta,\perp} + o_p(1)
\end{split}
\end{equation*}
Furthermore, since the errors follow a multivariate Gaussian distribution,
\begin{equation*}
    \frac{s_\delta}{T^2} \bB_{S_\delta,\perp}^\prime\bC_{S_\delta}\bE^\prime\bU^\prime\bM\bU\bE\bC_{S_\delta}^\prime\bB_{S_\delta,\perp} \overset{d}{=} \frac{s_\delta}{T^2}\bR^{1/2}\bE_G^\prime\bU^\prime\bM\bU\bE_G\bR^{1/2},
\end{equation*}
where $\bR = \bB_{S_\delta,\perp}^\prime\bC_{S_\delta}\bSigma_\epsilon\bC_{S_\delta}^\prime\bB_{S_\delta,\perp}$, and $\bE_G = (\bepsilon_{G,1},\ldots,\bepsilon_{G,T})^\prime$ with $\bepsilon_{G,s} \overset{i.i.d.}{\sim} \mathcal{N}\left(\bm{0},\bI_{s_\delta}\right)$ being an $s_\delta$-dimensional Gaussian white noise process. Furthermore, for any $\bx \in \mathbb{R}^{s_\delta}$ with $\bx^\prime\bx = 1$,
\begin{equation*}
\bx^\prime \bB_{S_\delta,\perp}^\prime\bC_{S_\delta}\bSigma_\epsilon\bC_{S_\delta}^\prime\bB_{S_\delta,\perp}\bx = \by^\prime \bSigma_\epsilon \by \geq \phi_\min,
\end{equation*}
where $\by \neq \bm{0}$, because $\bC_{S_\delta}^\prime\bB_{S_\delta,\perp}$ has full column rank by construction, and the inequality follows by Assumption \ref{Ass:moments}. Thus,
\begin{equation*}
\begin{split}
&\lambda_\min\left(\bA_1\right) \geq \lambda_\min\left(\bR\right)\lambda_\min\left(\frac{s_\delta}{T^2}\bE_G^\prime\bU^\prime\bM\bU\bE_G\right) + o_p(1) \geq \phi_\min\lambda_\min\left(\frac{s_\delta}{T^2}\bE_G^\prime\bU^\prime\bM\bU\bE_G\right) + o_p(1).
\end{split}
\end{equation*}
Letting $\bP = \bD\left(\bD^\prime\bD\right)^{-1}\bD^\prime$, we may rewrite $\bU^\prime\bU = \bU^\prime\bM\bU + \bU^\prime\bP\bU$. If $\bD$ is empty, because the researcher does not wish to incorporate deterministic components, we simply define $\bP = \bm{0}$. Then, by application of Theorem A.8 in \citet{Bai2010} (originally Theorem 2 in \citet{Fan1951}),
\begin{equation}\label{eq:UU_ineq}
    \lambda_{i+1}\left(\bU^\prime\bM\bU\right) \geq \lambda_{i+j+1}\left(\bU^\prime\bU\right) - \lambda_{j+1}\left(\bU^\prime\bP\bU\right).
\end{equation}
Depending on the construction of $\bD$, the matrix $\bU^\prime\bP\bU$ has at most rank two, such that we may set $j=2$ in \eqref{eq:UU_ineq} to obtain
\begin{equation}\label{eq:UU_ineq2}
\lambda_{i+1}\left(\bU^\prime\bM\bU\right) \geq \lambda_{i+3}\left(\bU^\prime\bU\right).
\end{equation}
By the spectral decomposition, we may write $\bU^\prime\bM\bU = \bV\bLambda\bV^\prime$ where $\bLambda=\diag(\lambda_1,\ldots,\lambda_T)$, such that
\begin{equation*}
\bE_G^\prime\bU^\prime\bU\bE_G = \bE_G^\prime\bV\bLambda\bV^\prime\bE_G = \tilde{\bE}\bLambda\tilde{\bE},
\end{equation*}
where $\tilde{\bE} = \bV^\prime\bE_G$. By the rotational invariance of the multivariate normal distribution, $\tilde{\bE}$ is again an $(T \times s_\delta)$-dimensional matrix with independent standard normal entries. Define $\mathbb{R}_G = \left\lbrace \bx \in \mathbb{R}^{s_\delta} \ : \ \bx^\prime\bx = 1\right\rbrace$. Let $\by_x = \bV^\prime\bE_G\bx$. Then, similar to the proof of Remark 3.5 in \citet{Zhang2018b},
\begin{equation}\label{eq:Zhang_bound1}
\begin{split}
&\lambda_\min\left(\frac{s_\delta}{T^2}\bE_G^\prime\bU^\prime\bM\bU\bE_G\right) =  \frac{s_\delta}{T^2}\lambda_\min\left(\bE_G^\prime\bV\bLambda\bV^\prime\bE_G\right) = \frac{s_\delta}{T^2}\underset{\bx \in \mathbb{R}_G}{\text{min}}\bx^\prime\bE_G^\prime\bV\bLambda\bV^\prime\bE_G\bx\\
&\quad = \frac{s_\delta}{T}\underset{\bx \in \mathbb{R}_G}{\text{min}}\frac{\by_x^\prime\by_x}{T}\frac{\by_x^\prime\bLambda\by_x}{\by_x^\prime\by_x} \geq \frac{s_\delta}{T}\left(\underset{\bx \in \mathbb{R}_G}{\text{min}}\frac{\by_x^\prime\by_x}{T}\right)\left(\underset{\bx \in \mathbb{R}_G}{\text{min}}\frac{\by_x^\prime\bLambda\by_x}{\by_x^\prime\by_x}\right)\\
&\quad = \frac{s_\delta}{T}\lambda_\min\left(\frac{\tilde{\bE}^\prime\tilde{\bE}}{T}\right)\left(\underset{\bx \in \mathbb{R}_G}{\text{min}}\frac{\by_x^\prime\bLambda\by_x}{\by_x^\prime\by_x}\right) \geq \frac{s_\delta}{T}\lambda_\min\left(\frac{\tilde{\bE}^\prime\tilde{\bE}}{T}\right)\left(\underset{\bx \in \mathbb{R}_G}{\text{min}}\frac{\sum_{j=1}^{s_\delta} y_{x,j}^2\lambda_j}{\by_x^\prime\by_x}\right)\\
&\quad \geq \frac{ks_\delta}{T^2}\frac{\lambda_\min\left(\tilde{\bE}^\prime\tilde{\bE}/T\right)}{\lambda_\max\left(\tilde{\bE}^\prime\tilde{\bE}/T\right)}\underset{\bx \in \mathbb{R}_G}{\text{min}}\frac{1}{k}\sum_{j=1}^k y_{x,j}^2\lambda_{k+1},
\end{split}
\end{equation}
for any $k\leq T$. By Lemma 1 in \citet{Akesson1998}, it holds that
\begin{equation}\label{eq:eig_UU1}
\lambda_t\left(\bU^\prime\bU\right)^{-1} = 4\sin^2\left(\frac{\omega_t}{2}\right) = 2(1-\cos \omega_t),
\end{equation}
with $\omega_t = \frac{(2t - 1)\pi}{2T+1}$, where the second equality in \eqref{eq:eig_UU1} is based on the identity $\cos(2\alpha) = 1 - 2\sin^2(\alpha)$. Combining \eqref{eq:UU_ineq2} and \eqref{eq:eig_UU1}, it holds that
\begin{equation}\label{eq:lambda_k}
\lambda_{k+1} \geq \lambda_{k+3}\left(\bU^\prime\bU\right) = \frac{1}{2(1 - \cos \omega_{k+3})} \geq \frac{1}{2\omega_{k+3}^2} = \frac{(2T+1)^2}{2(2k+5)^2\pi^2} \geq \frac{2T^2}{(2k+5)^2\pi^2}.
\end{equation}
In addition, by Theorem 2.1 in \citet{Chen2013} it follows that, for any $\epsilon>0$, $\Prob\left(\norm{\frac{\tilde{\bE}^\prime\tilde{\bE}}{T} - \bI_{s_\delta}}_2 > \epsilon\right) \to 0$ as $s_\delta,T \to \infty$ with $\frac{s_\delta}{T^{1/2}} \to 0$. As a consequence of Weyl's inequality in Lemma \ref{Lemma:eig_transl}, this implies that $\frac{\lambda_\min\left(\tilde{\bE}^\prime\tilde{\bE}/T\right)}{\lambda_\max\left(\tilde{\bE}^\prime\tilde{\bE}/T\right)} \overset{p}{\to} 1$
as $s_\delta,T \to \infty$ with $\frac{s_\delta}{T^{1/2}} \to 0$. Finally, define $\bV^k = \left(\bv_1,\ldots,\bv_k\right)$, and note that $\left(\bV^k\right)^\prime\bE_G$ is a $(k \times s_\delta)$-dimensional matrix with independent standard normal entries. Then, choosing $k$ such that $\frac{s_\delta}{k} \to y \in (0,1)$, it follows from Theorem 1 in \citet{Bai1993} that
\begin{equation}\label{eq:Bai_lim}
\lim \lambda_\min\left(\frac{\bE_G^\prime\bV^K\left(\bV^k\right)^\prime\bE_G}{k}\right) = (1-\sqrt{y})^2, \text{ a.s.}
\end{equation}
Then, plugging \eqref{eq:lambda_k}-\eqref{eq:Bai_lim} into \eqref{eq:Zhang_bound1}, we obtain
\begin{equation}\label{eq:Zhang_bound2}
\begin{split}
&\lambda_\min\left(\frac{s_\delta}{T^2}\bE_G^\prime\bU^\prime\bM\bU\bE_G\right) \geq \frac{2ks_\delta}{(2k+5)^2\pi^2}\frac{\lambda_\min\left(\tilde{\bE}^\prime\tilde{\bE}/T\right)}{\lambda_\max\left(\tilde{\bE}^\prime\tilde{\bE}/T\right)}\underset{\bx \in \mathbb{R}_G}{\text{min}}\frac{1}{k}\sum_{j=1}^k y_{x,j}^2\\
&= \frac{2ks_\delta}{(2k+5)^2\pi^2}\frac{\lambda_\min\left(\tilde{\bE}^\prime\tilde{\bE}/T\right)}{\lambda_\max\left(\tilde{\bE}^\prime\tilde{\bE}/T\right)}\lambda_\min\left(\frac{\bE_G^\prime\bV^K\left(\bV^k\right)^\prime\bE_G}{k}\right) \to \frac{y(1-\sqrt{y})^2}{2\pi^2},
\end{split}
\end{equation}
in probability as $s_\delta,k,T \to \infty$ with $\frac{s_\delta}{k} \to y$ and $\frac{s_\delta}{T^{1/2}} \to 0$.

It remains to show that, as defined in \eqref{eq:As}, $\norm{\bA_2}_2,\norm{\bA_3}_2 = o_p(1)$ as $s_\delta,T \to \infty$. First, by the BN decomposition of $\bC_{S_\delta}(L) = \bC_{S_\delta}(1) + \bC_{S_\delta}^*(L)(1-L)$, letting $\boeta_t = \bC_{S_\delta}^*(L)\bepsilon_t$, we can rewrite
\begin{equation}\label{eq:BS_partial}
\begin{split}
&\bA_2 = \bB_{S_\delta,\perp}^\prime\bC_{S_\delta}\left(\frac{s_\delta }{T^2}\sum_{t=1}^T \tilde{\bs}_{t-1}\tilde{\bu}_{S_\delta,t-1}^\prime\right)\bB_{S_\delta,\perp} = \bB_{S_\delta,\perp}^\prime\bC_{S_\delta}\left(\frac{s_\delta}{T^2}\sum_{t=1}^T \tilde{\bs}_{t-2}\tilde{\bepsilon}_{t-1}^\prime\right)\bC_{S_\delta}(1)\bB_{S_\delta,\perp}\\
&\quad + \bB_{S_\delta,\perp}^\prime\bC_{S_\delta}\left(\frac{s_\delta}{T^2}\sum_{t=1}^T \tilde{\bepsilon}_{t-1}\tilde{\bepsilon}_{t-1}^\prime\right)\bC_{S_\delta}(1)\bB_{S_\delta,\perp} + \bB_{S_\delta,\perp}^\prime\bC_{S_\delta}\left(\frac{s_\delta}{T^2}\sum_{t=1}^T \tilde{\bs}_{t-1}\left(\boeta_{t-1}-\boeta_{t-2}\right)^\prime\right)\bB_{S_\delta,\perp}.\\
\end{split}
\end{equation}
Furthermore, using summation by parts, we can further simplify the last term on the RHS of \eqref{eq:BS_partial} to
\begin{equation}\label{eq:summ_part_Bs}
\begin{split}
&\bB_{S_\delta,\perp}^\prime\bC_{S_\delta}\left(\frac{s_\delta}{T^2}\sum_{t=1}^T \tilde{\bs}_{t-1}\left(\boeta_{t-1}-\boeta_{t-2}\right)^\prime\right)\bB_{S_\delta,\perp}\\
&\quad = \bB_{S_\delta,\perp}^\prime\bC_{S_\delta}\left(\frac{s_\delta}{T^2}\tilde{\bs}_{T-1}\boeta_{T-1}^\prime\right)\bB_{S_\delta,\perp}  - \bB_{S_\delta,\perp}^\prime\bC_{S_\delta}\left(\frac{s_\delta}{T^2}\sum_{t=1}^{T-1}\tilde{\bepsilon}_t\boeta_{t-1}^\prime\right)\bB_{S_\delta,\perp}.
\end{split}
\end{equation}
Then, plugging \eqref{eq:summ_part_Bs} into \eqref{eq:BS_partial}, we obtain the lengthy expression
\begin{equation}\label{eq:Bs}
\begin{split}
&\bB_{S_\delta,\perp}^\prime\bC_{S_\delta}\left(\frac{s_\delta}{T^2}\sum_{t=1}^T \tilde{\bs}_{t-1}\tilde{\bu}_{S_\delta,t-1}^\prime\right)\bB_{S_\delta,\perp}= \bB_{S_\delta,\perp}^\prime\bC_{S_\delta}\left(\frac{s_\delta}{T^2}\sum_{t=1}^T \tilde{\bs}_{t-2}\tilde{\bepsilon}_{t-1}^\prime\right)\bC_{S_\delta}(1)\bB_{S_\delta,\perp}\\
&\quad\quad + \bB_{S_\delta,\perp}^\prime\bC_{S_\delta}\left(\frac{s_\delta}{T^2}\sum_{t=1}^T \tilde{\bepsilon}_{t-1}\tilde{\bepsilon}_{t-1}^\prime\right)\bC_{S_\delta}(1)\bB_{S_\delta,\perp} + \bB_{S_\delta,\perp}^\prime\bC_{S_\delta}\left(\frac{s_\delta}{T^2}\tilde{\bs}_{T-1}\boeta_{T-1}^\prime\right)\bB_{S_\delta,\perp}\\ 
&\quad\quad - \bB_{S_\delta,\perp}^\prime\bC_{S_\delta}\left(\frac{s_\delta}{T^2}\sum_{t=1}^{T-1}\tilde{\bepsilon}_t\boeta_{t-1}^\prime\right)\bB_{S_\delta,\perp} =: \sum_{i=1}^4\bD_i,
\end{split}
\end{equation}
with each $\bD_i$ corresponding to the $i$-th term on the RHS of the first equation. Thus, the convergence rate of $\bA_2$ in \eqref{eq:As} follows from the rates of the terms $\bD_i$ in \eqref{eq:Bs}. Let $\ba_i = \bC_{S_\delta}^\prime\bbeta_{S_\delta,\perp,i}$ and $\bb_j = \bC_{S_\delta}(1)\bbeta_{S_\delta,\perp,j}$. For $\bD_1$ we obtain, by application of part (2) of Lemma \ref{Prop:idemp} and Markov's inequality, that
\begin{equation}\label{eq:B1_1}
\begin{split}
&\Prob\left(\norm{\bD_1}_2 \geq \zeta \right)  \leq \Prob\left(\norm{\bB_{S_\delta,\perp}^\prime\bC_{S_\delta}\left(\frac{s_\delta}{T^2}\sum_{t=1}^T \bs_{t-2}\bepsilon_{t-1}^\prime\right)\bC_{S_\delta}(1)\bB_{S_\delta,\perp}}_2 \geq \zeta\right) + o(1)\\
&\quad = \Prob\left(\sum_{i,j=1}^{s_\delta}\left(\frac{s_\delta}{T^2}\sum_{t=3}^T \ba_i^\prime\bs_{t-2}\bepsilon_{t-1}^\prime\bb_j\right)^2 \geq \zeta^2\right)  + o(1) \leq \frac{s_\delta^2 \sum_{i,j=1}^{s_\delta}\E\left(\sum_{t=3}^T\ba_i^\prime \bs_{t-2}\bepsilon_{t-1}^\prime\bb_j\right)^2}{T^4\zeta^2} + o(1).
\end{split}
\end{equation}
Then, noting that $\lbrace \ba_i^\prime \bs_{t-2}\bepsilon_{t-1}^\prime\bb_j\rbrace$ is a martingale difference sequence, it follows from Burkholder's inequality in combination with the $C_r$-inequality that we can bound the expectation by
\begin{equation}\label{eq:B1_2}
\begin{split}
&\E\left(\sum_{t=3}^T\ba_i^\prime \bs_{t-2}\bepsilon_{t-1}^\prime\bb_j\right)^2 \leq K\sum_{t=3}^T\E\left(\ba_i \bs_{t-2}\bepsilon_{t-1}^\prime\bb_j\right)^2 = K\sum_{t=3}^T\E\left(\ba_i^\prime \bs_{t-2}\right)^2\E\left(\bb_j^\prime\bepsilon_{t-1}\right)^2\\
&\quad = K\left(\bb_j^\prime\bSigma_\epsilon\bb_j\right)\left(\ba_i^\prime\bSigma_\epsilon\ba_i\right)\sum_{t=1}^{T-2} t \leq  T^2K\norm{\ba_i}_2^2\norm{\bb_j}_2^2\phi_{\max}^2 \leq T^2K\norm{\bC_{S_\delta}}_2^2\norm{\bC_{S_\delta}(1)}_2^2\phi_{\max}^2,
\end{split}
\end{equation}
where we use that, by the column normalization on $\bB_{S_\delta,\perp}$, we have
\begin{align*}
\norm{\ba_i}_2^2 &= \norm{\bC_{S_\delta}^\prime\bbeta_{S_\delta,\perp,i}}_2^2 \leq \norm{\bC_{S_\delta}}_2^2\norm{\bbeta_{S_\delta,\perp,i}}_2^2 \leq  \norm{\bC_{S_\delta}}_2^2,\\
\norm{\bb_j}_2^2 \leq &= \norm{\bC_{S_\delta}(1)\bbeta_{S_\delta,\perp,j}}_2^2\leq \norm{\bC_{S_\delta}(1)}_2^2\norm{\bbeta_{S_\delta,\perp,j}}_2^2 \leq \norm{\bC_{S_\delta}(1)}_2^2.
\end{align*}
Plugging \eqref{eq:B1_2} into \eqref{eq:B1_1}, we conclude that $\Prob\left(\norm{\bD_1}_2 \geq \zeta \right) \to 0$ based on Assumption \ref{Ass:Dependence} and the assumption that $\frac{s_\delta}{T^{1/2}} \to 0$.

In a similar fashion, we bound $\bD_2$ in \eqref{eq:Bs}. By part (2) of Lemma \ref{Prop:idemp}, Markov's and Minkowski's inequality,
\begin{equation*}
\begin{split}
&\Prob\left(\norm{\bD_2}_2 > \zeta\right) \leq \frac{s_\delta^2\sum_{i,j=1}^{s_\delta} \left(\sum_{t=1}^T\sum_{s_1,s_2=1}^N\abs{a_{i,s_1}}\abs{b_{j,s_2}}\left(\E\left(\epsilon_{s_1,t-1}\epsilon_{s_2,t-1}\right)^2\right)^{1/2}\right)^2}{T^4\zeta^2} + o(1)\\
&\quad \leq \frac{s_\delta^2K\sum_{i,j=1}^{s_\delta} \norm{a_{i,s_1}}_1^2\norm{b_{j,s_2}}_1^2}{T^2\zeta^2} + o(1) \leq \frac{s_\delta^4 K\norm{\bC_{S_\delta}}_\infty^2\norm{\bC_{S_\delta}(1)}_\infty^2}{T^2\zeta^2} + o(1) \to 0,
\end{split}
\end{equation*}
where we have used that $\E\left(\epsilon_{s_1,t-1}\epsilon_{s_2,t-1}\right)^2 \leq K$ by Assumption \ref{Ass:moments} and the boundedness of $\norm{\bC_{S_\delta}}_\infty^2\norm{\bC_{S_\delta}(1)}_\infty^2$ follows from Assumption \ref{Ass:Dependence}. We omit repeating this argument in the following bounds.

\bigskip
Next, we bound $\bD_3$. Recall that $\boeta_{t} = \bC^*_{S_\delta}(L)\bepsilon_t$, where $\bC^*_{S_\delta}(z) = \sum_{l=0}^\infty\bC^*_{S_\delta,l}z^l$ with $\sum_{l=0}^\infty\norm{\bC^*_{S_\delta,l}}_\infty \leq K$ by Assumption \ref{Ass:Dependence}. Defining $\bb_{j,l} = \bC^*_{S_\delta,l}\bbeta_{S_\delta,\perp,j}$, we follow a similar strategy to obtain
\begin{equation*}
\begin{split}
&\Prob\left(\norm{\bD_3}_2 > \zeta\right) \leq \frac{s_\delta^2\sum_{i,j=1}^{s_\delta} \left(\sum_{k=1}^{T-1}\sum_{l=0}^\infty\sum_{s_1,s_2=1}^N \abs{a_{i,s_1}}\abs{b_{j,l,s_2}}\left(\E\left(\epsilon_{s_1,k}\epsilon_{s_2,T-1-l}\right)^2\right)^{1/2}\right)^2}{T^4\zeta^2} + o(1)\\
&\quad \leq \frac{s_\delta^2K\sum_{i,j=1}^{s_\delta}\left(\sum_{l=0}\norm{\bb_{j,l}}_1\right)^2\norm{\ba_i}_1^2}{T^2\zeta^2} + o(1) \leq \frac{s_\delta^4 K\left(\sum_{l=0}^\infty \norm{\bC^*_{S_\delta,l}}_\infty\right)^2\norm{\bC_{S_\delta}}_\infty^2}{T^2\zeta^2} + o(1) \to 0.
\end{split}
\end{equation*}
Finally, for $\bD_4$, we proceed analogously, to obtain
\begin{equation*}
\begin{split}
&\Prob\left(\norm{\bD_4}_2 > \zeta\right) \leq \frac{s_\delta^2\sum_{i,j=1}^{s_\delta}\left(\sum_{t=1}^{T-1}\sum_{l=0}^\infty \sum_{s_1,s_2=1}^N \abs{a_{i,s_1}}\abs{b_{j,s_2}}\left(\E\left(\epsilon_{s_1,t}\epsilon_{s_2,t-1-l}\right)^2\right)^{1/2}\right)^2}{T^4\zeta^2} + o(1)\\
&\quad \leq \frac{s_\delta^2K\sum_{i,j=1}^{s_\delta}\left(\sum_{l=0}^\infty \norm{\bb_j,l}_1\right)^2\norm{\ba_i}_1^2}{T^2\zeta^2} + o(1) \leq \frac{s_\delta^4K\left(\sum_{l=0}^\infty \norm{\bC^*_{S_\delta,l}}_\infty\right)^2\norm{\bC_{S_\delta}}_\infty^2}{T^2\zeta^2} + o(1) \to 0.\\
\end{split}
\end{equation*}
Combining the results for $\bD_1$ to $\bD_4$, it follows that $\Prob\left(\norm{\bA_2}_2 \geq \zeta\right) \to 0$ as $s_\delta,T \to \infty$.

\bigskip
The last term to derive the stochastic order for is $\bA_3$ in \eqref{eq:As}. Define $\ba_{i,l} = \bC_{S_\delta,l}^\prime\bbeta_{S_\delta,\perp,i}$. Then, by a combination of part (1) of Lemma \ref{Prop:idemp}, Markov's inequality and Minkowski's inequality,
\begin{equation*}
\begin{split}
&\Prob\left(\norm{\bA_3}_2 > \zeta\right) \leq \frac{s_\delta^2\sum_{i,j=1}^{s_\delta}\left(\sum_{t=1}^T\sum_{l_1,l_2=0}^\infty\sum_{s_1,s_2=1}^N\abs{a_{i,l_1,s_1}}\abs{a_{j,l_2,s_2}}\left(\E\left(\epsilon_{s_1,t-1-l_1}\epsilon_{s_2,t-1-l_2}\right)^2\right)^{1/2}\right)^2}{T^4\zeta^2}\\
&\quad\leq \frac{s_\delta^2K\sum_{i,j=1}^{s_\delta}\left(\sum_{l_1=0}^\infty\norm{a_{i,l_1}}_1\right)^2\left(\sum_{l_2=0}^\infty \norm{a_{j,l_2}}_1\right)^2}{T^2\zeta^2} \leq \frac{s_\delta^4K\left(\sum_{l=0}^\infty\norm{\bC_{S_\delta,l_1}}_\infty\right)^4}{T^2\zeta^2}.
\end{split}
\end{equation*}
Hence, $\Prob\left(\norm{\bA_3}_2 \geq \zeta\right) \to 0$ as $s_\delta,T \to \infty$ with $\frac{s_\delta}{T^{1/2}}\to 0$, thereby completing the proof.
\end{proof}

\begin{proof}[\textbf{Proof of Theorem \ref{Lemma:Sigma_22_gen}}]
Define $\bs_{u,t} = \sum_{s=1}^t \bepsilon_{u,s}$, such that $\bs_t = \bD\bs_{u,t}$. The proof uses a Gaussian approximation of $\bu_t$, similar to \citet{Zhang2018b} in their proof of Remark 3.4. By the martingale version of the Skorokhod representation theorem \citep[][Thm 4.3]{Strassen1967}, on an extended  probability space a standard Brownian motion $W(t)$ and non-negative stopping times $\lbrace \tau_{i,j}\rbrace$ exist, such that for all $i$ and $t \geq 1$, $s_{u,it} = W\left(\sum_{j=1}^t \tau_{i,j}\right) \text{ and } \E\left[\tau_{i,t}\vert \mathcal{F}_{i,t-1}\right] = \E\left[\epsilon_{u,t,i}^2\vert \mathcal{F}_{i,t-1}\right]$, where $\mathcal{F}_{i,t}$ is the natural filtration of the stochastic process $\lbrace \epsilon_{u,s,i}, s\leq t\rbrace$. By the proof of Remark 3.4 in \citet{Zhang2018b}, under Assumption \ref{Ass:moments}, for every sequence $\lbrace \epsilon_{u,t,i}\rbrace$, an independent and standard normal sequence $\lbrace v_{i,t}\rbrace$ exists such that
\begin{equation*}\label{eq:s_ut}
\underset{1 \leq i \leq N}{\text{max}} \ \underset{0 \leq t \leq T}{\text{max}}\E\left(\sum_{s=1}^{[Tt]} (\epsilon_{u,s,i} - \sigma_{u,ii}v_{s,i})\right)^2 = O\left(T^{1/2}\right).
\end{equation*}

Define $\bE_v = \left(\bepsilon_{v,1},\ldots,\bepsilon_{v,T}\right)^\prime$, where $\bepsilon_{v,t} = (\epsilon_{v,t,1},\ldots,\epsilon_{v,t,N})^\prime$ with $\epsilon_{v,t,i} = \sigma_{ii}v_{t,i}$ and $\bE_u = (\bepsilon_{u,1},\ldots,\bepsilon_{u,T})^\prime$. As in the proof of Theorem \ref{Lemma:Zhang_LB}, let $\bU$ be a $(T\times T)$ lower-triangular matrix with ones on and below the diagonal, and let
\begin{equation*}
\tilde{\bA}_1 = \frac{s_\delta}{T^2}\bB_{S_\delta,\perp}^\prime\bC_{S_\delta}\bD\bE_v^\prime\bU^\prime\bU\bE_v\bD^\prime\bC_{S_\delta}^\prime\bB_{S_\delta,\perp}.
\end{equation*}
By the proof of Theorem \ref{Lemma:Zhang_LB}, there exist a $\zeta>0$ such that $\Prob\left(\lambda_\min\left(\tilde{\bA}_1\right) > \zeta\right) \to 1$, as $s_\delta,T \to \infty$. Recall the decomposition $\hat{\bSigma}_{22} = \bA_1 + \bA_2 + \bA_2^\prime + \bA_3$, given by \eqref{eq:As}, such that
\begin{equation}\label{eq:sigma_22_gen}
\begin{split}
\lambda_\min\left(\hat{\bSigma}_{22}\right) &\geq \lambda_\min\left(\tilde{\bA}_1\right) - \norm{\bA_1 - \tilde{\bA}_1}_2 - 2 \norm{\bA_2}_2 - \norm{\bA_3}_2.
\end{split}
\end{equation}
In Theorem \ref{Lemma:Zhang_LB}, we show that for any $\zeta > 0$, $\Prob\left(\norm{\bA_2}_2 > \zeta\right) \to 0$ and $\Prob\left(\norm{\bA_3}_2 > \zeta\right)$, as $s_\delta,T \to \infty$. Therefore, it remains to show that $\Prob\left(\norm{\bA_1 - \tilde{\bA}_1}_2 > \zeta\right) \to 0$, as $s_\delta,T \to \infty$ on the extended probability space. Define $\bs_{v,t} = \sum_{j=1}^t\bepsilon_{v,j}$ and note that by application of Lemma \ref{Prop:idemp} and norm-consistency
\begin{equation*}
\begin{split}
&\norm{\bA_1 - \tilde{\bA}_1}_2 \leq \frac{s_\delta}{T^2}\norm{
\bB_{S_\delta,\perp}^\prime\bC_{S_\delta}\bD\left(\bE_u^\prime\bU^\prime\bU\bE_u - \bE_v^\prime\bU^\prime\bU\bE_v \right)\bD^\prime\bC_{S_\delta}^\prime\bB_{S_\delta,\perp}}_2\\
& \leq \frac{s_\delta}{T^2}\norm{\bD^\prime\bC_{S_\delta}^\prime\bB_{S_\delta,\perp}}_2^2\norm{\bE_u^\prime\bU^\prime\bU\bE_u - \bE_v^\prime\bU^\prime\bU\bE_v}_2= \frac{s_\delta}{T^2}\norm{\bD^\prime\bC_{S_\delta}^\prime\bB_{S_\delta,\perp}}_2^2\norm{\sum_{t=1}^T\bs_{u,t}\bs_{u,t}^\prime - \sum_{t=1}^T\bs_{v,t}\bs_{v,t}^\prime}_2,
\end{split}
\end{equation*}
where $\norm{\bD^\prime\bC_{S_\delta}^\prime\bB_{S_\delta,\perp}}_2^2 \leq \norm{\bD}_2^2\norm{\bC_{S_\delta}}_2^2\norm{\bB_{S_\delta,\perp}}_2^2 < \infty$, as $\norm{\bD}_2 \leq K$, by Assumption \ref{Ass:Dependence} and the normalization on $\bB_{S_\delta,\perp}$. By the proof of Lemma 9 in the supplement of \citet[][p.~3]{Zhang2018b},
\begin{equation*}
\norm{\sum_{t=1}^T\bs_{u,t}\bs_{u,t}^\prime - \sum_{t=1}^T\bs_{v,t}\bs_{v,t}^\prime}_2 = O_p\left(NT^{7/4}\right),
\end{equation*}
such that $\norm{\bA_1 - \tilde{\bA}_1}_2 = O_p\left(\frac{s_\delta N}{T^{1/4}}\right)$. Then by \eqref{eq:sigma_22_gen} a $\zeta > 0$ exists such that\\
$\Prob\left(\lambda_\min\left(\hat{\bSigma}_{22}\right) > \zeta\right) \to 1$, as $s_\delta,N,T \to \infty$ with $\frac{s_\delta N}{T^{1/4}} \to 0$.
\end{proof}

\subsection{Data Description}\label{App: Variables}
\small
\begin{longtable}{ccccc}
\hline 
Variable & groups & Translation & Inclusion & Differenced\tabularnewline
\hline 
\hline 
vakantiebaan & Job Search & holiday job & 100\% & N\tabularnewline
Unemployment & Y & Unemployment & 80\% & Y\tabularnewline
uwv vacatures & Job Search & uwv vacancies & 78\% & Y\tabularnewline
werkloos & Social Security & unemployed & 76\% & Y\tabularnewline
ww uitkering & Social Security & ww benefits & 72\% & Y\tabularnewline
Ww & Social Security & Ww & 69\% & Y\tabularnewline
nationale vacaturebank & RA & nationale vacaturebank & 59\% & Y\tabularnewline
cv maken & Application training & CV write & 57\% & Y\tabularnewline
indeed & RA & indeed & 52\% & Y\tabularnewline
jobtrack & RA & jobtrack & 52\% & Y\tabularnewline
motivatiebrief & Application training & motivation letter & 52\% & Y\tabularnewline
sollicitatiebrief schrijven & Application training & write application letter & 50\% & Y\tabularnewline
voorbeeld cv & Application training & example cv & 48\% & Y\tabularnewline
tempo team & RA & tempo team & 48\% & Y\tabularnewline
ontslagvergoeding & Social Security & severance pay & 46\% & Y\tabularnewline
ww uitkering aanvragen & Social Security & request unemployment benefits & 46\% & Y\tabularnewline
aanvragen uitkering & Social Security & request benefits & 44\% & N\tabularnewline
interin & RA & interin & 44\% & Y\tabularnewline
manpower & RA & manpower & 44\% & Y\tabularnewline
randstad & General & randstad (geographical area) & 44\% & Y\tabularnewline
werkzoekende & Social Security & job seeker & 43\% & Y\tabularnewline
job & General & job & 43\% & Y\tabularnewline
uwv & Social Security & uwv & 43\% & Y\tabularnewline
werk.nl & Job Search & werk.nl & 41\% & Y\tabularnewline
job vacancy & Job Search & job vacancy & 41\% & Y\tabularnewline
uitkering & Social Security & benefits & 41\% & Y\tabularnewline
ontslag & Social Security & resignation & 41\% & N\tabularnewline
vacature & Job Search & vacancy & 41\% & Y\tabularnewline
sollicitatiebrief voorbeeld & Application training & application letter example & 39\% & Y\tabularnewline
sollicitatie & Application training & application & 39\% & Y\tabularnewline
sollicitatiebrief & Application training & application letter & 39\% & Y\tabularnewline
uitzendbureau & RA & employment agency & 39\% & Y\tabularnewline
vakantiewerk & Job Search & holiday job & 37\% & N\tabularnewline
tence & RA & tence & 37\% & Y\tabularnewline
vacaturebank & Job Search & vacaturebank & 37\% & Y\tabularnewline
sollicitatiegesprek & Application training & application interview & 37\% & N\tabularnewline
tempo team uitzendbureau & RA & tempo team employment agency & 35\% & N\tabularnewline
motivatiebrief voorbeeld & Application training & motivation letter example & 35\% & Y\tabularnewline
bijstand & Social Security & social benefits & 35\% & Y\tabularnewline
open sollicitatiebrief & Application training & open application letter & 35\% & Y\tabularnewline
vrijwilligerswerk & General & volunteer work & 35\% & N\tabularnewline
werk nl & Job Search & werk nl & 35\% & N\tabularnewline
adecco & RA & adecco & 33\% & N\tabularnewline
creyfs & RA & creyfs & 33\% & Y\tabularnewline
randstad uitzendbureau & Job Search & randstad employment agency & 33\% & Y\tabularnewline
cv maken voorbeeld & Application training & write CV example & 31\% & Y\tabularnewline
werkbedrijf & Job Search & werkbedrijf & 31\% & Y\tabularnewline
tempo-team & RA & tempo-team & 31\% & Y\tabularnewline
werkloosheidsuitkering & Social Security & unemployment benefits & 31\% & N\tabularnewline
tempo team vacatures & RA & tempo team vacancies & 31\% & Y\tabularnewline
curriculum vitae voorbeeld & Application training & CV Example & 31\% & Y\tabularnewline
cv & Application training & cv & 31\% & N\tabularnewline
solliciteren & Application training & applying & 31\% & Y\tabularnewline
indeed jobs & RA & indeed jobs & 30\% & Y\tabularnewline
motivation letter & Application training & motivation letter & 30\% & N\tabularnewline
resume example & Application training & resume example & 28\% & N\tabularnewline
olympia uitzendbureau & RA & olympia employment agency & 28\% & Y\tabularnewline
tempoteam & RA & tempoteam & 28\% & Y\tabularnewline
randstad vacatures & Job Search & randstad vacancies & 26\% & Y\tabularnewline
banen & General & jobs & 26\% & N\tabularnewline
vrijwilliger & General & volunteer & 26\% & N\tabularnewline
baan & General & job & 26\% & N\tabularnewline
start uitzendbureau & RA & start employment agency & 24\% & Y\tabularnewline
jobnet & RA & jobnet & 24\% & N\tabularnewline
monsterboard & Job Search & monsterboard & 24\% & Y\tabularnewline
baan zoeken & Job Search & job search & 20\% & N\tabularnewline
functieomschrijving & General & job position description & 20\% & N\tabularnewline
resume template & Application training & resume template & 19\% & N\tabularnewline
omscholen & Application training & retraining & 19\% & Y\tabularnewline
job interview & Application training & job interview & 19\% & N\tabularnewline
werken bij & General & working at & 19\% & Y\tabularnewline
vacatures & Job Search & vacancies & 19\% & Y\tabularnewline
uwv uitkering & Social Security & uwv benefits & 17\% & Y\tabularnewline
job description & General & job description & 17\% & Y\tabularnewline
werk zoeken & General & job search & 17\% & Y\tabularnewline
jobs & General & jobs & 17\% & Y\tabularnewline
resumé & Application training & resume & 15\% & Y\tabularnewline
bijscholen & Application training & retraining & 15\% & N\tabularnewline
curriculum vitae template & Application training & CV Template & 13\% & N\tabularnewline
curriculum vitae & Application training & CV & 11\% & Y\tabularnewline
sollicitaties & Application training & applications & 9\% & Y\tabularnewline
werkeloos & Social Security & unemployed & 9\% & N\tabularnewline
werkloosheid & Social Security & unemployment & 4\% & N\tabularnewline
resume & Application training & resume & 2\% & N\tabularnewline
arbeidsbureau & RA & employment office & 2\% & N\tabularnewline
uitzendbureaus & RA & employment agencies & 2\% & Y\tabularnewline
werkloosheidswet & Social Security & unemployment law & 0\% & N\tabularnewline
\hline 
\end{longtable}

\end{small}
\end{appendices}
\end{document}